\documentclass[%
 aip,
 amsmath, amssymb,
 reprint
]{revtex4-1}

\usepackage{graphicx}
\usepackage{booktabs}
\usepackage{makecell}
\usepackage{bm}
\usepackage[utf8]{inputenc}
\usepackage[T1]{fontenc}
\usepackage{mathptmx}
\usepackage{etoolbox}
\usepackage{xcolor}
\usepackage{hyperref}
\usepackage{nicematrix}
\usepackage{listings}                 
\lstdefinestyle{jsonschema}{%
  basicstyle=\ttfamily\scriptsize,
  breaklines=true,
  columns=fullflexible,
  keepspaces=true,
  showstringspaces=false,
  mathescape=false,
  texcl=false,
  frame=single,
  framesep=3pt,
  xleftmargin=3pt,
  aboveskip=4pt,
  belowskip=2pt,
}
\hypersetup{colorlinks=true, allcolors=black}
\graphicspath{{./}{../}}

\begin{document}

\preprint{AIP/APL-ML}

\title[APS-RAG]{A corrective agentic hybrid RAG and an operations-grounded\texorpdfstring{\\}{ }evaluation for a scientific facility} 

\author{Rajat Sainju}
\email[Author to whom correspondence should be addressed: ]{rsainju@anl.gov}
\author{Dariusz Jarosz}
\author{Hairong Shang}
\author{Michael Prince}
\author{Ryan M. Aydelott}
\author{Mathew J. Cherukara}
\author{Yine Sun}
\author{Michael Borland}
\affiliation{Advanced Photon Source, Argonne National Laboratory, Lemont, Illinois 60439, USA}

\date{\today}

\begin{abstract}
Scientific user facilities accumulate decades of operational knowledge that no single search index covers: electronic logbooks, technical documents, internal wikis, operations chat messages, maintenance records, and live control-system data. We present \textsc{APS-RAG}---Advanced Photon Source Retrieval Augmented Generation---a deployed platform that makes the institutional knowledge at the \text{Advanced Photon Source} (APS) accessible to staff through natural-language queries, along with an operations-grounded evaluation. The retrieval engine fuses dense, sparse, and knowledge-graph (KG) channels with query-type-adaptive reciprocal-rank fusion, adds a corrective agentic loop, and runs a native-tool ReAct executor over a Model Context Protocol (MCP) tooling layer. We construct APS-Bench, a 50-question, question-answering (QA) dataset with auditable gold answers. Every retrieval-augmented variant numerically improves strict vital-nugget recall over a naive BM25 baseline (63.8\%~$\rightarrow$~65.5--70.3\%), with the full corrective Agentic GraphRAG scoring (70.3\%). The cross-encoder reranker contributes significantly to the answer quality: removing it and allowing the LLM to score the relevance drastically reduces strict vital recall by 32.8\% (95\% CI $[-47.4,-19.1]$; $p<10^{-4}$). The graph channel and corrective loop contribute positively as expected, but the performance gains are marginal. Additionally, we also compare the performance of open-source and closed-source LLMs in final answer synthesis. We release the APS-Bench construction methodology, the six-layer evaluation harness, and the underlying codebase, along with the \textit{/aps-rag} retrieval agent skill framework, to support reproduction and adoption at other facilities. Together, the deployed platform and its operations-grounded evaluation present a promising workflow for trustworthy, statistically grounded AI assistance in facility operations, transferable to other large scientific instruments.
\end{abstract}

\maketitle 
\section{Introduction}\label{sec:intro}
Large-scale scientific user facilities operate continuously and accumulate operational knowledge in forms that are usually not available through a unified search interface. These include scientific logbooks, internal technical documentation, manuals, policy documents, maintenance work requests, operations chat messages, downtime records, internal group wikis, and live control-system process variables (PVs). For an X-ray light source such as the Advanced Photon Source (APS)\cite{galayda:pac95-mad02,borl:ipac18-thxgbd1}, which operates over decades, the volume of knowledge required for operation, maintenance, and improvement accumulates at a rate that exceeds the capacity of any individual to absorb it. Much of the knowledge is tacit, residing within the records of specialized groups (e.g., Controls, Diagnostics, Radio Frequency, Magnetics, Beamline Controls and Data Acquisition). For example, a beam-loss event may be documented by a operations staff in BELY, the APS's scientific logbook~\cite{jarosz_logging_nodate}; the subsystem fault that caused it may have been filed in a maintenance work request days earlier; the resolution lives in an operator's wiki; the subsystem specification is a technical document in the Integrated Content Management System (ICMS); and the relevant setpoint is a live process variable accessed via the archiving system like EPICS~\cite{shankar2015epics}. This fragmentation imposes high costs. When solutions to recurring problems cannot be found quickly, downtime lengthens; when experienced staff members rotate off shift or retire, the reasoning behind past decisions can leave with them. Effective knowledge management is therefore essential for minimizing downtime and preserving institutional knowledge at large-scale scientific facilities.

\begin{figure*}[t]
    \centering
    \includegraphics[width=0.95\linewidth]{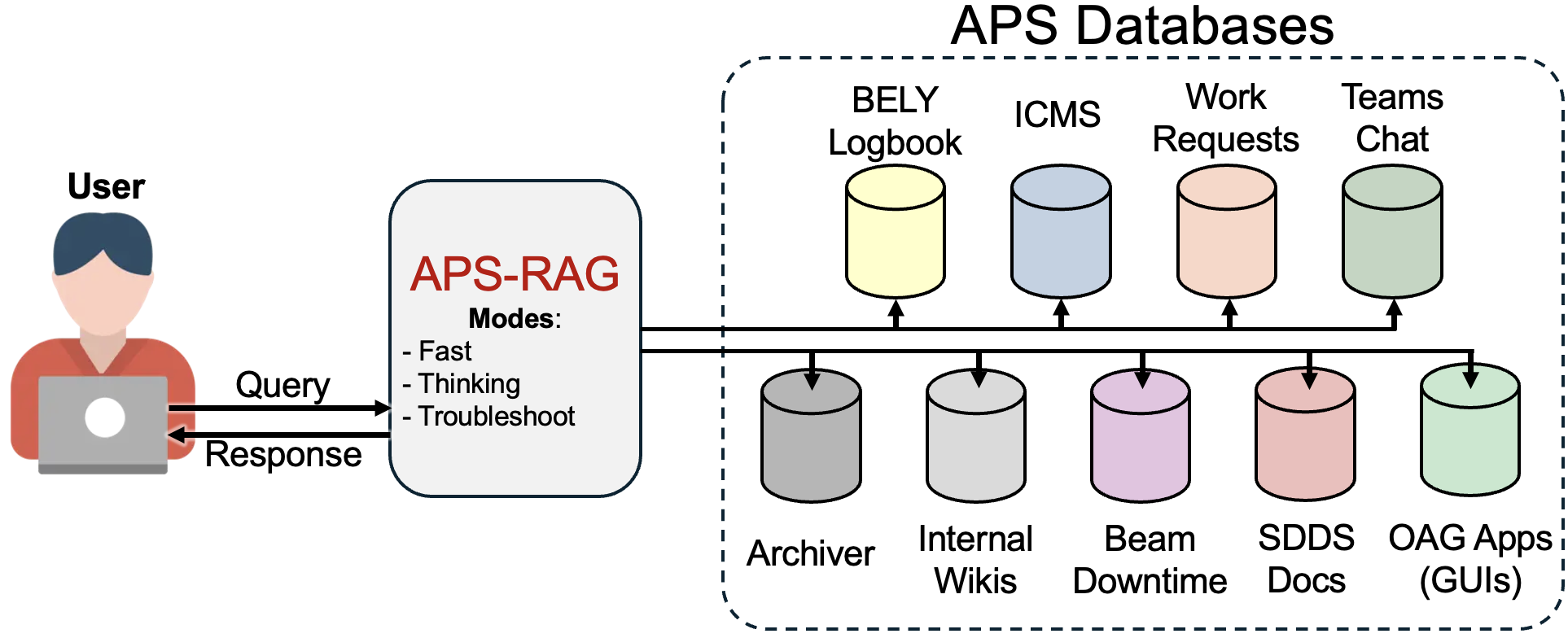}
    \caption{APS-RAG deployment overview. Through a single natural-language interface, APS staff can query the facility's dispersed operational record. The platform pools information from nine distinct APS data sources: eight indexed knowledge bases are kept current by a daily diff-and-upsert pipeline, and the EPICS Archiver is not pre-indexed and is accessed at query time for live and historical process-variable data through the Model Context Protocol (MCP) tooling layer. Three user-selectable query modes are available: \emph{APS-RAG (Fast)} for quicker responses, \emph{APS-RAG (Thinking)}, which enables the corrective agentic loop, and \emph{AI Troubleshoot}, which retrieves knowledge-graph-targeted fault-resolution steps.}
    \label{fig:fig1_overview}
\end{figure*}

Different research groups at APS require tailored access to operational knowledge. For example, scientists preparing machine studies require provenance information and configuration details of previous measurements. Engineers designing subsystem upgrades depend on controlling specifications and the history of related modifications. Technicians diagnosing faults at 3~a.m. need to know how the same symptom was resolved the last three times it occurred. Although such information is recorded, locating it necessitates familiarity with the relevant systems and their query mechanisms. A large language model (LLM) interface could address this challenge by enabling natural language queries~\cite{gu2024survey, brown_language_2020}; however, an LLM alone lacks facility-specific knowledge and may generate fluent yet unsupported responses, which is unsuitable for operational decision-making. Retrieval-augmented generation (RAG) addresses this limitation by grounding responses in passages retrieved from facility records rather than relying solely on LLM's memory~\cite{lewis_retrieval-augmented_nodate,gao_retrieval-augmented_2024}. This approach supplies essential operations knowledge, enhances factual accuracy, and enables source citation for each LLM-synthesized answer. Thus, minimizing unsupported or speculative claims.

In this work, we present APS-RAG (Fig.~\ref{fig:fig1_overview}), a deployed agentic knowledge-retrieval platform that transforms natural language queries from staff into a source-cited answer drawn from the facility's dispersed databases. To evaluate the system's performance, we constructed a benchmark dataset, APS-Bench, using a methodology similar to InPars~\cite{bonifacio_inpars_2022}. This enables us to generate question-and-answer pairs from the operational corpus and to go beyond the usually reported recall-style or qualitative results. During the last two-month deployment, the system processed an average of eight queries per day. The system was iteratively improved based on user feedback. The primary contribution of this study is the operational integration of established techniques, including hybrid retrieval that combines dense~\cite{karpukhin2020dpr}, sparse~\cite{robertson_probabilistic_2009}, and knowledge graph~\cite{edge2024graphrag} retrieval; adaptive reciprocal-rank fusion (RRF)~\cite{cormack_reciprocal_2009}, corrective retrieval~\cite{yan2024crag}, a persistent Model Context Protocol (MCP) layer that combines retrieved documents with live facility data and generates code for statistics-related queries (e.g., "How many work requests entered in year 2026 exist for kickers?"), and nugget-based evaluation~\cite{pradeep2024autonuggetizer}. We also report controlled ablations of each component. We find that agentic RAG shows great promise for fast, accurate, and reproducible actionable knowledge retrieval in operational settings.

\section{Related work}\label{sec:related}
APS-RAG sits at the intersection of three rapidly developing lines. \emph{General RAG.} We build on retrieval-augmented generation~\cite{lewis2020rag,gao2023ragsurvey} with dense retrieval~\cite{karpukhin2020dpr}, BM25~\cite{robertson_probabilistic_2009}, reciprocal-rank fusion~\cite{cormack_reciprocal_2009}, and cross-encoder reranking~\cite{nogueira2020rerank}; on knowledge-graph-augmented RAG~\cite{edge2024graphrag,guo2024lightrag,gutierrez2024hipporag} (surveyed in Ref.~\cite{peng2024graphragsurvey}; Ref.~\cite{xiang_when_2026} analyzes when graph retrieval helps); and on agentic and self-correcting RAG --- corrective RAG~\cite{yan2024crag}, Self-RAG~\cite{asai2024selfrag}, Adaptive-RAG~\cite{jeong_adaptive-rag_2024}, and the ReAct paradigm~\cite{yao_reac_2023} (surveyed in Ref.~\cite{singh2025agenticsurvey}) --- over a Model-Context-Protocol tooling layer~\cite{anthropic2024mcp}. We evaluate with nugget recall~\cite{pradeep2024autonuggetizer}, a claim-level faithfulness metric inspired by FaithJudge~\cite{tamber2025faithjudge} with an HHEM-2.1 backstop~\cite{vectara2024hhem}, and a RAGAS-style proxy~\cite{es2024ragas}, with benchmark questions generated following the InPars ~\cite{bonifacio_inpars_2022} method. Similar automated self-data curation has been shown to be effective for adapting and evaluating RAG systems in a new domain~\cite{mao_rag-studio_2024}.

\emph{Accelerator and scientific-facility RAG.} Machine learning methods are now well established for accelerator control and diagnostics, as documented in recent lterature~\cite{edelen2018opportunities,kaiser2023rlbo,humble_resilient_2024,blokland_uncertainty_2022,rajput2023errant,lobach_recurrent_2024,edelen_anomaly_2021}. More recently, large language models have been applied to facility operations, with several systems pairing an LLM with document retrieval and tool integration. Notable examples include CALMS at Argonne~\cite{prince_opportunities_2024}, GAIA at DESY/European XFEL~\cite{mayet_gaia_2024}, a RAG summarization agent for the Electron-Ion Collider~\cite{suresh_towards_2024}, a multi-facility logbook-RAG effort~\cite{sulc_towards_2024} with companion logbook-retrieval and analytics studies~\cite{maldonado_enhancing_nodate}, and a RAG demo for technical documentation ~\cite{stuhlmann_intelligent_nodate}. Relative to all of these, APS-RAG adds query-type-adaptive dense+sparse+knowledge-graph fusion, a corrective loop with a knowledge-graph-grounding trigger and a self-critique gate, and --- most importantly --- a domain-specific operational evaluation, where prior facility systems report recall-style or qualitative results. A second, adjacent line targets control rather than knowledge access: agentic experiment execution at the Advanced Light Source~\cite{hellert2026agentic}, agents that operate scientific instruments~\cite{vriza_operating_2026}, the agentic-control vision of Ref.~\cite{sulc2024agentic}, and natural-language tuning~\cite{kaiser_large_2025}. We share the plan-first, bounded-tool design philosophy of Ref.~\cite{hellert2026agentic}, but answer questions and read instruments rather than steer the machine. The nearest methodological analog, agentic hybrid RAG for muon-collider analysis~\cite{jiang_agentic_2026}, likewise combines hybrid retrieval with agentic query decomposition and a benchmark that carries key points and unanswerable questions, but targets the offline high-energy-physics literature. The systems above do not contain a knowledge-graph fault-chain channel and do not report claim-level faithfulness under cross-family judging, which is part of what we show in this work. To our knowledge, APS-RAG is the first accelerator-operations RAG system to add cross-database search, allowing users to access multiple institutional knowledge bases through one natural language interface. 

\begin{figure*}[bt]
    \centering
    \includegraphics[width=0.95\linewidth]{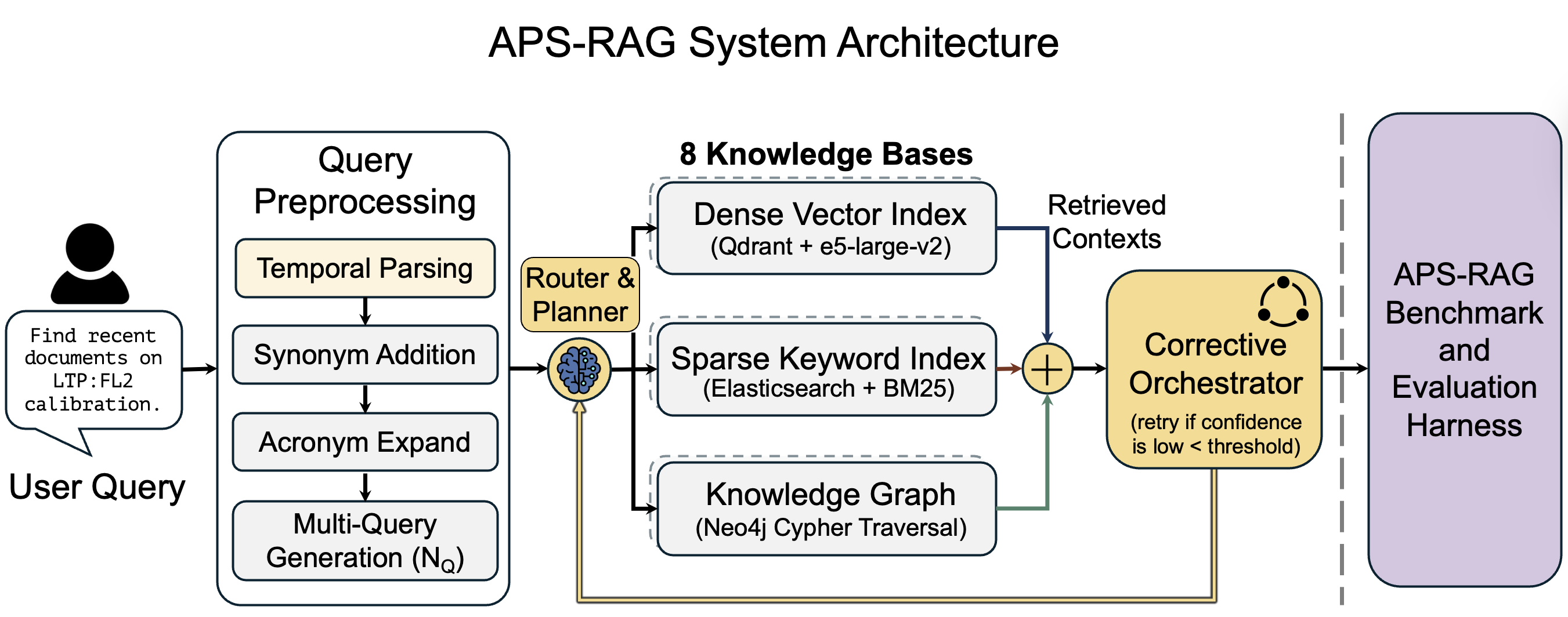}
    \caption{APS-RAG system architecture. A user query first undergoes a preprocessing stage --- temporal parsing, synonym addition, acronym expansion, and multi-query generation --- after this a router and planner which includes query intent detector (e.g., factual, procedure, multi-hop) distributes the query across eight knowledge bases through three retrieval channels: a dense vector index (Qdrant~\cite{qdrant}, e5-large-v2 embeddings), a sparse keyword index (Elasticsearch~\cite{elasticsearch}, BM25), and a Neo4j~\cite{neo4j} knowledge graph queried via Cypher. Contexts retrieved from these channels are merged through query-type-adaptive reciprocal-rank fusion, reranked by a cross-encoder, and forwarded to a corrective orchestrator, which retries retrieval with a broadened query whenever answer confidence falls below threshold. This same pipeline is exercised end-to-end by the APS-RAG benchmark and evaluation harness.}
    \label{fig:fig2_arch}
\end{figure*}

\section{The APS-RAG platform and system architecture}\label{sec:platform}
\subsection{APS databases}\label{sec:corpus}
APS-RAG connects to nine databases that collectively encompass the facility's written operational knowledge. An overview of these connected sources is shown in Fig.~\ref{fig:fig1_overview}. Five of these databases correspond to live operational systems that staff interact with daily. \emph{BELY}~\cite{jarosz_logging_nodate} serves as APS's scientific logbook and maintains a time-ordered record of shift entries, beam-loss events, machine experiments, and troubleshooting notes. The \emph{Integrated Content Management System (ICMS)} functions as the facility's controlled-document repository for procedures, specifications, technical notes, and reports. The \emph{Work Request system} tracks maintenance and equipment-replacement jobs from request through completion; the operational \emph{Microsoft Teams chat history} captures the routine operational communication among operators, engineers, and scientists, and the \emph{internal wikis} which are group- and subsystem-maintained reference web pages.) Three additional sources provide supporting documentation: the \emph{MCR beam downtime records} (a curated, structured record of downtime events and their descriptions, complementary to the BELY entries), the \emph{SDDS documentation} set~\cite{borland_users_1995} (current SDDS user guides), and the \emph{OAG codebase and application documentation}, which describes operator-facing graphical user interface and tool descriptions. The ninth source, the EPICS Archiver, contains live and historical process-variable data. This source is accessed at query time through the Model-Context-Protocol tooling layer, as its data are continuous and timestamped rather than document-based, and thus are not pre-indexed. Table~\ref{tab:corpus-size} summarizes the document, character, and token counts for each source.

\begin{table}[t]
    \centering
    \caption{Composition of the APS-RAG retrieval corpus. Each row corresponds to one institutional data source ingested and stored as an indexed collection. Character counts reflect raw UTF-8 text extracted after document preprocessing; token counts are computed with the GPT-4/cl100k byte-pair-encoding tokenizer.}
    \vspace{2mm}
    \label{tab:corpus-size}
    \begin{tabular}{lccc}
        \hline
        Collection & \makecell{Unique documents \\ (ct)} &  \makecell{Characters \\ (ct)} & \makecell{Tokens \\ (GPT-4/cl100k)} \\
        \hline
        BELY & 4.5k & 14.0M & 4.2M \\
        ICMS & 5.8k & 72.0M & 20.6M \\
        Work Requests & 17.3k & 3.0M & 0.8M \\
        Teams Chat & 9.4k & 2.4M & 0.8M \\
        Internal Wikis & 0.2k & 0.8M & 0.1M \\
        Beam Downtime & 103.0k & 7.9M & 1.9M \\
        SDDS & 0.6k & 1.4M & 0.4M \\
        OAG Apps & 0.7k & 0.8M & 0.2M \\
        \hline
    \end{tabular}
\end{table}

\subsection{Data ingestion and indexing}\label{sec:ingestion}
Database-backed sources such as BELY, ICMS, Work Requests, and Beam downtime records are pulled automatically via REST APIs and export libraries. The APS wikipage contents are acquired using a dedicated web-crawler that traverses the internal wiki's page-and-attachment graph, removing navigation boilerplate. We have a separate in-house pipeline that performs optical character recognition (OCR) on Teams conversation screenshots to extract the conversation thread ~\cite{pymupdf, smith2007overview}. SDDS has standard documents, while the OAG Apps codebase and usage description were included as markdown. Each ingested record is converted into a collection-specific representation with source-specific metadata (shown in Figs.~S1 and~S2 of the supplementary material). Attachments, such as figures and schematics, are linked to their parent records. The deployed ingestion layer is broader than those included in the evaluation. The benchmark corpus and all results in this paper cover six of these collections (BELY, ICMS, Work Requests, Teams, SDDS, OAG Apps), whereas the APS wiki and beam-downtime records were added to the production index only after the evaluation corpus was frozen. Consequently, these sources are present in the deployed system but are excluded from the benchmark, thereby preserving the frozen snapshot that underpins the reported results.

\paragraph{Per-source schemas:}\label{schema_para}
A normalized schema was developed for each source to preserve the fields required for filtering, temporal scoping, citation, and to preserve the recording conventions specific to each contributing team. Then, it maps the free-text fields to the indexed body (see Fig.~S1 of the supplementary material). The schemas retain native identifiers, such as BELY\_ID, WRQ\_ID, ICMS\_ID, and likewise for other sources, ensuring that an inline citation resolves back to the source record and its canonical web URL. The inclusion of inline citations enables any staff member to verify the answer at their original sources.

\paragraph{Data freshness via daily diffs and upserts:}\label{fresh_para}
A scheduled daily pipeline maintains index freshness without requiring a full corpus rebuild. For each data source, the pipeline performs an incremental diff operation, comparing the content hash of each record against a persisted per-document synchronization state. This process ensures that only new or modified records are re-fetched. Updated records are upserted, using the document ID as the key, into both the dense vector store (Qdrant~\cite{qdrant} and the keyword-based store (Elasticsearch~\cite{elasticsearch}). Simultaneously, an incremental update is applied to the Neo4j knowledge graph for the corresponding documents. Deletions are similarly propagated based on hash comparison. Consequently, the computational cost of re-indexing is determined by the daily change set rather than the entire corpus. This design ensures that time-sensitive information, such as recent shift logs, open work requests, and current downtime, remains consistent and up-to-date across all three storage backends.

\paragraph{Parent-child indexing and embeddings:}\label{parent-child_para}
Retrieval is conducted over two complementary channels on a shared chunk store. Documents use a two-level parent-child scheme that decouples the unit embedded or scored from the unit returned~\cite{sarthi2024raptorrecursiveabstractiveprocessing}. A parent record retains the full context of a complete record or section, such as a complete logbook file or technical report, while each parent is split into atomic children through recursive character-based splitting, using the uniform two-stage sizes given in Sec.~\ref{sec:adaptive} and a 200-character overlap. Each child is prepended with a concise header containing the source file\_id, document title, and date, ensuring self-descriptiveness. Both retrieval channels utilize the same set of children. In the dense retrieval channel, each child is encoded as a 1024-dimensional vector using e5-large-v2~\cite{wang2022e5} and stored in Qdrant for approximate nearest neighbor search. The e5 model was selected for its robust zero-shot retrieval performance on out-of-domain text, which is particularly relevant for the terse, acronym-rich language characteristic of accelerator operations. In the sparse retrieval channel, the same children are indexed in Elasticsearch and scored using BM25. This redundancy is intentional: dense retrieval enables recovery of paraphrased or semantically related entries, while BM25 provides reliable matching for exact identifiers such as process-variable names (e.g., \textit{LTS:H1:CurrentAO}, \textit{S-DCCT:CurrentM}), sector names, and equipment labels, which are often not captured by dense encoders. As a result, matching is achieved at fine granularity, while the enclosing parent record is supplied to the generator to provide necessary context for synthesis.

\begin{figure*}[t!]
    \centering
    \includegraphics[width=0.8\textwidth]{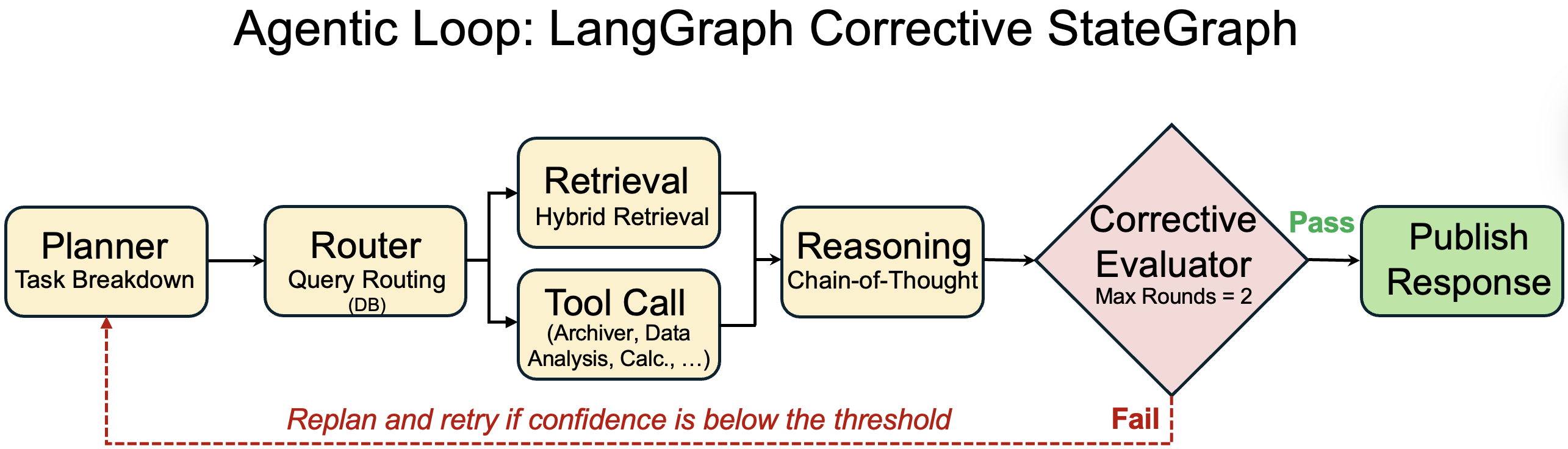}
    \caption{State-machine representation of the corrective agentic loop (LangGraph). Nodes are atomic operations: a planner decomposes the task, a router dispatches it to hybrid retrieval and, when needed, MCP tool calls (Archiver, data analysis, image search), and a chain-of-thought reasoning step drafts the answer, which the corrective evaluator scores via Eq.~\eqref{eq:score}. On a pass, the response is published; on a fail (dashed edge), control returns to the planner, which replans and retries with an escalated retrieval strategy if confidence falls below the configured threshold or more than 50\% of the retrieved documents fail the relevance evaluation. The schematic shows the state machine's maximum-rounds parameter at five; the deployed configuration caps corrective rounds at two (Sec.~\ref{sec:corrective}).}
    \label{fig:fig3_corrective}
\end{figure*}

\subsection{APS-RAG architecture overview}\label{sec:arch}
APS-RAG, as shown in Fig.~\ref{fig:fig2_arch}, turns a user's natural-language query into a source-cited answer drawn from the facility's dispersed operational databases. A question is routed through one of three service modes: \textit{APS-RAG (Fast)}, a single pass through the pipeline; \textit{APS-RAG (Thinking)}, which enables bounded corrective retrieval rounds (Sec.~\ref{sec:corrective}); and \textit{AI Troubleshoot}, for fault-resolution queries (Sec.~\ref{sec:kg}). An entered query is first preprocessed by (i) detecting the query intent - factoid (single value), range (temporal), multi-hop, procedural, comparison, or diagnostic; (ii) resolving the search date range; (iii) adding synonyms for domain keywords; (iv) expanding acronyms while retaining the original acronym; and (v) generating $N_q$ paraphrases (multi-query generation; $N_q=5$ in the deployed configuration).
Depending on the query and selected mode, the retrieval channels run in parallel across the indexed corpus: a dense vector store, a sparse keyword store, and optionally a knowledge graph. Their ranked lists are merged by reciprocal-rank fusion~\cite{cormack_reciprocal_2009} with query-type-adaptive weights, reranked by a cross-encoder over the top 50 fused candidates (each channel retrieves top-$k=100$ before fusion; \textit{reranker\_top\_k}$=50$ in the deployed configuration), and passed to a frontier LLM served through Argonne's ARGO gateway (provides data-secure programmable access to multiple LLMs exclusively for internal users), which synthesizes an answer whose inline citations link directly to the original records. The dense-plus-sparse hybrid with adaptive fusion and reranking is the deployed default; the corrective orchestration loop and the Neo4j knowledge-graph channel are deployed as optional modes and are described in Secs.~\ref{sec:corrective} and~\ref{sec:kg}.

\subsection{Adaptive three-way retrieval fusion}\label{sec:adaptive}
\begin{figure}[htpb]
    \centering
    \includegraphics[width=0.5\linewidth]{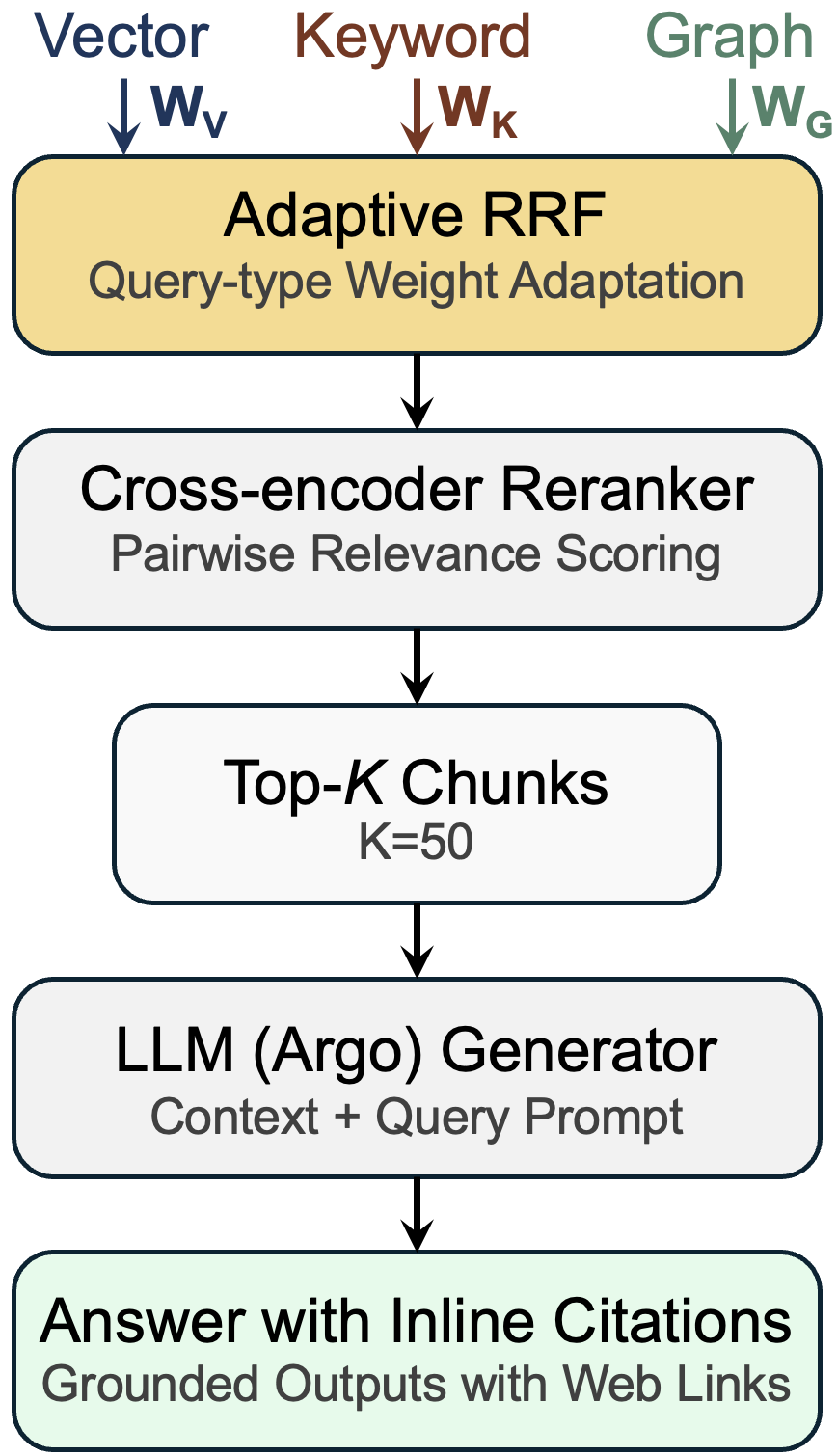}
    \caption{The APS-RAG retrieval and generation pipeline comprises three retrieval channels: a dense vector channel implemented with Qdrant and e5-large-v2, a sparse keyword channel based on Elasticsearch and BM25, and an optional knowledge-graph channel utilizing Neo4j. Each channel returns candidate passages, which are merged using reciprocal-rank fusion with query-type-adaptive weights. The merged candidates are subsequently reranked by a cross-encoder that jointly evaluates each passage with the query. Controlled ablation experiments indicate that this reranking stage strongly contributes to APS-RAG's performance. The highest-ranked passages, together with the query, are then provided to a frontier large language model accessed via Argonne's ARGO gateway, GPT-5.4. This model synthesizes an answer that includes inline citations, linking each claim to its corresponding source passage and web-accessible record.}
    \label{fig:fig4_retrive-generate-pipeline}
\end{figure}

The dense, sparse, and graph channels are combined by reciprocal-rank fusion~\cite{cormack_reciprocal_2009,bruch_analysis_2024}: a document $d$ at rank $r_c(d)$ in channel $c$ receives the fused score
\begin{equation}
\mathrm{RRF}(d)=\sum_{c}\frac{w_c}{k+r_c(d)}, \qquad k=60,
\label{eq:rrf}
\end{equation}
where the sum runs over the channels in which $d$ is retrieved, and $k$ is the smoothing constant. The three channels combining via RRF is shown in Fig.~\ref{fig:fig4_retrive-generate-pipeline}. What distinguishes APS-RAG is that the channel weights $W_V$ (vector), $W_K$ (keyword), and $W_G$ (knowledge graph) are not fixed. A query router classifies each request into one of eight operational intents, and each intent maps to its own $(\text{dense}/\text{BM25}/\text{graph})$ profile: causal, multi-hop, and troubleshooting queries up-weight the graph channel (weights $0.3/0.2/0.5$), factual, and specific value queries up-weight BM25 ($0.3/0.5/0.2$, exploiting the literal date and shift tokens that lexical search handles well), and mathematical or code queries up-weight the dense channel. The actual weight setting for various query intents is tabulated in Table~S3. The weights are hand-set from operational experience rather than learned. The performance on different query types is presented in Fig.~S3 of the supplementary material. It shows that troubleshooting and multi-hop questions still seem to be areas where the answers have alleviated hallucination and lower faithfulness scores. Note: A re-retrieval with an expanded time range is triggered when no records are found within the defined time range.

\subsection{Corrective agentic loop and self-critique gate}\label{sec:corrective}
The orchestrator is a LangGraph~\cite{langgraph2024} state machine (Fig.~\ref{fig:fig3_corrective}) whose corrective cycle scores each draft answer as
\begin{equation}
s = 0.4\,\rho + 0.3\,g + 0.3\,c - 0.05\,n_{\text{KG}},
\label{eq:score}
\end{equation}
where $\rho$ is the relevance of the retrieved evidence, $g$ the grounding of the draft in that evidence, $c$ its completeness, and $n_{\text{KG}}$ the number of answer entities that cannot be verified against the knowledge graph. Similar to the fusion weights of Sec.~\ref{sec:adaptive}, the coefficients are manually tuned from operational experience: relevance weighted highest, grounding and completeness equally, and a small per-entity penalty for unverifiable knowledge-graph mentions rather than learned or tuned; they were fixed before either benchmark was scored.

Re-retrieval is triggered when the score $s$ falls below a configured threshold, with $\tau=0.4$ used in the deployed configuration. The default threshold and the impact of the corrective trigger on an ablation subset are presented in Fig.~S4 of the supplementary material. Substantive improvements in RAG answer quality are observed at the threshold and after $\tau=0.3$. The corrective trigger is activated when the supported fraction of retrieved evidence is less than one-half, or when at least three knowledge-graph entities remain unverified.

Upon trigger activation, the retrieval strategy advances to a more intensive loop. For instance, a stalled three-way search is replaced by a dense-only retrieval strategy, and the query is broadened using an LLM-generated reformulation for up to two additional rounds. After these iterations, the orchestrator returns the best available draft with a floored confidence value, thereby preventing indefinite looping. Three answer metrics were evaluated: strict vital recall ($V_{\text{strict}}$), and faithfulness~\cite{pradeep2024autonuggetizer}. These metrics increase modestly from $\tau = 0.3$ and plateau near $\tau = 0.7$ to $0.8$, while the answer change rate saturates at $\tau=0.5$. Prior to this comparatively expensive corrective loop, a self-critique gate is applied. A faster model (Claude Haiku~4.5, served through ARGO) assesses whether the draft appropriately answers the query, is well-cited, and is concise. If the draft fails this assessment, a single targeted rewrite is triggered. This gate is designed to resolve straightforward failures at a fraction of the computational cost required for a full corrective round.

\begin{figure*}[htbp]
    \centering
    \includegraphics[width=1.0\linewidth]{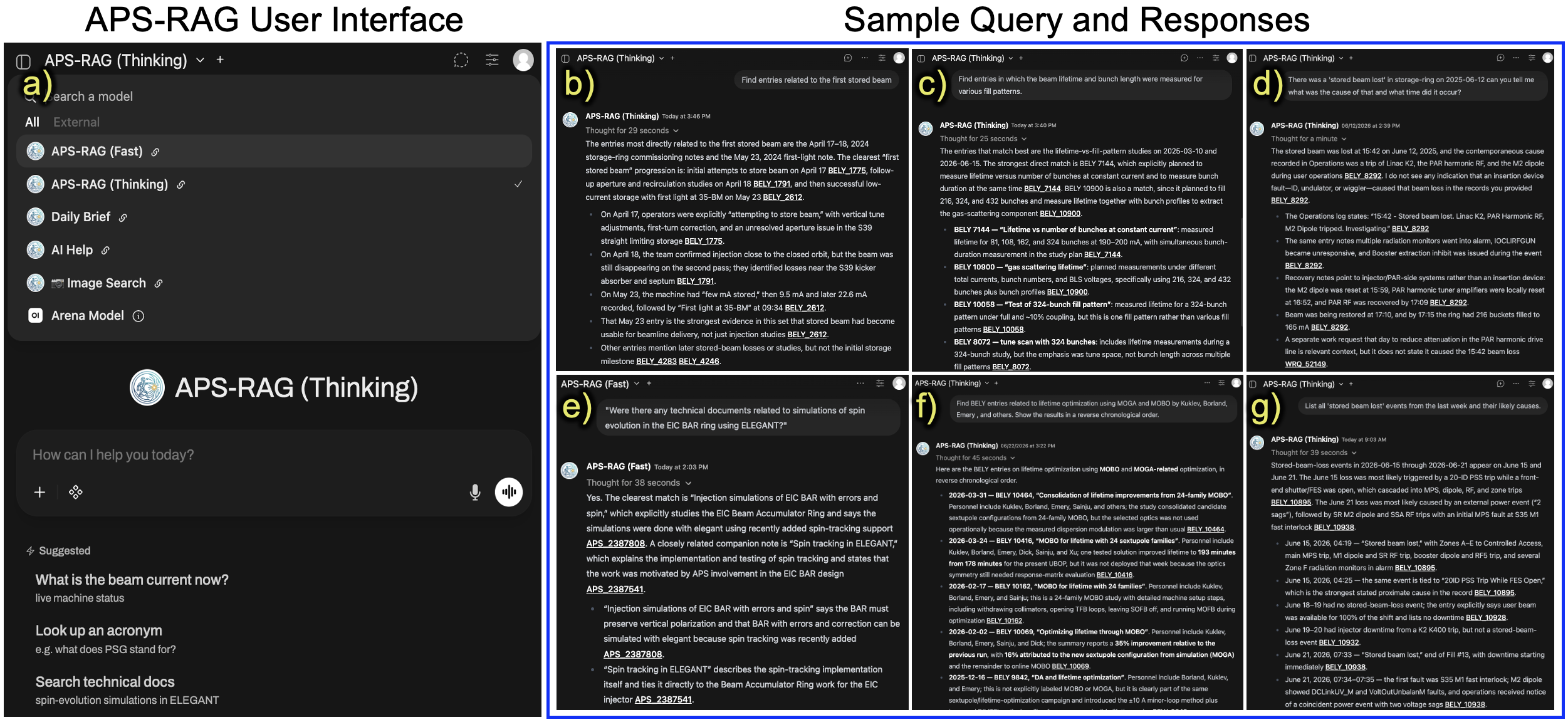}
    \caption{The APS-RAG interface offers multiple retrieval modes, including Fast and Thinking, as well as image-search capabilities and scheduled-summary and help assistants. The interface provides suggested entry points for tasks such as live machine-status lookups, acronym resolution, and technical-document searches. Six representative query–response pairs demonstrate the system's performance on operational queries, such as (b) tracing the first-stored-beam commissioning sequence, (c) identifying beam-lifetime-versus-fill-pattern studies, (d) diagnosing stored-beam-loss events, (e) retrieving technical documentation, (f) history of lifetime optimization, and (g) finding the causes of stored beam lost over a week's time. Each response contains inline citations to underlined record identifiers, including BELY, Work Request, and ICMS file IDs, which link directly to the relevant logbook entry, work request, or technical document. This design allows for systematic verification of the provenance of each response. The outputs presented are generated by the deployed system and illustrate the response format and citation traceability.}
    \label{fig:fig5_interface}
\end{figure*}

\subsection{A troubleshooting-oriented knowledge graph}\label{sec:kg}
We built an offline pipeline to extract entities and relations from six collections with a custom two-stage LLM extractor, loading the results into Neo4j. The graph contains 96,517 nodes and 84,222 edges across 14 types, including Fault, Component, System, and LogEntry. Retrieval supports entity lookups, path-based traversals, and shortest-path searches to reconstruct cause-to-effect-to-resolution chains beyond what flat vector retrieval can achieve. Graph analytics use native Cypher queries in Neo4j Community Edition, with degree-centrality and community-membership proxies and exact shortest-path traversal \cite{edge2024graphrag}. These serve as proxies, not a full analytics stack. The graph is purpose-built to reconstruct troubleshooting procedures from historical records. We connect LogEntry, Fault, Component, System, Procedure, WorkRequest, and beam-downtime nodes via cause-and-resolution relations (CAUSED\_BY, RESOLVED\_BY, HAS\_FAULT, ABOUT\_SYSTEM, FOLLOWS\_PROCEDURE, CONTACT\_PERSON). This enables tracing symptom-to-cause-to-action-to-resolution chains, mirroring how experienced staff resolve issues. The system retrieves similar historical fault-resolution chains and generates actionable, citation-backed troubleshooting steps based on proven solutions. This is especially effective for causal, multi-hop, and troubleshooting queries where evidence is distributed across records and vector similarity is insufficient. We assess this using the fault-chain-completeness metric (Sec.~\ref{sec:results}), prioritizing fault-chain coverage over complete procedural correctness.

\subsection{Native-tool ReAct and the MCP tooling layer}\label{sec:mcp}
APS-RAG executes tools using a dual-path ReAct executor, which prioritizes a native function-calling approach while retaining a prompt-based fallback for legacy compatibility. The system integrates eight tools, provided by four FastMCP servers (archiver, graph, compute, and image; see Table~S1 in the supplementary material), each operating behind a standardized error envelope and supported by a persistent session pool. This design amortizes the substantial index loading overhead, which would otherwise significantly increase per-call latency. The resulting architecture allows APS-RAG to synthesize live PV readings with document-based evidence to generate unified responses. As this capability is not reflected in the primary evaluation metrics, further details are provided in Sec.~S1 of the supplementary material. Deployment within operational facility infrastructure necessitated ongoing reliability engineering to address infrastructure drift, including gateway migrations, modifications to response envelopes, and silently disabled preprocessing chains. A summary of representative incidents and the corresponding regression safeguards is presented in Sec.~S3 of the supplementary material.

\subsection{The deployed interface and representative query and responses}\label{sec:interface}
The pipeline of Secs.~\ref{sec:arch}--\ref{sec:kg} reaches staff through the OpenWebUI ~\cite{baek2025openwebuiopenextensible} chat interface shown in Fig.~\ref{fig:fig5_interface}(a): a mode selector exposes \textit{APS-RAG (Fast)} and \textit{APS-RAG (Thinking)} alongside companion assistants (image search, a scheduled daily-brief summarizer, and a help assistant), and suggested entry points advertise the most common needs live machine status (answered through the tooling layer of Sec.~\ref{sec:mcp}), acronym resolution (the query-enhancement chain of Sec.~\ref{sec:arch}), and technical-document search. Panels~(b)-(g) of Fig.~\ref{fig:fig5_interface} show six live query--response pairs that span the query-type taxonomy of Sec.~\ref{sec:benchmarks} and make the preceding subsections concrete. Panel~(b) reconstructs the first-stored-beam commissioning sequence as a dated narrative stitched from multiple logbook entries, from the April 2024 injection attempts through ``first light at 35-BM'' (BELY\_2612). Panel~(c) answers a parametric study question by retrieving and differentiating four beam-lifetime-versus-fill-pattern studies (81--432 bunch configurations). Panel~(d) is the fault-chain behavior Sec.~\ref{sec:kg} targets: for a stored-beam-loss query, it returns the proximate cause and timestamp, the step-by-step recovery timeline, and a linked maintenance record (WRQ\_52149) that it flags as \emph{related context while declining to assert it as the cause}, alongside an explicit statement that the records do not support an insertion-device attribution---the citation discipline the self-critique gate of Sec.~\ref{sec:corrective} is designed to enforce. Panel~(e) shows Fast-mode technical-document lookup returning two linked ICMS reports on ELEGANT spin-evolution simulations. Panel~(f) executes a personnel- and method-scoped search (MOGA/MOBO lifetime optimization) and honors the requested reverse-chronological ordering with per-entry quantitative outcomes. Panel~(g) resolves the relative window ``last week'' into absolute dates (the temporal-parsing step of Sec.~\ref{sec:arch}), aggregates the stored-beam-loss events in it with their likely proximate causes, and states plainly which days had none. The common thread is traceability: every claim carries an inline citation to a native record identifier (BELY, WRQ, or ICMS ID) that resolves to the source record, so a staff member can verify each fact where it lives, and the visible reasoning-time indicators expose the Thinking mode's deliberation cost. These outputs are from the deployed system and illustrate response format and citation traceability. Lastly, APS-RAG can be directly reached through Claude Code or Codex using the \textit{/aps-rag} skill (see Sec.~S4 for details).

\begin{figure*}[t]
    \centering
    \includegraphics[width=0.8\linewidth]{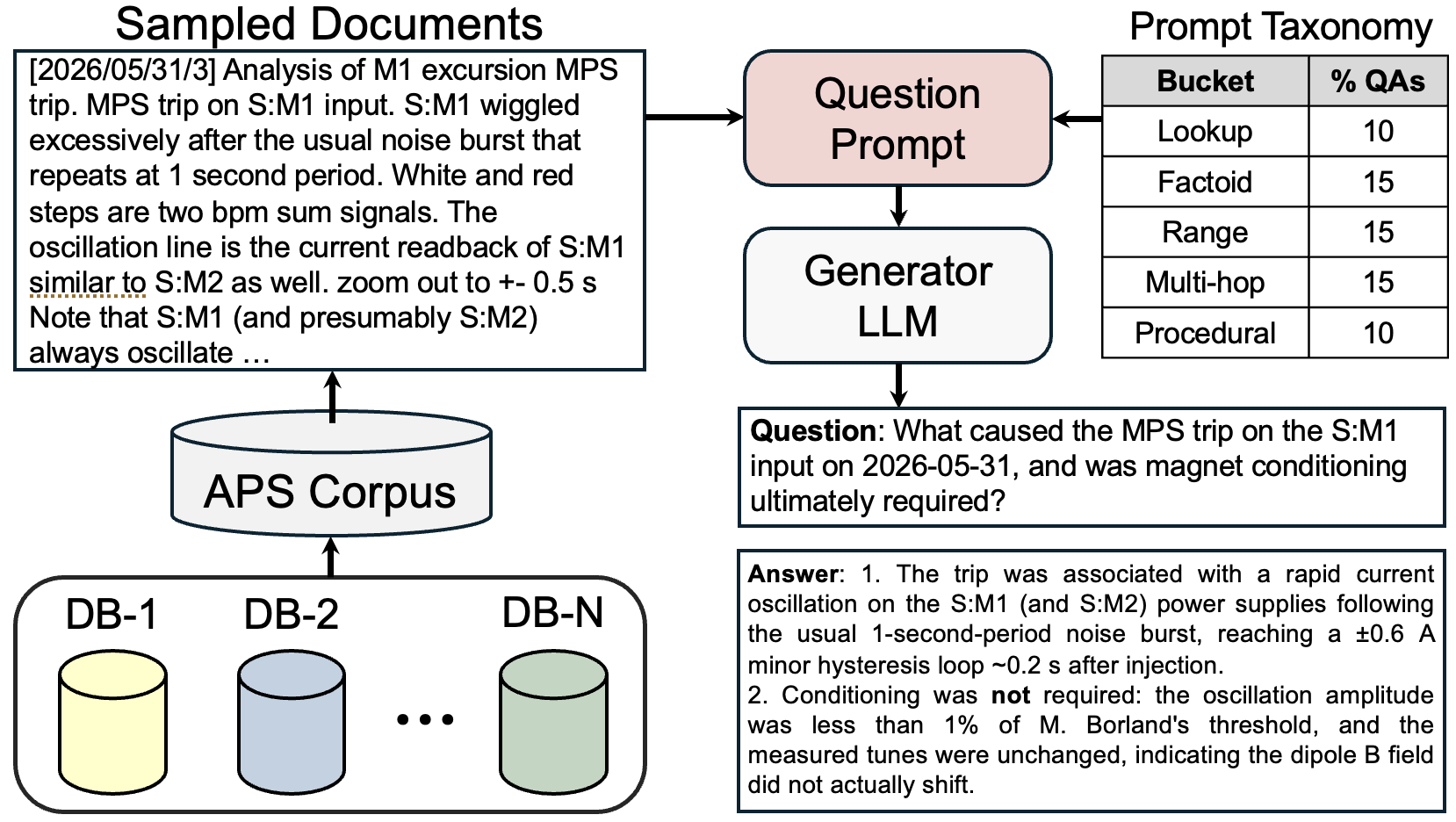}
    \caption{Construction of the corpus-derived benchmark (APS-Bench). The questions are generated from source passages from the APS Corpus. Documents are sampled from the indexed corpus, the union of the per-database collections (Sec.~\ref{sec:ingestion}), and passed, with a question-type taxonomy controlling the mix of query types, to a generator LLM that produces a question paired with a reference answer grounded in the sampled passage; each item is retained only if its gold passage is retrievable from the corpus. The example shown is illustrative: a multi-hop causal question generated from a logbook entry on an M1-excursion machine-protection-system trip, with an answer that both attributes the cause and resolves whether a follow-up action was required. The taxonomy shown is schematic; the realized question-type and collection composition of the 50-question benchmark are summarized in Table~S4, as well as its difficulty strata, and its vital-nugget annotation is reported in Sec.~S9 of the supplementary material.}
    \label{fig:fig6_benchgen}
\end{figure*}

\section{Benchmarks and evaluation methodology}\label{sec:methods}
Evaluation is an integral part of model development, crucial for new domain adaptation, and critical for safety-first applications such as accelerator operations, so we describe it in some detail. It is designed to answer three questions: whether retrieval augmentation improves answer quality on the facility's institutional knowledge, which pipeline components contribute reliably to that improvement, and how performance on a corpus-derived benchmark transfers to verbatim staff questions. The results in Sec.~\ref{sec:results} are organized around these questions.

\subsection{Real operations data benchmark}\label{sec:benchmarks}
APS-Bench (corpus-derived, $n=50$; 49 answerable, one abstention item) is built from the live corpus, not handwritten (Fig.~\ref{fig:fig6_benchgen}). For each (query-type~$\times$~collection) cell, we sample a full record from a collection which is part of the APS corpus and generate type-conditioned questions with an answer set using a GPT-family model (\texttt{GPT 5.4} via ARGO, temperature 0.0); gold answers are then constructed from the sampled document, with grounded inline citations, and each answer is decomposed into 3--10 atomic nuggets labeled \emph{vital} or \emph{okay} (at least two vital). This method is similar to the technique described in the InPars~\cite{bonifacio_inpars_2022}. In Fig.~\ref{fig:fig6_benchgen}, a sampled logbook titled 'Analysis of M1 excursion MPS trip' is sent to a generator LLM with a question-answer generation prompt to create a question and answer pair from the full logbook record. The percentage of each question type (e.g., lookup, factoid, or other) to be selected in APS-Bench is preset. Every item carries an auditable provenance trail (prompt, questions, verified answers, original document IDs, and prompt-template hashes; see Sec.~S9 of the supplementary material), and APS-Bench's gold remains highly BM25-retrievable (mean gold-document minimum rank 0.41; Table~S4 of the supplementary material, which reports the composition of the benchmark dataset). Construction details and the benchmark dataset are given in Sec.~S8--~S9 of the supplementary material. One thing to note here is that, in the question-prompt for benchmark dataset creation, five real APS staff questions were used as few-shot examples for the question-generation prompt. The benchmark carries one abstention item - a question the knowledge base cannot answer - that tests whether a system declines rather than fabricates; abstention items are excluded from nugget recall and faithfulness and scored only by the abstention-correctness sub-metric.

We deliberately use $n=50$ because each question requires several LLM inference calls beyond a single-pass RAG query. These include intent classification, temporal parsing, multi-query generation, planning and routing, synthesis, a self-critique gate with targeted rewrite, and up to two corrective rounds. The five RAG variants and three-seed design then multiply this by $3\times$ end-to-end runs before judging. Nugget-presence judging adds roughly six calls per question, based on an average of 6.3 nuggets, and faithfulness evaluation requires a separate call to decompose the answer into claims. Altogether, a single 50-question, five-system evaluation costs on the order of $\sim$ 5,000 LLM inferences.

\subsection{A six-layer evaluation}\label{sec:layers}
Five pipeline variants are evaluated in an ablation ladder, beginning with a naive BM25 baseline (Naive RAG), followed by Hybrid RAG, Agentic RAG, and GraphRAG, culminating in the full corrective Agentic GraphRAG which is deployed in production as APS-RAG across six layers: (1) retrieval quality (MRR@10, graded NDCG@10, Recall@$k$); (2) nugget answer quality~\cite{pradeep2024autonuggetizer} metrics, including strict vital recall $V_{\text{strict}}$, vital recall $V$, and all-nugget weighted recall $W$ (\emph{headline}); (3) claim-level faithfulness using an HHEM-2.1 backstop~\cite{vectara2024hhem} (\emph{headline}), which employs a decompose-then-verify hallucination rate inspired by FaithJudge~\cite{tamber2025faithjudge} but omits the human-annotated peer-response pool required by FaithJudge, and is therefore not presented as FaithJudge itself; (4) a RAGAS-style~\cite{es2024ragas} lexical-overlap \emph{proxy} (for comparability only); (5) the domain-specific metrics for institutional knowledge described above; and (6) paired statistics. Two answer-match metrics (token-level F1 and semantically judged containment accuracy) link these results to the short-form QA literature and are defined and reported in Sec.~S5 of the supplementary material. Traditional reference-based metrics (such as BLEU, ROUGE, and BERTScore) are intentionally excluded from the primary evaluation, as they assess surface-level text similarity rather than the accurate conveyance of underlying institutional knowledge.

\subsection{Judging, tiering, and statistics}\label{sec:judging}
Answers are generated by a GPT-family model (GPT 5.4 via ARGO) and subsequently evaluated by a large language model judge. This evaluation protocol is used instead of reference-based lexical overlap metrics, which are insufficient for determining whether a long-form operational answer accurately reflects the intended institutional fact. Automated evaluation using large language models as judges has become standard for open-ended and retrieval-augmented generation tasks, with robust judges demonstrating high correlation with human preference assessments~\cite{zheng2023judging}. In this protocol, the judge receives the question, the gold answer annotated with atomic nuggets, and the system's response, and produces per-nugget presence decisions for vital-nugget recall as well as per-claim support decisions based on the retrieved evidence for the faithfulness metric~\cite{es2024ragas}. It is important to note that large language model judges are subject to systematic biases, including position and verbosity effects~\cite{zheng2023judging}, and, in this context, a self-enhancement bias in which a model tends to assign higher scores to outputs from its own family. This bias persists even when evaluators can identify their own generations~\cite{panickssery2024selfpreference}.

To avoid confounding answer quality with self-preference, the evaluation protocol enforces a strict separation between generation and judgment model families. For each evaluation, the generator and judge are drawn from distinct model families. Answers produced by the GPT family are evaluated by Claude, which serves as the primary judge. A separate Gemini model family is used for secondary verification. Judge assignments are determined by call site: Opus~4.6 is used for tasks requiring heavy reasoning, while Sonnet~4.6 is used for lighter or bulk evaluation steps. These assignments are fixed across all five systems, as detailed in Table~\ref{tab:main}. This approach isolates architectural effects and maintains a single, fixed judge per headline metric. Absolute differences are reported with 95\% paired bootstrap confidence intervals, computed using 10,000 resamples per question, and exact McNemar tests are applied for paired binary outcomes. Evaluation is performed using seeds $(42,123,456)$. Because RAG inference is cached per (system, question) and scoring layers operate deterministically at temperature 0, the standard deviations reported are near zero by construction. These values should not be interpreted as evidence of low run-to-run variance. The appropriate measure of uncertainty is the bootstrap confidence interval over questions, rather than the standard deviation over seeds. Since a single judge, even when cross-family, may still introduce idiosyncratic biases, the recommended practice of assessing inter-judge agreement is also followed~\cite{gu2024survey}. The full system's APS-Bench answers were re-scored with a fully independent third-family judge (Gemini), disjoint from both the generator and the primary judge, and substantial per-nugget agreement ($\kappa\approx0.8$) was observed, with the headline strict vital recall only marginally affected under the swap. This procedure bounds, but does not eliminate, single-judge sensitivity.

\begin{table*}[htbp]
  \centering
  \caption{Generation quality of the five RAG variants on APS-Bench ($n=50$): strict vital recall $V_{\text{strict}}$, claim-level hallucination rate (lower is better), with 95\% paired-bootstrap CIs of the difference versus the naive baseline. Accuracy and F1 scores are also measured. Values are means over three seeds: generation, cross-family judging, and test protocols in Sec.~\ref{sec:judging}. Details on $V_{\text{strict}}$, hallucination rate, and the answer-match metrics (Acc./F1) are in the supplementary materials. Note: Agentic GraphRAG is deployed in production as APS-RAG.}
  \vspace{2mm}
  \label{tab:main}
  \small
  \begin{tabular}{l @{\hspace{4mm}} l cccccc}
    \hline
    System & Retrieval Strategy & \makecell{Strict Vital Recall \\ $V_{\text{strict}}$ (\%)} & \makecell{$\Delta V_{\text{strict}}$ vs.\ Naive \\ {}[95\% CI]} & \makecell{Hallucination \\ Rate (\%)} & \makecell{$\Delta$ Halluc.\ vs.\ Naive \\ {}[95\% CI]} & \makecell{Accuracy \\ (\%)} & \makecell{F1 \\ (\%)}\\
    \hline
    Naive RAG & BM25 & 63.8 & -- & 11.2 & -- & 58.8 & 63.1 \\
    Hybrid RAG & Vector + BM25 & 65.5 & $+1.7$ $[-4.1, +7.8]$ & 9.4 & $-1.7$ $[-8.5, +4.3]$ & 65.0 & 73.3 \\
    Agentic RAG & Vector + BM25  & 66.1 & $+2.3$ $[-4.4, +9.2]$ & 7.9 & $-3.3$ $[-15.5, -1.9]$ & 60.6 & 67.1\\
    GraphRAG & Vector + BM25 + KG & 69.1 & $+5.3$ $[-2.5, +13.6]$ & 11.3 & $+0.1$ $[-7.8, +7.9]$ & 61.9 & 69.2\\
    Agentic GraphRAG (ours) & Vector + BM25 + KG & 70.3 & $+6.4$ $[-2.1, +15.6]$ & 8.7 & $-2.5$ $[-10.2, +4.4]$ & 64.1 & 71.2 \\
    \hline
  \end{tabular}
\end{table*}

\begin{table*}[htbp]
  \centering
  \caption{Retrieval quality was evaluated on APS-Bench ($n=50$). NDCG was calculated using graded relevance (primary $=2$, supporting $=1$), and rankings were deduplicated to address document granularity. Reported values represent means over three random seeds. According to Holm correction applied across each metric's ten system pairs, no comparison reached statistical significance. Nominally, the hybrid method (MRR@10 $+8.3$ $[+1.6,+16.6]$ versus naive) and GraphRAG (MRR@10 $+6.3$ $[+0.3,+14.2]$) outperformed the naive baseline, while NDCG@10 and Recall@10/20 did not exhibit nominal differences.}
  \vspace{2mm}
  \label{tab:retrieval}
  \small
  \begin{tabular}{lccccc}
    \hline
    System & \makecell{MRR@10 (\%)} & \makecell{NDCG@10 (\%)} & \makecell{Recall@5 (\%)} & \makecell{Recall@10 (\%)} & \makecell{Recall@20 (\%)} \\
    \hline
    Naive RAG & 80.7 & 69.6 & 63.5 & 67.3 & 71.2 \\
    Hybrid RAG & 89.0 & 73.6 & 62.6 & 70.4 & 76.3 \\
    Agentic RAG & 83.8 & 71.0 & 67.8 & 76.5 & 74.0 \\
    GraphRAG & 87.0 & 73.3 & 61.9 & 69.1 & 76.7 \\
    Agentic GraphRAG (ours) & 85.6 & 72.2 & 62.1 & 69.1 & 75.1 \\
    \hline
  \end{tabular}
\end{table*}

\section{Results and discussion}\label{sec:results}
Unless stated otherwise, all results are on APS-Bench ($n=50$), evaluated against the triangulated gold; table values are computed from a single frozen evaluation source. The single most important framing point comes first: at this benchmark size, with most ablation CIs spanning zero (Table~\ref{tab:ablation}), the only component contribution that is statistically reliable on the current data is the reranker, and that finding survives multiplicity correction: under Holm/Benjamini-Hochberg adjustment the reranker remains significant (paired-bootstrap $p<10^{-4}$; Bonferroni across all tests), while every other component- and system-level difference is not detected --- a count consistent with the ${\approx}0.4$ false positives expected by chance across the ablation family. Accordingly, we describe differences without characterizing small gaps as definitive wins.

\subsection{Answer quality and faithfulness}
The primary result indicates that all retrieval-augmented variants achieve higher strict vital recall compared to the naive BM25 baseline, increasing from 63.8\% to a range of 65.5\%--70.3\%, corresponding to gains of $+1.7$ to $+6.4$ points (Table~\ref{tab:main}). The full Agentic GraphRAG configuration attains the highest recall at 70.3\%, followed by GraphRAG at 69.1\%. However, paired statistical tests reported in Table~\ref{tab:main} show that none of the ten pairwise $V_{\text{strict}}$ comparisons reach statistical significance, either nominally or after Holm correction. Similarly, exact McNemar tests on binary per-question outcomes do not identify significant differences (minimum Holm $p=0.92$), indicating a lack of detection at this benchmark size rather than equivalence. Weighted recall demonstrates three nominal improvements over the naive baseline (hybrid, GraphRAG, and the full system), but none remain significant after Holm correction (see supplementary material). Component ablation (Table~\ref{tab:ablation}) reveals that neither the graph channel nor the corrective loop alone is individually decisive at this benchmark size. Claim-level hallucination rates range from 7.9\%--11.3\% across systems, with the lowest value observed for the corrective Agentic RAG (7.9\%). Analysis of the four augmented variants shows that the corrective loop reduces hallucination relative to its hybrid base (9.4\%~$\rightarrow$~7.9\%), while adding the graph channel alone maintains hallucination at the naive level (11.3\% versus 11.2\%). Combining the graph channel with the corrective loop reduces hallucination from 11.3\%~$\rightarrow$~8.7\%. The only starred interval in Table~\ref{tab:main} (Agentic RAG versus naive) is nominal and does not remain significant after Holm correction. The answer-match metrics are consistent with these findings: accuracy ranges from 58.8\%--65.0\% and best-span F1 from 63.1\%--73.3\%, with the hybrid configuration achieving the highest values (65.0\%/73.3\%) and the full system second (64.1\%/71.2\%). All retrieval-augmented variants outperform the naive baseline on every answer-quality metric, although no ordering among the augmented variants is statistically supported. These results suggest that the primary benefit of the corrective loop is improved faithfulness, while the advantage of the graph channel is conditional. Further discussion is provided in Sec.~\ref{sec:graphbenefit}.

\subsection{Answer quality across difficulty categories}\label{sec:strata}
Analysis of the full configuration's answer quality by difficulty stratum (easy $n=15$, medium $n=24$, hard $n=10$, $n=1$ is abstention; out of the 50 questions).  The question composition in Table~S4 reveals a non-monotonic pattern. Strict vital recall is highest for the easy band (77.9\%), remains similar for the hard band (72.6\%), and is lowest for the medium band (64.5\%). Mean claim-level hallucination follows the inverse order (3.4\%, 10.1\%, and 21.6\%, respectively). Given the sample sizes of 10-24 questions per category, no significance tests are attached to these differences; the following analysis is descriptive. The easy band behaves as intended, consisting of single-hop factual questions anchored to precise operational identifiers, where both the sparse channel and the reranker demonstrate optimal performance. For example, when queried about which after-scan PVs must be zeroed to prevent 9-BM 2D scans from hanging (APS-BENCH-006), the system returns both PV names verbatim (\texttt{9bmc:scan1.AWAIT}, \texttt{9bmc:scan2.AWAIT}), the ``NoWait'' qualifier, and the failure mode, citing the correct entry [BELY\_10635] (strict vital recall 100\%, hallucination 0\%). This level of grounding also extends to multi-entry lookups requiring negative discrimination. For the mispolarized-BTS-quadrupole discovery (APS-BENCH-036), the system retrieves both 2024-04-12 entries with accurate timestamps and the cable-swap resolution [BELY\_1651, BELY\_1653] while explicitly excluding an earlier magnet-testing entry as related but not the discovery itself [BELY\_721].

The medium band is empirically the most challenging, with its performance deficit attributable to composition effects rather than a decline in system competence. This stratum is characterized by terse operational comparisons and temporal texts, where evidence is dispersed across sparse logbook and work-request records, and reference answers frequently hedge or abstain appropriately. When evidence is well localized, medium-band questions are addressed as effectively as those in the easy band. For instance, the linac macro-pulse question (APS-BENCH-042) yields the complete quantitative result—a single macro pulse at 13 kV that splits into two at 14–15 kV [BELY\_5290], along with the companion RG2 measurement of two pulses of approximately 2 and 9 ns separated by approximately 7 ns at 15.2 kV [BELY\_7482]—achieving 100\% strict vital recall and zero hallucination. The cross-beamline comparison (APS-BENCH-029) retrieves the 7 mrad Pt-stripe HR-mirror setting and calculates the 3 mrad differential relative to the 20-BM AuPd setup at 4 mrad, again without hallucination. However, its supporting citation refers to an adjacent entry from the same test series rather than the gold-annotated record, resulting in content accuracy but citation imprecision. The band's 21.6\% mean hallucination rate is primarily driven by items with scattered evidence, where retrieval provides only partial context, and the generator compensates with under-supported specifics. This is the failure mode that inexpensive gating components are designed to suppress in ablation section.

The hard band performs better than its designation implies (62.6\%), primarily due to its composition. It is dominated by multi-hop and causal questions that are grounded in stable, well-documented reference materials such as the SDDS Toolkit and ICMS. Although the evidence is dispersed, it remains accessible. Two key behaviors are observed. The first is causal discipline: when queried about whether a three-screen emittance measurement improved linac-to-PAR charge transmission (APS-BENCH-041), the system avoids over-attribution. It reports the improvement, enumerates the confounded co-interventions (gun-front-end optimization, RF and trajectory tuning, kicker timing) [BELY\_4799, BELY\_8520], explicitly states that the log does not isolate the measurement as the cause, and cites the standard procedure to clarify its actual role as beam matching rather than a standalone charge booster [APS\_1684426]. This matches the reference at zero hallucination, demonstrating the guarded behavior that the self-critique gate and knowledge-graph grounding are intended to produce. The second behavior is the characteristic hard-band failure: the issue is coverage rather than fabrication. For example, the two-month spin-tracking synthesis (APS-BENCH-037) recovers the correct three-thread structure (implementation, validation, applications) and the correct time window, but omits vital per-document details and cites a partially different document subset, resulting in 25\% strict vital recall and only 8\% hallucination. Thus, hard multi-hop errors are typically omissions rather than fabrications; recall declines before faithfulness. This mirrors, at the question level, the recall/faithfulness decoupling and aligns with the conditional, per-type interpretation of the graph channel's benefit.

\subsection{Retrieval quality}
Hybrid dense+sparse fusion gives the best first-stage ranking quality: MRR@10 89.0\% versus 80.7\% for naive BM25 (a nominal gain of $+8.3$ points $[+1.6,+16.6]$) and NDCG@10 73.6\% versus 69.6\%; GraphRAG follows at MRR@10 87.0\% ($+6.3$ $[+0.3,+14.2]$, likewise nominal), and neither gain survives Holm correction (Table~\ref{tab:retrieval}). The graph-bearing variants rank slightly below hybrid (MRR@10 85.6--87.0\% versus 89.0\%), consistent with the graph channel adding relational candidates that dilute ranking precision even as they feed downstream reasoning, while the corrective agentic variant recovers recall at deeper cutoffs (Recall@5 67.8\% and Recall@10 76.5\%, the best of the five; Recall@20 is led by GraphRAG at 76.7\%). NDCG@10 and Recall@10/20 show no nominal differences across the ten system pairs. That the full Agentic GraphRAG nonetheless posts the highest answer quality (70.3\%, Table~\ref{tab:main}) despite trailing hybrid on first-stage ranking (85.6\% versus 89.0\% MRR@10) suggests its corrective reasoning compensates downstream for first-stage ranking it does not itself lead.

\subsection{Component ablation}\label{sec:ablation}
The ablation experiments utilize a hard-biased, type-stratified subset of APS-Bench, referred to as the ablation subset. Each question receives a score of $+2$ if its difficulty label is hard and $+2$ if its query type is multi-hop, causal, or troubleshooting. The subset is populated in a round-robin manner across all eight query types, prioritizing the most difficult questions. This process results in 19 questions (8 hard, 4 medium, 7 easy), ensuring representation from all query types. The subset serves as a stratified sample, distinct from the difficulty stratum referenced in Table~S4. On this subset (Table~\ref{tab:ablation}), one component is particularly influential: substituting the Jina~v3 cross-encoder~\cite{wang2025jinav3} with the legacy LLM scorer leads to a reduction of 32.8 points in strict vital recall (95\% CI $[-47.4,-19.1]$). Two important qualifications constrain this result. First, the per-query distribution indicates that the legacy scorer not only ranks less effectively but often fails entirely: NDCG@10 falls below 0.02 (indicating no usable ranking) for 63\% of the subset questions and below 0.10 for 68\%. On a per-question basis, strict vital recall is lower for 12 out of 19 questions, unchanged for 7, and never higher, resulting in an aggregate NDCG@10 of 13.4 compared to 69.5 for the full configuration. This row should therefore be interpreted as a comparison between a functioning cross-encoder and a degraded fallback, rather than as a general estimate of the marginal value of reranking. The comparison among competent rerankers, by contrast, is a single-digit lever: on a fixed 10k-document diagnostic index constructed to isolate reranker choice, swapping the Jina~v3 cross-encoder for bge-reranker-v2-m3 or Qwen3-Reranker-4B shifts MRR@10 only from 86.9 to 83.3--84.7 over a BM25 first stage (Tables~S6--~S7, and Fig.~S6 of the supplementary material, which also report the no-rerank baseline). Second, the validity of this finding is not compromised by multiple comparisons. The reranker effect remains significant under all applied corrections: its paired-bootstrap $p<10^{-4}$ (none of 100,000 resamples crossed zero), and it persists after Bonferroni correction, whether applied to the nine ablation rows alone or to the nine rows plus the ten system pairs of Table~\ref{tab:main}, as well as after Benjamini–Hochberg correction.

Removal of the corrective loop ($-6.6$, $[-21.4,6.2]$), the graph retriever ($-6.4$, $[-20.7,5.0]$), or the adaptive weights ($-3.5$, $[-16.8,6.7]$) results in directional decreases in recall, though the confidence intervals remain wide at $n=19$. The self-critique gate and the query-enhancement steps do not exhibit individually significant effects on this subset. These results indicate that, at the current benchmark size, recall performance is primarily determined by retrieval and reranking quality. Any recall contribution from agentic components appears limited and will require a larger, pre-registered study for definitive assessment.

Excluding the reranker outlier, the per-ablation changes in strict vital recall and hallucination rate are positively correlated ($r=+0.63$): components that best preserve recall are also those whose removal most increases hallucination. For example, removal of the self-critique gate leaves strict vital recall essentially unchanged ($+0.4$, CI $[-15.8,14.9]$) but increases the hallucination rate by 3.7 points ($7.7\%\rightarrow11.4\%$). The adaptive RRF weights and the synonym resolver exhibit similar patterns ($-3.5$ and $-6.3$ recall, both CIs spanning zero, with hallucination increases of $+1.8$ and $+1.4$ points, respectively). Temporal parsing is the only component whose removal slightly reduces hallucination ($7.7\%\rightarrow6.9\%$) while decreasing recall ($-6.7$), consistent with the hypothesis that date-scoped retrieval elicits more specific and thus more falsifiable claims, whereas its absence leads to vaguer responses that miss nuggets but assert less. These results indicate that several inexpensive gating and preprocessing components primarily contribute to faithfulness rather than recall. A recall-only interpretation of Table~\ref{tab:ablation} would underestimate the value of the self-critique gate, which functions as a faithfulness guard by pruning weakly grounded claims without recovering new evidence. The corrective loop demonstrates a similar effect at full-benchmark scale, directionally lowering hallucination from 9.4\% to 7.9\% over its hybrid base (Table~\ref{tab:main}). This decoupling is reported as directional, since the hallucination-rate deltas do not include per-row confidence intervals at $n=19$. The $\Delta$-recall-versus-$\Delta$-hallucination scatter, provided in the supplementary material, is consistent across four of the nine rows. These findings support the practical recommendation to prioritize reranking and to retain inexpensive faithfulness gates, as their removal has minimal impact on recall but increases unsupported claims by 6-9 points.

\begin{table*}
  \centering
  \caption{Component ablation was conducted on the \textbf{19-question} stratified ablation subset (selection rule in Sec.~\ref{sec:ablation}; 8 hard, 4 medium, 7 easy; all eight query types) by removing or swapping a single component of the full configuration, evaluated on the triangulated gold standard. The $\Delta$ column reports the absolute change in strict vital recall compared to the full configuration (percentage points), with a 95\% paired bootstrap confidence interval. The reranker row is the only interval that excludes 0, and this result remains significant under both Bonferroni and Benjamini–Hochberg correction (paired-bootstrap $p<10^{-4}$). Reported values are means over three seeds (deterministic, temperature = 0, cached scoring). Confidence intervals are computed only for the strict-vital-recall delta; the remaining columns, including hallucination rates discussed in Sec.~\ref{sec:ablation}, are directional at this subset size. The LLM-scorer row represents a degraded fallback, not a comparison to a competent reranker (Sec.~\ref{sec:ablation}).}
  \vspace{2mm}
  \label{tab:ablation}
  \small
 \begin{tabular}{@{}l cccc c@{} cc}
    \hline
    Configuration & \makecell{Strict Vital Recall \\ $V_{\text{strict}}$ (\%)} & \makecell{Weighted Recall \\ $W$ (\%)} & \makecell{Hallucination \\ Rate (\%)} & \makecell{NDCG@10 \\ (\%)} & \makecell{$\Delta$ Strict Vital Recall (pts) \\ {}[95\% CI]} & \makecell{Accuracy \\ (\%)} & \makecell{F1 \\ (\%)}\\
    \hline
    Agentic GraphRAG (full, ours)                  & 64.5 & 70.2 & 7.7 & 69.5 & \textemdash{} (reference) & 78.9 & 72.6 \\
    \hline
    $-$ Corrective loop (single-pass)        & 57.9 & 63.3 & 7.1 & 63.1 & $-6.6$ $[-21.4, 6.2]$ & 73.7 & 73.0 \\
    $-$ Knowledge-graph channel              & 58.1 & 64.1 & 8.2 & 58.6 & $-6.4$ $[-20.7, 5.0]$ & 78.9 & 71.6 \\
    $-$ Adaptive RRF weights (fixed)         & 61.0 & 61.4 & 9.5 & 61.5 & $-3.5$ $[-16.8, 6.7]$ & 73.7 & 73.0 \\
    $-$ Self-critique gate                   & 64.9 & 66.5 & 11.4 & 60.1 & $+0.4$ $[-15.8, 14.9]$ & 78.9 & 72.4 \\
    $-$ Reranker                             & 31.8 & 43.9 & 30.0 & 13.4 & $-32.8$ $[-47.4, -19.1]$ & 42.1 & 61.0 \\
    $-$ Temporal parsing                     & 57.9 & 60.4 & 6.9 & 61.5 & $-6.7$ $[-20.7, 3.9]$ & 78.9 & 73.2 \\
    $-$ Synonym resolver                     & 58.2 & 63.2 & 9.1 & 62.7 & $-6.3$ $[-19.7, 3.9]$ & 84.2 & 72.5 \\
    $-$ Acronym resolver                     & 58.2 & 62.3 & 9.4 & 60.3 & $-6.4$ $[-18.9, 2.9]$ & 68.4 & 72.0 \\
    $-$ Multi-query generator                & 61.8 & 63.0 & 10.0 & 62.2 & $-2.8$ $[-15.8, 6.4]$ & 78.9 & 71.9 \\
   \hline
 \end{tabular}

\end{table*}

\subsection{When graph retrieval helps, and what it costs}\label{sec:graphbenefit}
The honest reading of these results is nuanced rather than triumphal, and we think the nuance is the contribution. The full corrective agentic GraphRAG achieves the highest strict vital recall on the full benchmark, with a score of 70.3\% (Table~\ref{tab:main}). However, the recommendation to deploy this system is based on a numerical advantage that is not statistically supported by the paired tests in Table~\ref{tab:main}. For organizations implementing a knowledge-management system of this type, the primary recommendation is to prioritize investment in reranking mechanisms before introducing agentic components. The inexpensive gating components should be retained, as their contribution is primarily to faithfulness rather than recall (Sec.~\ref{sec:ablation}).

The benefit of the knowledge graph is directional. The graph-bearing variants achieve the highest answer quality on the full benchmark, with scores of 69.1\% and 70.3\% (Table~\ref{tab:main}). However, the ablation removing the graph is inconclusive, with a difference of -6.4 points and a 95\% confidence interval of [-20.7, 5.0] (Table~\ref{tab:ablation}). The graph channel does not improve first-stage ranking precision, as indicated by MRR@10 values of 85.6\% to 87.0\% compared to 89.0\% for the hybrid system (Table~\ref{tab:retrieval}). Without the corrective loop, the graph-based system exhibits the highest hallucination rate among the five systems, at 11.3\% (Table~\ref{tab:main}). The value of the knowledge graph appears to be concentrated in causal, multi-hop, and troubleshooting query types, where it is associated with increased fault-chain completeness, expert finding, and procedural correctness. Figure~S1 of the supplementary material provides the per-question faithfulness breakdown by query type. This observation is hypothesis-generating rather than established. The per-query-type weights were fixed prior to evaluation, which reduces but does not eliminate the risk of post-hoc stratum selection. The conditions under which graph retrieval is beneficial remain an open question in general-domain RAG, as noted by Xiang et al.~\cite{xiang_when_2026}. A pre-registered, adequately powered per-stratum test is required to resolve this, and such an evaluation is identified as future work rather than a current claim.

\subsection{Effect of the embedding model and corrective rounds}
We evaluated four embedding models on a fixed 10,000-document diagnostic index (Fig.~S7 and Table~S5 of the supplementary material), observing a clear hierarchy in performance. The bge-m3 model yielded the highest vector-only MRR@10 at 79.2\%, with the deployed e5-large-v2 achieving 70.2\%. The all-MiniLM-L12-v2 and domain-pretrained accphysbert models performed less effectively, with MRR@10 values between 58.5\% and 60.6\%. Hybrid fusion narrowed the performance gap, producing MRR@10 values in the range of 74.6\% to 85.1\%. The first-stage BM25 baseline also achieved strong results (82.1\%). Despite overlapping bootstrap confidence intervals, two principal observations can be made. First, the domain-pretrained encoder does not surpass general-purpose models on this corpus. Second, retrieval performance is influenced more by the cross-encoder reranker than by the choice of embedding model (Table 1; Tables~S6 and S7 of the supplementary material). A two-dimensional t-SNE~\cite{vandermaaten2008tsne} projection of e5-large-v2 embeddings (Fig.~S5 of the supplementary material) shows the source collections forming distinct yet partially overlapping clusters, with no single dense neighborhood covering the facility's full documented knowledge. This heterogeneity motivates the hybrid, graph-augmented architecture, in which the knowledge graph supplies cross-collection links, such as those connecting a fault reported in a logbook entry to its resolution in a work request, which are not guaranteed by embedding proximity and would otherwise require manual reconstruction across disparate systems.

A hyperparameter sweep (Fig. S9(a) of the supplementary material) was conducted to assess the impact of varying the maximum number of corrective rounds on APS-Bench. The primary performance improvement is observed after the first corrective round, with additional rounds yielding no consistent further benefit. This pattern of diminishing returns supports using a two-round cap in deployment. Full-benchmark results indicate that the corrective loop predominantly improves faithfulness rather than recall, reducing hallucination from 9.4\% to 7.9\% relative to the hybrid baseline (Table~\ref{tab:main}). Most questions are adequately grounded after a single pass, so the observed faithfulness gains are concentrated among a subset of complex, multi-hop questions. This improvement is accompanied by increased latency: in the May 2026 deployment study (Table~S2 and Fig.~S9(b) of the supplementary material), end-to-end wall time for a 10-query advanced-reasoning workload ranged from 40 to 178 seconds, depending on the planner model. Retrieval fan-out accounted for the majority of this time (55–104 seconds), with reranking contributing $\sim$18 seconds.

\subsection{Effect of the synthesis model}
 Replacing the synthesis large language model (LLM) across ten generators sourced from four vendors, while keeping retrieval and fusion components constant, as well as reranking and evaluation criteria fixed, results in a change in strict vital recall of only approximately 12 percentage points (58.1--70.3\%)---the best open-weight generator (GPT-OSS~120B, 67.7\%) sits within 2.6 points of the deployed GPT-5.4 (70.3\%)---while the hallucination rate spreads more than fivefold (5.6--30.6\%) and tracks neither recall nor parameter count (Fig.~S8 of the supplementary material). The 31B open-weight Gemma4 is the most faithful generator tested (5.6\% hallucination at 62.3\% recall) and is small enough to serve and fine-tune, on facility hardware, whereas the larger open models reach near-frontier recall at hallucination rates (23.8--30.6\%) we would not deploy against a safety-relevant record without the pipeline's faithfulness guards. The component-level decoupling as noted in Sec.~\ref{sec:ablation} thus reappears at the generator level---recall behaves like a retrieval property, faithfulness like a generator property---so hallucination rate, rather than recall, should be the primary criterion for generator selection in operational retrieval-augmented generation (RAG) systems.

\subsection{Limitations}\label{sec:limitations}
Several factors constrain the scope of the present study and inform the direction for subsequent deployment. The current evaluation is limited to a single facility, the APS, and the extent to which these findings generalize to other accelerator environments remains to be established. Benchmark questions are generated automatically, as detailed in Sections~S8 and ~S9 of the supplementary material. The primary nugget-recall and faithfulness metrics are based on gold nuggets extracted by large language models and on the nugget-presence decisions of an LLM judge. Validation against human judgment, particularly given the brevity and technical nature of the domain text, is planned for future work. As an interim measure, cross-family inter-judge agreement on per-nugget support decisions is substantial, with Cohen's kappa approximately $k=0.8$ when compared to an independent Google-family judge, and the strict vital recall metric remains unchanged under this judge substitution. While this reduces the risk of artifacts arising from reliance on a single judge, it does not eliminate it. The expert audit therefore remains the definitive standard.

Citation precision on APS-Bench is approximately 73\%, which merits discussion. At the level of base-document granularity, precision is defined as the proportion of cited documents that are included in the gold evidence pool, while recall is the proportion of gold documents that are cited. This precision is partly attributable to the fact that each gold answer is supported by only two to three evidence passages. Nevertheless, this result indicates that a substantial number of non-gold passages are retrieved as supporting evidence, which has implications for traceability when staff seek to verify facts at their source. Citation recall, which ranges from 57\% to 66\%, is additionally reported as a more informative complementary metric. Future work will prioritize claim-level citation scoring using an expanded evidence pool.

Comparative claims in this study are intentionally limited to internal systems. The reported standard deviations correspond to deterministic replication and do not capture run-to-run variability. Although the tooling layer has been deployed, it has not yet been evaluated using the primary metrics, and its systematic assessment will proceed as deployment continues. These limitations do not detract from the main contributions of this work: the establishment of a benchmark and metric suite for institutional knowledge, a systematic comparison of five systems, and a reranker result that remains robust after correction for multiple comparisons. Collectively, these considerations define the set of questions that ongoing in-service operation is expected to address.

\section{Conclusion}\label{sec:conclusion}
We have introduced APS-RAG, a knowledge-management platform designed to make a large accelerator facility's dispersed institutional knowledge accessible to all staff, including scientists, engineers, operators, and technicians, through natural-language queries. The platform integrates a corrective, agentic GraphRAG engine, adaptive three-channel retrieval, and an evaluation framework tailored for operational deployment. On APS-Bench, a corpus-derived, nugget-annotated benchmark with auditable gold standards, all retrieval-augmented variants demonstrate clear improvements in answer quality over naive RAG. The Agentic GraphRAG integration achieves the highest point estimate (70.3\% versus 63.8\% in strict vital recall compared to the naive baseline), although per-system paired tests do not distinguish this lead at the current benchmark size, so we do not present it as definitive. The necessity of a functioning cross-encoder reranker is evident in the component-level recall analysis, while the inexpensive gating components primarily contribute to faithfulness rather than recall. We are releasing the code, benchmark dataset, evaluation harness, and a reusable agentic retrieval agent skill. Effective knowledge management transforms a facility's accumulated experience into a resource that reduces downtime and mitigates staff turnover. Our goal is for the benchmark creation methodology and metric family to be broadly applicable, supporting any organization or facility seeking to maximize the utilization and value of its institutional knowledge.

\section*{Supplementary Material}\label{sec:supp}
The supplementary material details the experimental protocols, extended analyses, and reference tables that support the main text. Section~S1 outlines the implementation of the native-tool ReAct executor and the Model Context Protocol (MCP) tooling layer. Section~S2 presents the results of the component-latency measurements and the reranker A/B evaluation. Section~S3 summarizes the reliability-engineering effort. Section~S4 describes the design and deployment of the reusable /aps-rag retrieval skill. Section~S5 specifies the exact-match, F1, and accuracy metrics applied in the answer-match evaluations reported in Table~\ref{tab:main}. Section~S6 compiles supplementary figures, including the faithfulness breakdown by query type (Fig.~S1) and the sensitivity of model answer evaluation between open-source and closed-source models  (Fig.~S8). Sections~S7 through ~S9 document the corpus schema, the APS-Bench construction protocol, and the provenance of the benchmark dataset.

\begin{acknowledgments}
R.S would like to thank AOP Group Members, Greg Fystro, Randy Flood, Brendan Chandler, Elaine Chandler, and Felix M. Lacap for productive discussions and support in developing APS-RAG. The research is supported by the U. S. Department of Energy, Office of Science, under Contract No. DE-AC02-06CH11357. We gratefully acknowledge the computing resources provided on Improv and Swing, a high-performance computing cluster operated by the Laboratory Computing Resource Center at Argonne National Laboratory.
\end{acknowledgments}

\section{Data Availability Statement}
The APS-RAG code is available on GitHub (\href{https://github.com/rajatsainju2025/aps-rag}), benchmark (APS-Bench), the evaluation harness, and the \emph{/aps-rag} retrieval skill (see Sec.~S4), which lets the platform's institutional knowledge be queried from Claude Code and Codex, will be released publicly; the derived metrics that support the findings are reported in the article and its supplementary material. The underlying institutional-knowledge data were generated at the Advanced Photon Source, a large-scale user facility, and derived data are available from the corresponding author upon reasonable request, subject to the facility's data-handling policy.

\section{References}
\bibliography{3_references}

@inproceedings{galayda:pac95-mad02,
    author = {J. N. Galayda},
    title = {{The Advanced Photon Source}},
    booktitle = {Proc. PAC'95},
    pages = {4--8},
    paper = {MAD02},
    venue = {Dallas, TX, USA, May 1995},
    publisher = {JACoW Publishing, Geneva, Switzerland},
}

@inproceedings{borl:ipac18-thxgbd1,
    author = {M. Borland and others},
    title = {{The Upgrade of the Advanced Photon Source}},
    booktitle = {Proc. IPAC'18},
    pages = {2872--2877},
    paper = {THXGBD1},
    venue = {Vancouver, Canada, Apr.-May 2018},
    publisher = {JACoW Publishing, Geneva, Switzerland},
    doi = {10.18429/JACoW-IPAC2018-THXGBD1},
    url = {http://accelconf.web.cern.ch/ipac2018/papers/THXGBD1.pdf},
}

@article{robertson_probabilistic_2009,
	address = {Hanover, MA, USA},
	title = {The {Probabilistic} {Relevance} {Framework}: {BM25} and {Beyond}},
	volume = {3},
	issn = {1554-0669},
	url = {https://doi.org/10.1561/1500000019},
	doi = {10.1561/1500000019},
	number = {4},
	journal = {Found. Trends Inf. Retr.},
	publisher = {Now Publishers Inc.},
	author = {Robertson, Stephen and Zaragoza, Hugo},
	month = apr,
	year = {2009},
	pages = {333--389},
}

@techreport{borland_users_1995,
	title = {User`s guide for {SDDS} toolkit {Version} 1.4},
	url = {http://www.osti.gov/servlets/purl/179270-NWQdeR/webviewable/},
	doi = {10.2172/179270},
	number = {ANL/ASD/RP--88322, 179270},
	urldate = {2026-06-24},
	author = {Borland, M.},
	month = jul,
	year = {1995},
	pages = {ANL/ASD/RP--88322, 179270},
}

@misc{jiang_agentic_2026,
	title = {Agentic {Hybrid} {RAG} for {Evidence}-{Grounded} {Muon} {Collider} {Analysis}},
	url = {http://arxiv.org/abs/2606.10381},
	doi = {10.48550/arXiv.2606.10381},
	urldate = {2026-06-23},
	publisher = {arXiv},
	author = {Jiang, Ruobing and Fu, Dawei and Jiang, Cheng and Yang, Tianyi and Wang, Zijian and Wu, Youpeng and Ban, Yong and Mao, Yajun and Li, Qiang},
	month = jun,
	year = {2026},
	note = {arXiv:2606.10381 [hep-ex]},
}

@article{stuhlmann_intelligent_nodate,
	title = {Intelligent {Retrieval} {Systems} for {Accelerator} {Physics}: {Matching} {Retrieval} {Approaches} to {Use} {Cases}},
	author = {Stuhlmann, L and Sapinski, M and Dai, Q and Ischebeck, R},
}

@inproceedings{mao_rag-studio_2024,
	address = {Miami, Florida, USA},
	title = {{RAG}-{Studio}: {Towards} {In}-{Domain} {Adaptation} of {Retrieval} {Augmented} {Generation} {Through} {Self}-{Alignment}},
	shorttitle = {{RAG}-{Studio}},
	url = {https://aclanthology.org/2024.findings-emnlp.41},
	doi = {10.18653/v1/2024.findings-emnlp.41},
	urldate = {2026-06-01},
	booktitle = {Findings of the {Association} for {Computational} {Linguistics}: {EMNLP} 2024},
	publisher = {Association for Computational Linguistics},
	author = {Mao, Kelong and Liu, Zheng and Qian, Hongjin and Mo, Fengran and Deng, Chenlong and Dou, Zhicheng},
	year = {2024},
	pages = {725--735},
}

@article{vriza_operating_2026,
	title = {Operating advanced scientific instruments with {AI} agents that learn on the job},
	volume = {12},
	issn = {2057-3960},
	url = {https://www.nature.com/articles/s41524-026-02005-0},
	doi = {10.1038/s41524-026-02005-0},
	number = {1},
	urldate = {2026-06-01},
	journal = {npj Computational Materials},
	author = {Vriza, Aikaterini and Prince, Michael H. and Zhou, Tao and Chan, Henry and Cherukara, Mathew J.},
	month = mar,
	year = {2026},
	pages = {160},
}

@article{bruch_analysis_2024,
	title = {An {Analysis} of {Fusion} {Functions} for {Hybrid} {Retrieval}},
	volume = {42},
	issn = {1046-8188, 1558-2868},
	url = {http://arxiv.org/abs/2210.11934},
	doi = {10.1145/3596512},
	number = {1},
	urldate = {2026-05-20},
	journal = {ACM Transactions on Information Systems},
	author = {Bruch, Sebastian and Gai, Siyu and Ingber, Amir},
	month = jan,
	year = {2024},
	note = {arXiv:2210.11934 [cs.IR]},
	pages = {1--35},
}

@inproceedings{jeong_adaptive-rag_2024,
	address = {Mexico City, Mexico},
	title = {Adaptive-{RAG}: {Learning} to {Adapt} {Retrieval}-{Augmented} {Large} {Language} {Models} through {Question} {Complexity}},
	shorttitle = {Adaptive-{RAG}},
	url = {https://aclanthology.org/2024.naacl-long.389},
	doi = {10.18653/v1/2024.naacl-long.389},
	urldate = {2026-05-20},
	booktitle = {Proceedings of the 2024 {Conference} of the {North} {American} {Chapter} of the {Association} for {Computational} {Linguistics}: {Human} {Language} {Technologies} ({Volume} 1: {Long} {Papers})},
	publisher = {Association for Computational Linguistics},
	author = {Jeong, Soyeong and Baek, Jinheon and Cho, Sukmin and Hwang, Sung Ju and Park, Jong},
	year = {2024},
	pages = {7036--7050},
}

@misc{xiang_when_2026,
	title = {When to use {Graphs} in {RAG}: {A} {Comprehensive} {Analysis} for {Graph} {Retrieval}-{Augmented} {Generation}},
	shorttitle = {When to use {Graphs} in {RAG}},
	url = {http://arxiv.org/abs/2506.05690},
	doi = {10.48550/arXiv.2506.05690},
	urldate = {2026-05-20},
	publisher = {arXiv},
	author = {Xiang, Zhishang and Wu, Chuanjie and Zhang, Qinggang and Chen, Shengyuan and Hong, Zijin and Huang, Xiao and Su, Jinsong},
	month = feb,
	year = {2026},
	note = {arXiv:2506.05690 [cs.CL]},
}

@misc{yao_reac_2023,
	title = {{ReAct}: Synergizing reasoning and acting in language models},
	note = {International Conference on Learning Representations (ICLR); arXiv:2210.03629},
	author = {Yao, Shunyu and Zhao, Jeffrey and Yu, Dian and Du, Nan and Shafran, Izhak and Narasimhan, Karthik and Cao, Yuan},
	year = {2023},
}

@article{jarosz_logging_nodate,
	title = {Logging a new era at the {APS} using {BELY}},
	issn = {2226-0358},
	doi = {10.18429/JACoW-ICALEPCS2025-THMG007},
	publisher = {20th Int. Conf. Accel. Large Exp. Phys. Control Syst. (ICALEPCS'25)},
	author = {Jarosz, D and Chandler, E and Shen, G and Xiao, L and Arnold, N and Veseli, S},
	pages = {1482--1486},
}

@techreport{edelen_anomaly_2021,
	title = {Anomaly {Detection} in {Particle} {Accelerators} using {Autoencoders}},
	url = {http://arxiv.org/abs/2112.07793},
	doi = {10.26024/p6mv-en77},
	urldate = {2026-05-11},
	author = {Edelen, Jonathan P. and Cook, Nathan M.},
	month = mar,
	year = {2021},
	note = {arXiv:2112.07793 [physics.acc-ph]},
}

@article{blokland_uncertainty_2022,
	title = {Uncertainty aware anomaly detection to predict errant beam pulses in the {Oak} {Ridge} {Spallation} {Neutron} {Source} accelerator},
	volume = {25},
	issn = {2469-9888},
	url = {https://link.aps.org/doi/10.1103/PhysRevAccelBeams.25.122802},
	doi = {10.1103/PhysRevAccelBeams.25.122802},
	number = {12},
	urldate = {2026-05-11},
	journal = {Physical Review Accelerators and Beams},
	author = {Blokland, Willem and Rajput, Kishansingh and Schram, Malachi and Jeske, Torri and Ramuhalli, Pradeep and Peters, Charles and Yucesan, Yigit and Zhukov, Alexander},
	month = dec,
	year = {2022},
	pages = {122802},
}

@misc{mayet_gaia_2024,
	title = {{GAIA}: {A} {General} {AI} {Assistant} for {Intelligent} {Accelerator} {Operations}},
	shorttitle = {{GAIA}},
	url = {http://arxiv.org/abs/2405.01359},
	doi = {10.48550/arXiv.2405.01359},
	urldate = {2026-05-11},
	publisher = {arXiv},
	author = {Mayet, Frank},
	month = may,
	year = {2024},
	note = {arXiv:2405.01359 [cs.CL]},
}

@misc{sulc_towards_2024,
	title = {Towards {Unlocking} {Insights} from {Logbooks} {Using} {AI}},
	url = {http://arxiv.org/abs/2406.12881},
	doi = {10.48550/arXiv.2406.12881},
	urldate = {2026-05-11},
	publisher = {arXiv},
	author = {Sulc, Antonin and Bien, Alex and Eichler, Annika and Ratner, Daniel and Rehm, Florian and Mayet, Frank and Hartmann, Gregor and Hoschouer, Hayden and Tuennermann, Henrik and Kaiser, Jan and John, Jason St and Maldonado, Jennefer and Hazelwood, Kyle and Kammering, Raimund and Hellert, Thorsten and Wilksen, Tim and Kain, Verena and Hu, Wan-Lin},
	month = may,
	year = {2024},
	note = {arXiv:2406.12881 [physics.acc-ph]},
}

@misc{suresh_towards_2024,
	title = {Towards a {RAG}-based {Summarization} {Agent} for the {Electron}-{Ion} {Collider}},
	url = {http://arxiv.org/abs/2403.15729},
	doi = {10.48550/arXiv.2403.15729},
	urldate = {2026-05-11},
	publisher = {arXiv},
	author = {Suresh, Karthik and Kackar, Neeltje and Schleck, Luke and Fanelli, Cristiano},
	month = jun,
	year = {2024},
	note = {arXiv:2403.15729 [cs.CL]},
}

@article{kaiser_large_2025,
	title = {Large language models for human-machine collaborative particle accelerator tuning through natural language},
	volume = {11},
	issn = {2375-2548},
	url = {https://www.science.org/doi/10.1126/sciadv.adr4173},
	doi = {10.1126/sciadv.adr4173},
	number = {1},
	urldate = {2026-05-11},
	journal = {Science Advances},
	author = {Kaiser, Jan and Lauscher, Anne and Eichler, Annika},
	month = jan,
	year = {2025},
	pages = {eadr4173},
}

@article{lewis_retrieval-augmented_nodate,
	title = {Retrieval-{Augmented} {Generation} for {Knowledge}-{Intensive} {NLP} {Tasks}},
	author = {Lewis, Patrick and Perez, Ethan and Piktus, Aleksandra and Petroni, Fabio and Karpukhin, Vladimir and Goyal, Naman and Küttler, Heinrich and Lewis, Mike and Yih, Wen-tau and Rocktäschel, Tim and Riedel, Sebastian and Kiela, Douwe},
}

@misc{gao_retrieval-augmented_2024,
	title = {Retrieval-{Augmented} {Generation} for {Large} {Language} {Models}: {A} {Survey}},
	shorttitle = {Retrieval-{Augmented} {Generation} for {Large} {Language} {Models}},
	url = {http://arxiv.org/abs/2312.10997},
	doi = {10.48550/arXiv.2312.10997},
	urldate = {2026-05-11},
	publisher = {arXiv},
	author = {Gao, Yunfan and Xiong, Yun and Gao, Xinyu and Jia, Kangxiang and Pan, Jinliu and Bi, Yuxi and Dai, Yi and Sun, Jiawei and Wang, Meng and Wang, Haofen},
	month = mar,
	year = {2024},
	note = {arXiv:2312.10997 [cs.CL]},
}

@inproceedings{cormack_reciprocal_2009,
	address = {Boston MA USA},
	title = {Reciprocal rank fusion outperforms condorcet and individual rank learning methods},
	isbn = {978-1-60558-483-6},
	url = {https://dl.acm.org/doi/10.1145/1571941.1572114},
	doi = {10.1145/1571941.1572114},
	urldate = {2026-05-11},
	booktitle = {Proceedings of the 32nd international {ACM} {SIGIR} conference on {Research} and development in information retrieval},
	publisher = {ACM},
	author = {Cormack, Gordon V. and Clarke, Charles L A and Buettcher, Stefan},
	month = jul,
	year = {2009},
	pages = {758--759},
}

@inproceedings{bonifacio_inpars_2022,
	address = {Madrid Spain},
	title = {{InPars}: {Unsupervised} {Dataset} {Generation} for {Information} {Retrieval}},
	isbn = {978-1-4503-8732-3},
	shorttitle = {{InPars}},
	url = {https://dl.acm.org/doi/10.1145/3477495.3531863},
	doi = {10.1145/3477495.3531863},
	urldate = {2026-05-11},
	booktitle = {Proceedings of the 45th {International} {ACM} {SIGIR} {Conference} on {Research} and {Development} in {Information} {Retrieval}},
	publisher = {ACM},
	author = {Bonifacio, Luiz and Abonizio, Hugo and Fadaee, Marzieh and Nogueira, Rodrigo},
	month = jul,
	year = {2022},
	pages = {2387--2392},
}

@misc{zhang_qwen3_2025,
	title = {Qwen3 {Embedding}: {Advancing} {Text} {Embedding} and {Reranking} {Through} {Foundation} {Models}},
	shorttitle = {Qwen3 {Embedding}},
	url = {http://arxiv.org/abs/2506.05176},
	doi = {10.48550/arXiv.2506.05176},
	urldate = {2026-05-06},
	publisher = {arXiv},
	author = {Zhang, Yanzhao and Li, Mingxin and Long, Dingkun and Zhang, Xin and Lin, Huan and Yang, Baosong and Xie, Pengjun and Yang, An and Liu, Dayiheng and Lin, Junyang and Huang, Fei and Zhou, Jingren},
	month = jun,
	year = {2025},
	note = {arXiv:2506.05176 [cs]},
}

@misc{brown_language_2020,
	title = {Language {Models} are {Few}-{Shot} {Learners}},
	url = {http://arxiv.org/abs/2005.14165},
	doi = {10.48550/arXiv.2005.14165},
	urldate = {2025-04-26},
	publisher = {arXiv},
	author = {Brown, Tom B. and Mann, Benjamin and Ryder, Nick and Subbiah, Melanie and Kaplan, Jared and Dhariwal, Prafulla and Neelakantan, Arvind and Shyam, Pranav and Sastry, Girish and Askell, Amanda and Agarwal, Sandhini and Herbert-Voss, Ariel and Krueger, Gretchen and Henighan, Tom and Child, Rewon and Ramesh, Aditya and Ziegler, Daniel M. and Wu, Jeffrey and Winter, Clemens and Hesse, Christopher and Chen, Mark and Sigler, Eric and Litwin, Mateusz and Gray, Scott and Chess, Benjamin and Clark, Jack and Berner, Christopher and McCandlish, Sam and Radford, Alec and Sutskever, Ilya and Amodei, Dario},
	month = jul,
	year = {2020},
	note = {arXiv:2005.14165 [cs]},
}

@article{humble_resilient_2024,
	title = {Resilient {VAE}: {Unsupervised} {Anomaly} {Detection} at the {SLAC} {Linac} {Coherent} {Light} {Source}},
	volume = {295},
	copyright = {https://creativecommons.org/licenses/by/4.0/},
	issn = {2100-014X},
	shorttitle = {Resilient {VAE}},
	url = {https://www.epj-conferences.org/10.1051/epjconf/202429509033},
	doi = {10.1051/epjconf/202429509033},
	urldate = {2025-04-26},
	journal = {EPJ Web of Conferences},
	author = {Humble, Ryan and Colocho, William and O’Shea, Finn and Ratner, Daniel and Darve, Eric},
	editor = {De Vita, R. and Espinal, X. and Laycock, P. and Shadura, O.},
	year = {2024},
	pages = {09033},
}

@article{hellert_domain-specific_2025,
	title = {Domain-specific text embedding model for accelerator physics},
	volume = {28},
	issn = {2469-9888},
	url = {https://link.aps.org/doi/10.1103/PhysRevAccelBeams.28.044601},
	doi = {10.1103/PhysRevAccelBeams.28.044601},
	number = {4},
	urldate = {2025-04-26},
	journal = {Physical Review Accelerators and Beams},
	author = {Hellert, Thorsten and Montenegro, João and Venturini, Marco and Pollastro, Andrea},
	month = apr,
	year = {2025},
	pages = {044601},
}

@article{prince_opportunities_2024,
	title = {Opportunities for retrieval and tool augmented large language models in scientific facilities},
	volume = {10},
	issn = {2057-3960},
	url = {https://www.nature.com/articles/s41524-024-01423-2},
	doi = {10.1038/s41524-024-01423-2},
	number = {1},
	urldate = {2025-03-18},
	journal = {npj Computational Materials},
	author = {Prince, Michael H. and Chan, Henry and Vriza, Aikaterini and Zhou, Tao and Sastry, Varuni K. and Luo, Yanqi and Dearing, Matthew T. and Harder, Ross J. and Vasudevan, Rama K. and Cherukara, Mathew J.},
	month = nov,
	year = {2024},
	pages = {251},
}

@article{maldonado_enhancing_nodate,
	title = {Enhancing {Electronic} {Logbooks} {Using} {Machine} {Learning}},
	author = {Maldonado, J},
}

@article{lobach_recurrent_2024,
	title = {Recurrent neural networks for anomaly detection in magnet power supplies of particle accelerators},
	volume = {18},
	issn = {26668270},
	url = {https://linkinghub.elsevier.com/retrieve/pii/S2666827024000616},
	doi = {10.1016/j.mlwa.2024.100585},
	urldate = {2025-02-26},
	journal = {Machine Learning with Applications},
	author = {Lobach, Ihar and Borland, Michael},
	month = dec,
	year = {2024},
	pages = {100585},
}

@inproceedings{lewis2020rag,
  title = {Retrieval-augmented generation for knowledge-intensive {NLP} tasks},
  author = {Lewis, Patrick and Perez, Ethan and Piktus, Aleksandra and Petroni, Fabio and Karpukhin, Vladimir and Goyal, Naman and K{\"u}ttler, Heinrich and Lewis, Mike and Yih, Wen-tau and Rockt{\"a}schel, Tim and Riedel, Sebastian and Kiela, Douwe},
  booktitle = {Advances in Neural Information Processing Systems},
  volume = {33},
  pages = {9459--9474},
  year = {2020},
}

@misc{gao2023ragsurvey,
  title = {Retrieval-augmented generation for large language models: {A} survey},
  author = {Gao, Yunfan and Xiong, Yun and Gao, Xinyu and Jia, Kangxiang and Pan, Jinliu and Bi, Yuxi and Dai, Yi and Sun, Jiawei and Wang, Meng and Wang, Haofen},
  year = {2023},
  note = {arXiv:2312.10997},
}

@inproceedings{karpukhin2020dpr,
  title = {Dense passage retrieval for open-domain question answering},
  author = {Karpukhin, Vladimir and O{\u{g}}uz, Barlas and Min, Sewon and Lewis, Patrick and Wu, Ledell and Edunov, Sergey and Chen, Danqi and Yih, Wen-tau},
  booktitle = {Proceedings of the 2020 Conference on Empirical Methods in Natural Language Processing (EMNLP)},
  pages = {6769--6781},
  year = {2020},
}

@misc{nogueira2020rerank,
  title = {Passage re-ranking with {BERT}},
  author = {Nogueira, Rodrigo and Cho, Kyunghyun},
  year = {2019},
  note = {arXiv:1901.04085},
}

@misc{edge2024graphrag,
  title = {From local to global: {A} graph {RAG} approach to query-focused summarization},
  author = {Edge, Darren and Trinh, Ha and Cheng, Newman and Bradley, Joshua and Chao, Alex and Mody, Apurva and Truitt, Steven and Larson, Jonathan},
  year = {2024},
  note = {arXiv:2404.16130},
}

@misc{guo2024lightrag,
  title = {{LightRAG}: Simple and fast retrieval-augmented generation},
  author = {Guo, Zirui and Xia, Lianghao and Yu, Yanhua and Ao, Tu and Huang, Chao},
  year = {2024},
  note = {arXiv:2410.05779},
}

@inproceedings{gutierrez2024hipporag,
  title = {{HippoRAG}: Neurobiologically inspired long-term memory for large language models},
  author = {Jim{\'e}nez Guti{\'e}rrez, Bernal and Shu, Yiheng and Gu, Yu and Yasunaga, Michihiro and Su, Yu},
  booktitle = {Advances in Neural Information Processing Systems},
  volume = {37},
  year = {2024},
  note = {arXiv:2405.14831},
}

@misc{peng2024graphragsurvey,
  title = {Graph retrieval-augmented generation: {A} survey},
  author = {Peng, Boci and Zhu, Yun and Liu, Yongchao and Bo, Xiaohe and Shi, Haizhou and Hong, Chuntao and Zhang, Yan and Tang, Siliang},
  year = {2024},
  note = {arXiv:2408.08921},
}

@misc{yan2024crag,
  title = {Corrective retrieval augmented generation},
  author = {Yan, Shi-Qi and Gu, Jia-Chen and Zhu, Yun and Ling, Zhen-Hua},
  year = {2024},
  note = {arXiv:2401.15884},
}

@inproceedings{asai2024selfrag,
  title = {Self-{RAG}: Learning to retrieve, generate, and critique through self-reflection},
  author = {Asai, Akari and Wu, Zeqiu and Wang, Yizhong and Sil, Avirup and Hajishirzi, Hannaneh},
  booktitle = {International Conference on Learning Representations (ICLR)},
  year = {2024},
  note = {arXiv:2310.11511},
}

@misc{singh2025agenticsurvey,
  title = {Agentic retrieval-augmented generation: {A} survey on agentic {RAG}},
  author = {Singh, Aditi and Ehtesham, Abul and Kumar, Saket and Khoei, Tala Talaei},
  year = {2025},
  note = {arXiv:2501.09136},
}

@misc{anthropic2024mcp,
  title = {Introducing the {Model} {Context} {Protocol}},
  author = {{Anthropic}},
  year = {2024},
  note = {\url{https://www.anthropic.com/news/model-context-protocol}},
}

@misc{pradeep2024autonuggetizer,
  title = {Initial nugget evaluation results for the {TREC} 2024 {RAG} track with the {AutoNuggetizer} framework},
  author = {Pradeep, Ronak and Thakur, Nandan and Upadhyay, Shivani and Campos, Daniel and Craswell, Nick and Lin, Jimmy},
  year = {2024},
  note = {arXiv:2411.09607},
}

@inproceedings{tamber2025faithjudge,
  title = {Benchmarking {LLM} faithfulness in {RAG} with evolving leaderboards},
  author = {Tamber, Manveer Singh and Bao, Forrest Sheng and Xu, Chenyu and Luo, Ge and Kazi, Suleman and Bae, Minseok and Li, Miaoran and Mendelevitch, Ofer and Qu, Renyi and Lin, Jimmy},
  booktitle = {Proceedings of the 2025 Conference on Empirical Methods in Natural Language Processing: Industry Track},
  year = {2025},
  note = {arXiv:2505.04847},
}

@misc{vectara2024hhem,
  title = {{HHEM}-2.1-{Open}: Hughes hallucination evaluation model},
  author = {Bao, Forrest and Li, Miaoran and Luo, Rogger and Mendelevitch, Ofer},
  year = {2024},
  note = {Vectara; \url{https://huggingface.co/vectara/hallucination_evaluation_model}},
}

@inproceedings{es2024ragas,
  title = {{RAGAs}: Automated evaluation of retrieval augmented generation},
  author = {Es, Shahul and James, Jithin and Espinosa-Anke, Luis and Schockaert, Steven},
  booktitle = {Proceedings of the 18th Conference of the European Chapter of the Association for Computational Linguistics: System Demonstrations},
  pages = {150--158},
  year = {2024},
}

@misc{wang2022e5,
  title = {Text embeddings by weakly-supervised contrastive pre-training},
  author = {Wang, Liang and Yang, Nan and Huang, Xiaolong and Jiao, Binxing and Yang, Linjun and Jiang, Daxin and Majumder, Rangan and Wei, Furu},
  year = {2022},
  note = {arXiv:2212.03533},
}

@misc{edelen2018opportunities,
  title = {Opportunities in machine learning for particle accelerators},
  author = {Edelen, Auralee and Mayes, Christopher and Bowring, Daniel and Ratner, Daniel and Adelmann, Andreas and Ischebeck, Rasmus and Snuverink, Jochem and Agapov, Ilya and Kammering, Raimund and Edelen, Jonathan and Bazarov, Ivan and Valentino, Gianluca and Wenninger, Jorg},
  year = {2018},
  note = {arXiv:1811.03172},
}

@article{kaiser2023rlbo,
  title = {Reinforcement learning-trained optimisers and {Bayesian} optimisation for online particle accelerator tuning},
  author = {Kaiser, Jan and Xu, Chenran and Eichler, Annika and Santamaria Garcia, Andrea and Stein, Oliver and Br{\"u}ndermann, Erik and Kuropka, Willi and Dinter, Hannes and Mayet, Frank and Vinatier, Thomas and Burkart, Florian and Schlarb, Holger},
  journal = {Scientific Reports},
  volume = {14},
  pages = {15733},
  year = {2024},
}

@article{rajput2023errant,
  title = {Robust errant beam prognostics with conditional modeling for particle accelerators},
  author = {Rajput, Kishansingh and Schram, Malachi and Blokland, Willem and Alanazi, Yasir and Ramuhalli, Pradeep and Zhukov, Alexander and Peters, Charles and Vilalta, Ricardo},
  journal = {Machine Learning: Science and Technology},
  volume = {5},
  year = {2024},
  note = {DOI: 10.1088/2632-2153/ad2e18},
}

@inproceedings{sulc2024agentic,
  title = {Towards agentic {AI} on particle accelerators},
  author = {Sulc, Antonin and Hellert, Thorsten and Kammering, Raimund and Hoschouer, Hayden and St. John, Jason},
  booktitle = {Machine Learning and the Physical Sciences Workshop, 38th Conference on Neural Information Processing Systems (NeurIPS)},
  year = {2024},
  note = {arXiv:2409.06336},
}

@article{hellert2026agentic,
  title = {Agentic artificial intelligence for multistage physics experiments at a large-scale user facility particle accelerator},
  author = {Hellert, Thorsten and Bertwistle, Drew and Leemann, Simon C. and Sulc, Antonin and Venturini, Marco},
  journal = {Physical Review Research},
  volume = {8},
  pages = {L012017},
  year = {2026},
  note = {arXiv:2509.17255},
}

@article{vandermaaten2008tsne,
  title = {Visualizing data using t-{SNE}},
  author = {van der Maaten, Laurens and Hinton, Geoffrey},
  journal = {Journal of Machine Learning Research},
  volume = {9},
  pages = {2579--2605},
  year = {2008},
}

@inproceedings{zhai2023siglip,
  title = {Sigmoid loss for language image pre-training},
  author = {Zhai, Xiaohua and Mustafa, Basil and Kolesnikov, Alexander and Beyer, Lucas},
  booktitle = {Proceedings of the IEEE/CVF International Conference on Computer Vision (ICCV)},
  pages = {11975--11986},
  year = {2023},
}

@misc{douze2024faiss,
  title = {The {Faiss} library},
  author = {Douze, Matthijs and Guzhva, Alexandr and Deng, Chengqi and Johnson, Jeff and Szilvasy, Gergely and Mazar{\'e}, Pierre-Emmanuel and Lomeli, Maria and Hosseini, Lucas and J{\'e}gou, Herv{\'e}},
  year = {2024},
  note = {arXiv:2401.08281},
}

@inproceedings{chen2024bgem3,
  title = {{M3}-embedding: Multi-linguality, multi-functionality, multi-granularity text embeddings through self-knowledge distillation},
  author = {Chen, Jianlv and Xiao, Shitao and Zhang, Peitian and Luo, Kun and Lian, Defu and Liu, Zheng},
  booktitle = {Findings of the Association for Computational Linguistics: ACL 2024},
  year = {2024},
  note = {arXiv:2402.03216},
}

@inproceedings{wang2020minilm,
  title = {{MiniLM}: Deep self-attention distillation for task-agnostic compression of pre-trained transformers},
  author = {Wang, Wenhui and Wei, Furu and Dong, Li and Bao, Hangbo and Yang, Nan and Zhou, Ming},
  booktitle = {Advances in Neural Information Processing Systems},
  volume = {33},
  pages = {5776--5788},
  year = {2020},
}

@inproceedings{reimers2019sbert,
  title = {Sentence-{BERT}: Sentence embeddings using {Siamese} {BERT}-networks},
  author = {Reimers, Nils and Gurevych, Iryna},
  booktitle = {Proceedings of the 2019 Conference on Empirical Methods in Natural Language Processing and the 9th International Joint Conference on Natural Language Processing (EMNLP-IJCNLP)},
  pages = {3982--3992},
  year = {2019},
}

@misc{wang2025jinav3,
  title = {jina-reranker-v3: Last but not late interaction for listwise document reranking},
  author = {Wang, Feng and Li, Yuqing and Xiao, Han},
  year = {2025},
  note = {arXiv:2509.25085},
}

@misc{langgraph2024,
  title = {{LangGraph}},
  author = {{LangChain AI}},
  year = {2024},
  note = {\url{https://github.com/langchain-ai/langgraph}},
}

@misc{sarthi2024raptorrecursiveabstractiveprocessing,
      title={RAPTOR: Recursive Abstractive Processing for Tree-Organized Retrieval}, 
      author={Parth Sarthi and Salman Abdullah and Aditi Tuli and Shubh Khanna and Anna Goldie and Christopher D. Manning},
      year={2024},
      eprint={2401.18059},
      archivePrefix={arXiv},
      primaryClass={cs.CL},
      url={https://arxiv.org/abs/2401.18059}, 
}

@inproceedings{zheng2023judging,
  title     = {Judging {LLM}-as-a-Judge with {MT}-Bench and {Chatbot} {Arena}},
  author    = {Zheng, Lianmin and Chiang, Wei-Lin and Sheng, Ying and Zhuang, Siyuan
               and Wu, Zhanghao and Zhuang, Yonghao and Lin, Zi and Li, Zhuohan
               and Li, Dacheng and Xing, Eric P. and Zhang, Hao and Gonzalez, Joseph E.
               and Stoica, Ion},
  booktitle = {Advances in Neural Information Processing Systems 36 (NeurIPS 2023),
               Datasets and Benchmarks Track},
  year      = {2023},
  note      = {arXiv:2306.05685},
}

@inproceedings{panickssery2024selfpreference,
  title     = {{LLM} Evaluators Recognize and Favor Their Own Generations},
  author    = {Panickssery, Arjun and Bowman, Samuel R. and Feng, Shi},
  booktitle = {Advances in Neural Information Processing Systems 37 (NeurIPS 2024)},
  year      = {2024},
  note      = {arXiv:2404.13076},
}

@misc{gu2024survey,
  title         = {A Survey on {LLM}-as-a-Judge},
  author        = {Gu, Jiawei and Jiang, Xuhui and Shi, Zhichao and Tan, Hexiang
                   and Zhai, Xuehao and Xu, Chengjin and Li, Wei and Shen, Yinghan
                   and Ma, Shengjie and Liu, Honghao and Wang, Yuanzhuo and Guo, Jian},
  year          = {2024},
  eprint        = {2411.15594},
  archivePrefix = {arXiv},
  primaryClass  = {cs.CL},
}

@misc{baek2025openwebuiopenextensible,
      title={Open WebUI: An Open, Extensible, and Usable Interface for AI Interaction}, 
      author={Jaeryang Baek and Ayana Hussain and Danny Liu and Nicholas Vincent and Lawrence H. Kim},
      year={2025},
      eprint={2510.02546},
      archivePrefix={arXiv},
      primaryClass={cs.HC},
      url={https://arxiv.org/abs/2510.02546}, 
}

@inproceedings{shankar2015epics,
  title     = {The {EPICS} Archiver Appliance},
  author    = {Shankar, Murali V. and Li, Luofeng F. and Davidsaver, Michael A. and Konrad, Martin G.},
  booktitle = {Proceedings of the 15th International Conference on Accelerator and Large Experimental Physics Control Systems (ICALEPCS'15)},
  pages     = {761--764},
  year      = {2015},
  month     = oct,
  address   = {Melbourne, Australia},
  publisher = {JACoW},
  paper     = {WEPGF030},
  isbn      = {978-3-95450-148-9},
  doi       = {10.18429/JACoW-ICALEPCS2015-WEPGF030},
  url       = {https://jacow.org/ICALEPCS2015/papers/WEPGF030.pdf}
}

@software{pymupdf,
  author = {{Artifex Software}},
  title = {PyMuPDF: A high performance Python library for data extraction, analysis, conversion \& manipulation of PDF documents},
  url = {https://github.com/pymupdf/pymupdf},
  year = {2026}
}

@inproceedings{smith2007overview,
  title={An overview of the Tesseract OCR engine},
  author={Smith, Ray},
  booktitle={Ninth International Conference on Document Analysis and Recognition (ICDAR 2007)},
  volume={2},
  pages={629--633},
  year={2007},
  publisher={IEEE Computer Society},
  doi={10.1109/ICDAR.2007.4376991}
}

@misc{qdrant,
  author       = {{Qdrant}},
  title        = {Qdrant: Vector Database for the Next Generation of {AI} Applications},
  year         = {2026},
  howpublished = {\url{https://qdrant.tech}},
  note         = {Accessed: 2026-08-22}
}

@misc{elasticsearch,
  author       = {{Elastic N.V.}},
  title        = {Elasticsearch: The Official Distributed Search \& Analytics Engine},
  year         = {2026},
  howpublished = {\url{https://www.elastic.co/elasticsearch}},
  note         = {Accessed: 2026-08-22}
}

@misc{neo4j,
  author       = {{Neo4j, Inc.}},
  title        = {Neo4j Graph Database Platform},
  year         = {2026},
  howpublished = {\url{https://neo4j.com}},
  note         = {Accessed: 2026-08-22}
}

\clearpage
\onecolumngrid
\begin{center}
  {\large\bfseries Supplementary material\\[2pt]
   A corrective agentic hybrid RAG and an operations-grounded\\
   evaluation for a scientific facility\par}
  \vspace{6pt}
  Rajat Sainju,\footnote{Author to whom correspondence should be addressed: rsainju@anl.gov}
  Dariusz Jarosz, Hairong Shang, Michael Prince,\\
  Ryan M. Aydelott, Mathew J. Cherukara, Yine Sun, and Michael D. Borland\par
  \vspace{3pt}
  \textit{Advanced Photon Source, Argonne National Laboratory, Lemont, Illinois 60439, USA}
\end{center}
\twocolumngrid

\setcounter{totalnumber}{10}
\setcounter{topnumber}{4}
\setcounter{dbltopnumber}{4}
\renewcommand{\topfraction}{0.95}
\renewcommand{\bottomfraction}{0.7}
\renewcommand{\dbltopfraction}{0.95}
\renewcommand{\textfraction}{0.05}
\renewcommand{\floatpagefraction}{0.7}
\renewcommand{\dblfloatpagefraction}{0.7}
\renewcommand{\thetable}{S\arabic{table}}
\renewcommand{\thefigure}{S\arabic{figure}}
\renewcommand{\thesection}{S\arabic{section}}
\setcounter{section}{0}
\setcounter{figure}{0}
\setcounter{table}{0}

\section{Native-tool ReAct and the MCP tooling layer}\label{s:mcp}

APS-RAG executes tools using a dual-path ReAct executor. The primary execution path employs native function calling through a vendor-aware /chat/ client, NativeToolsArgoLLM, which generates structured tool calls compatible with OpenAI, Gemini, and Anthropic backbone models. This client accommodates three distinct response-envelope formats: object, list, and bare string. A legacy prompt-based TOOL\_CALL regular expression path is maintained as a fallback mechanism. Tool functionality is provided by four FastMCP servers, each registered in .mcp.json (see Table~\ref{tab:mcp}). The archiver server supports PV and device resolution as well as historical PV data access through channel\_find and channel\_read. The graph server enables knowledge-graph search. The compute server provides calculator operations, sandboxed code execution, and tabular analysis. The image server supports reverse-image and text-to-image search over logbook attachments using SigLIP~\cite{zhai2023siglip} and FAISS~\cite{douze2024faiss}. In total, eight tools are available.

Each tool returns a standardized envelope containing status, result, and error fields. In-tool exceptions are intercepted by a decorator, ensuring that exceptions do not propagate across the standard I/O boundary. A persistent session pool maintains a dedicated client session for each server throughout the host process lifetime. This approach amortizes the substantial memory requirements of SigLIP and FAISS models, as well as the initialization overhead of Argo HTTP, both of which would otherwise contribute significantly to per-call latency. If a transport failure occurs, the affected session is terminated and reestablished. The robustness of this layer is validated by 818 tests, comprising 754 unit tests and 64 integration tests. These include contract tests aligned with each stage for naive-RAG redirection, native function calling, MCP server operations, and the persistent session pool, as well as an integration test that verifies ReAct retry behavior in response to tool errors.

\begin{table}[htbp]
  \centering
  \caption{The four FastMCP servers and their eight tools.\label{tab:mcp}}
  \small
  \begin{tabular}{l l}
    \hline
    Server & Tools \\
    \hline
    archiver & channel\_find, channel\_read \\
    graph    & graph\_search \\
    compute  & calculator, code\_executor, data\_analysis \\
    image    & image\_search [reverse-image, text-to-image] \\
    \hline
  \end{tabular}
\end{table}

\begin{table*}[htbp]
  \centering
  \caption{Per-component wall time (s) by planner model, 10 advanced-reasoning queries (deployment study, May 2026, pre-freeze stack). Component times can sum to more than the wall time because retrieval fans out in parallel. The per-component wall time also depends on Argo latency, traffic, chain-of-thinking triggers, and corrective loops. Note: enhancer = query-preprocessing + router + planner. The added latency by embedding models is $\sim$30 ms.\label{tab:latency}}
  \small
  \begin{tabular}{l rrrrrr}
    \hline
    Planner & Wall (s) &  Enhancer (s) & Retrieval (s) & Reranker (s) & Synthesis (s) \\
    \hline
    Gemini 2.5 Flash & $88.8 \pm 24.3$  & $31.0 \pm 29.1$ & $65.1 \pm 21.6$  & $18.2 \pm 16.6$ & $6.5 \pm 1.5$ \\
    Gemini 2.5 Pro   & $96.0 \pm 29.1$  & $29.7 \pm 12.6$ & $56.1 \pm 29.8$  & $19.2 \pm 17.4$ & $8.9 \pm 3.3$ \\
    GPT-5.4          & $105.9 \pm 11.6$ & $28.4 \pm 13.5$ & $60.0 \pm 31.4$  & $18.8 \pm 17.0$ & $8.5 \pm 3.8$ \\
    Claude Opus 4.6  & $140.4 \pm 44.5$ & $71.7 \pm 59.6$ & $109.7 \pm 34.3$ & $18.8 \pm 17.1$ & $8.5 \pm 3.8$ \\
    \hline
  \end{tabular}
\end{table*}

\section{Component-latency study and reranker A/B}\label{s:latency} 
Timing instrumentation for each pipeline component was achieved using a non-invasive harness implemented via class-level monkey patching with \texttt{perf\_counter}. Query-level timing isolation was accomplished through a context variable. For the naive path, a module-global timing dictionary was utilized, with retrieval operations distributed across a thread pool. Table~\ref{tab:latency} summarizes deployment-latency measurements by planner model, evaluated on a 10-query advanced-reasoning workload. Measurements were collected in May 2026 using the pre-freeze deployment stack. These results characterize the deployed service and are not intended as benchmark measurements on a frozen stack.

The planner is the language model responsible for generating the structured multi-step retrieval and reasoning plan; it serves as the independent variable and is varied across rows, while all other stages use fixed models (claude-sonnet-4.6 for the router and enhancer, GPT-5.4 for synthesis, and Jina v3 for reranking), subject to the isolation caveat noted above. Wall time is the end-to-end query latency from submission to the final cited answer, computed per query and averaged over the query set; it is less than the sum of the component columns because sub-retrievers execute in parallel and reranking is nested within retrieval. The enhancer expands the input query into semantic variants and, in a single combined call, also performs intent routing and — for agentic queries — plan generation, consolidating functions previously handled by separate router and planner calls. Retrieval is hybrid, fusing concurrent dense-vector retrieval (Qdrant with e5-large-v2 query embeddings), lexical retrieval (BM25 in Elasticsearch), and multi-hop knowledge-graph retrieval (Neo4j) via reciprocal-rank fusion; it includes query embedding, and its wall time encompasses the reranker stage. The reranker applies cross-encoder reranking (Jina v3, GPU-local, top-k = 50) to the fused candidate pool to select the final context set, and although reported separately, executes within retrieval. Finally, synthesis composes a grounded, citation-annotated answer from the reranked evidence.

A five-batch reranker A/B evaluation compared the Jina v3 cross-encoder~\cite{wang2025jinav3} with the legacy LLM scorer on advanced-reasoning queries, using \texttt{reranker\_top\_k}=50 and active corrective rounds. Results show that both approaches have similar wall time, although the LLM scorer exhibits a high-variance tail. This comparison addresses latency only. In contrast, the quality comparison, as reported in Table~IV of the main text, strongly favors Jina v3, which constitutes the principal finding. Per-query analysis highlights the quality gap: the legacy scorer not only produces inferior rankings but also frequently fails to generate usable results. Specifically, NDCG@10 is less than 0.02 for 63\% of the 19 subset questions and less than 0.10 for 68\%, with strict vital recall per question worse on 12 of 19, unchanged on 7, and better on none. The comparison among competent rerankers is reflected in the single-digit spread of Tables~\ref{tab:reranker-bm25} and~\ref{tab:reranker-hybrid}. The LLM-scorer default row in main-text Table~IV should be interpreted as a degraded fallback rather than as a peer reranker.

A five-batch reranker A/B evaluation compared the Jina v3 cross-encoder~\cite{wang2025jinav3} with the legacy LLM scorer on advanced-reasoning queries, using \texttt{reranker\_top\_k}=50 and active corrective rounds. Results show that both approaches have similar wall time, although the LLM scorer exhibits a high-variance tail. This comparison addresses latency only. In contrast, the quality comparison, as reported in Table~IV of the main text, strongly favors Jina v3, which constitutes the principal finding. Per-query analysis highlights the quality gap: the legacy scorer not only produces inferior rankings but also frequently fails to generate usable results. Specifically, NDCG@10 is less than 0.02 for 63\% of the 19 subset questions and less than 0.10 for 68\%, with strict vital recall per question worse on 12 of 19, unchanged on 7, and better on none. The comparison among competent rerankers is reflected in the single-digit spread of Tables~\ref{tab:reranker-bm25} and~\ref{tab:reranker-hybrid}. The LLM-scorer default row in main-text Table~IV should be interpreted as a degraded fallback rather than as a peer reranker.

\section{Reliability-focused engineering effort}\label{s:reliability}
Deployment within live operational environments necessitated robust adaptation to ongoing infrastructure drift. The following representative events were identified, each detected or confirmed through the latency harness and safeguarded by the test suite. First, an Argo gateway migration was observed, which eliminated persistent 502 retry storms. Second, a response-envelope mismatch in the \texttt{/chat/} endpoint, characterized by a bare-string response shape, silently disrupted planner functionality until appropriate handling was implemented. Third, an acronym-resolution chain was silently disabled in production due to a path, import, or API drift bug. Fourth, phantom corrective rounds were triggered by scoring a post-formatted draft, which was resolved by instead scoring the raw draft. Finally, planner-sentinel false positives were observed in tool-skip accounting. These events collectively illustrate the need for comprehensive monitoring and rigorous testing to ensure system resilience amid infrastructure evolution.

\section{The APS-RAG retrieval skill}\label{s:skill}
The retrieval layer is implemented as a self-contained agent skill, \texttt{aps-rag}, which can be invoked directly from coding assistants such as Claude Code and Codex. The skill exposes the corpus as a retrieval-as-a-service primitive: rather than embedding a generator within the tool, it returns evidence passages for synthesis by the host model. Retrieval and generation are thus decoupled; the skill is responsible solely for surfacing relevant passages, while the calling model composes and cites the answer. This design enables agents that process tool outputs to reuse the indexed knowledge base without requiring the full agentic stack.

The skill is structured as three decoupled tiers. (i) A stateless FastAPI gateway performs lexical (BM25) retrieval directly against the live Elasticsearch indices; it contains no LLM and is strictly read-only. The gateway shares only the indices with the production agentic system, with no shared process, model, or datastore beyond Elasticsearch. As a result, the lightweight skill path and the corrective-agentic service are operationally independent and do not affect each other's stability. (ii) A thin MCP server fronts the gateway and exposes two tools to the host model: \texttt{query\_aps\_rag}, which issues a query and returns the top-$k$ ranked passages, each annotated with its source collection, document identifier, passage text, and retrieval score, and \texttt{aps\_rag\_health}, which reports gateway liveness and index
availability. The MCP layer contains no retrieval or generation logic and serves solely as a protocol bridge between the agent runtime and the gateway. (iii) The host model (Claude Code or Codex) consumes the returned passages and synthesizes the final response, attributing each claim to the (collection, document id) of the supporting passage, ensuring that the answer includes verifiable, corpus-grounded citations.

The design is intentionally minimal. Limiting the skill to BM25 retrieval over shared indices ensures low and predictable latency, deterministic behavior, and read-only safety, while avoiding redundant generation and any dependency on the embedding, reranking, knowledge-graph, or LLM components of the agentic system. The skill can be deployed wherever the Elasticsearch indices are accessible. Distribution is via an internal GitLab repository, accompanied by an install script that registers the MCP server in the agent configuration and specifies the gateway endpoint, enabling \texttt{/aps-rag} as a first-class command within existing agent sessions.

\section{Answer-match metrics: exact match, F1, and accuracy}\label{s:emf1acc}
To facilitate comparison with the short-form question answering literature, such as SQuAD and Natural Questions, which report exact match (EM), F1, and containment accuracy metrics as used in Adaptive-RAG, Self-RAG, FLARE, and related retrieval-augmented QA studies, we compute three corresponding answer-match metrics. Accuracy and F1 are presented alongside the main-text generation results in Tables II and IV, while EM is detailed below. Two methodological adaptations are necessary for this comparison. First, the APS-Bench gold standard consists of long-form answers, comprising a grounded paragraph and atomic nuggets, rather than a short factoid span. Therefore, we derive a concise reference answer for each question. Second, our systems generate long-form operational answers rather than short spans, which renders a verbatim whole-response match uninformative and limits the precision of token overlap when compared to a short reference. Both adaptations are addressed in the following sections.

\paragraph{Short reference answers.} For each answerable question we extract a short canonical answer $a^\star$ (a value with units, a PV/device name, a command, or a minimal phrase) together with an alias set $\mathcal{A}$ of equivalent surface forms (with/without units, numeric vs.\ spelled, common abbreviations) from the long-form gold answer and its vital nuggets, using the generator-family model (\texttt{gpt54}, temperature~0). These are LLM-extracted and carry the same
These short reference answers are subject to the same validation-pending status as the gold standard. We define the set $\mathcal{G} = \{a^\star\} \cup \mathcal{A}$, where $\hat{y}$ denotes the system's answer. All string comparisons are performed using SQuAD normalization $\mathcal{N}(\cdot)$, which includes lowercasing, removal of punctuation and articles (a, an, the), and whitespace normalization.

\paragraph{EM (exact match; strictest).} The exact match metric assigns $\mathrm{EM}=1$ if there exists a $g \in \mathcal{G}$ such that $\mathcal{N}(g)$ appears as a contiguous substring within $\mathcal{N}(\hat{y})$; otherwise, $\mathrm{EM}=0$. For paragraph-length answers, whole-response string equality, as used in standard EM, is nearly always zero. Therefore, we report exact-span containment, which represents the strictest form of meaningful containment. In this context, EM serves as a conservative lower bound on verbatim agreement and should not be interpreted as a measure of overall system accuracy.

\paragraph{F1 (token overlap).} The standard token-level $F1$ between the reference and the best-matching span of the response. For $\mathcal{N}(g)$ of token length $\ell$ we slide windows of length $\ell\pm2$ over the response tokens and take the maximum over windows and over $g\in\mathcal{G}$,
\[
F1=\max_{g\in\mathcal{G}}\ \max_{\text{window }w}\ \frac{2\,P\,R}{P+R},\quad
P=\frac{|w\cap g|}{|w|},\ \ R=\frac{|w\cap g|}{|g|},
\]
Here, $|\cdot \cap \cdot|$ denotes the multiset token overlap. Computing F1 over the best-matching span, rather than the entire response, mitigates the loss of precision that would otherwise result from comparing a long answer to a short reference.

\paragraph{Acc (containment accuracy).} Whether the response \emph{contains the correct answer}. Lexical matching systematically under-credits correct paraphrases, synonyms, and equivalent units (e.g.\ 14.5~nC'' vs.\ 14.5~nanocoulombs’‘, accumulator ring'' vs.\ Particle Accumulator Ring’‘); we verified that token $F1$/EM track the semantic vital-nugget recall while genuine correctness is higher. Accuracy is therefore judged \emph{semantically} by the cross-family primary judge (\texttt{claudeopus46}; Anthropic $\neq$ generator \texttt{gpt54}, OpenAI—the same generator$,\neq,$judge guard used for every other LLM-judged metric here). Given the question, the gold answer (and $a^\star$), and the response, the judge returns a binary decision on whether the response states the correct answer, explicitly accepting paraphrase, synonymy, equivalent units, and additional correct context; $\mathrm{Acc}=1$ for ``yes’'.

Aggregation and status. Each metric is computed as the average over all answerable questions and across three evaluation seeds, with standard deviation approximately zero at temperature zero. The metrics are additive, as they are derived from cached system outputs without requiring additional system runs, and they do not affect nugget-recall, faithfulness, or operational evaluation layers. On APS-Bench, the Agentic GraphRAG system achieves EM, F1, and Acc values of 32.7\%, 70.3\%, and 69.4\%, respectively, which are consistent with its semantic strict-vital-nugget recall. The concordance between the independent lexical (F1) and semantic (Acc, nugget recall) metrics indicates that the answer-correctness measurement is robust and not an artifact of any single matching approach. Since the systems generate long-form operational answers rather than short spans, EM is necessarily low and should not be interpreted as a measure of system accuracy. Instead, Acc, which reflects semantic containment, is the appropriate metric for determining whether the response contains the correct information. Both the short reference answers and the Acc judge are produced by large language models.

\begin{table*}[t]
  \centering
  \caption{Query-type-adaptive reciprocal-rank-fusion weights. The router recognizes eight operational intents; they carry a hand-set weight profile $(\text{dense}, \text{BM25}, \text{graph})$. Weights are set from operational experience, not learned.\label{tab:rrf-weights}}
  \small
  \begin{tabular}{@{}l ccc l@{}}
    \hline
    Intent & \makecell{Dense} & \makecell{BM25} & \makecell{Graph} & Emphasis \\
    \hline
    factual             & 0.2 &\textbf{0.5} & 0.3 & BM25 \\
    value               & 0.1 & \textbf{0.6} & 0.3 & BM25 (numerical)\\
    temporal            & 0.3 & \textbf{0.5} & 0.2 & BM25 (date/shift tokens) \\
    troubleshooting     & 0.3 & 0.2 & \textbf{0.5} & graph \\
    multi\_hop          & 0.3 & 0.2 & \textbf{0.5} & graph \\
    comparative         & \textbf{0.4} & 0.3 & 0.3 & balanced \\
    code                & 0.5 & 0.5 & 0.0 & balanced \\
    general             & \textbf{0.4} & 0.3 & 0.3 & balanced \\
    \hline
  \end{tabular}
\end{table*}

\begin{table*}[tp]
  \centering
  \caption{Construction-time composition of the two benchmarks ($n=50$ each), reported side by side and
  never pooled. ``Gold min-rank'' is the mean rank at which the first gold document is retrieved (lower
  is easier).}\label{tab:abcompose}
  \small
  \begin{tabular}{lr}
    \hline
    Property & APS-Bench \\
    \hline
    Answerable / abstention            & 50 / 0 \\
    Mean nuggets per question          & 6.27 \\
    Mean gold-doc min-rank             & 0.41 \\
    \hline
    \multicolumn{2}{l}{\emph{Query types}} \\
    \quad factual / temporal           & 7 / 6 \\
    \quad multi-hop / procedural       & 8 / 8 \\
    \quad causal / comparative         & 6 / 8 \\
    \quad troubleshooting / parametric & 5 / 2 \\
    \hline
    \multicolumn{2}{l}{\emph{Difficulty}} \\
    \quad easy / medium / hard         & 15 / 25 / 10 \\
    \hline
  \end{tabular}
\end{table*}

\begin{table*}[htbp]
  \centering
  \caption{Effect of embedding model on retrieval (APS-Bench; 10k-index diagnostic, first-stage,
  no reranker; mean over subsamples). Models: e5-large-v2~\cite{wang2022e5},
  bge-m3~\cite{chen2024bgem3}, all-MiniLM-L12-v2~\cite{wang2020minilm,reimers2019sbert}, and the
  domain-pretrained accphysbert~\cite{hellert_domain-specific_2025}.\label{tab:embedder-effect}}
  \small
  \begin{tabular}{l cccc}
    \hline
    Embedding Model        & Model Size  & MRR@10 (\%) & NDCG@10 (\%) & Recall@10 (\%) \\
    \hline
    BM25                   &  \textemdash & 82.1 & 69.4 & 63.1 \\
        e5-large-v2 vector &  335M   & 70.2 & 57.2 & 52.8 \\
        e5-large-v2 hybrid & 335M   & 79.2 & 67.5 & 61.9 \\
        bge-m3 vector      & 568M    & 79.2 & 65.6 & 60.6 \\
        bge-m3 hybrid      & 568M    & 85.1 & 72.3 & 65.7 \\
        all-MiniLM-L12-v2 vector & 33M & 60.6 & 48.6 & 43.9 \\
        all-MiniLM-L12-v2 hybrid & 33M & 77.8 & 64.1 & 59.9 \\
        accphysbert vector      & $\sim$ 110M & 58.5 & 43.1 & 37.1 \\
        accphysbert hybrid      & $\sim$ 110M& 74.6 & 60.9 & 57.0 \\
    \hline
  \end{tabular}
\end{table*}

\begin{table*}[htbp]
  \centering
  \caption{Effect of re-ranker model over BM25 first-stage retrieval (APS-Bench; 10k-index
  diagnostic). Rerankers: jina-reranker-v3~\cite{wang2025jinav3},
  bge-reranker-v2-m3~\cite{chen2024bgem3}, and qwen3-reranker~\cite{zhang_qwen3_2025}. Precision@10
  is not reported because each gold answer spans 2--3 passages.\label{tab:reranker-bm25}}
  \small
  \begin{tabular}{l cccc}
    \hline
    Reranker Model        & Model Size  & MRR@10 (\%) & NDCG@10 (\%) & Recall@10 (\%) \\
    \hline
    (no rerank)           & \textemdash & 62.1 & 49.4 & 43.1 \\
    jina-reranker-v3      & 0.6B        & 86.9 & 73.8 & 64.1 \\
    bge-reranker-v2-m3    & 0.6B       & 83.3 & 71.5 & 64.6 \\
    qwen3-reranker-4B     & 4B       & 84.7 & 71.9 & 66.4 \\
    \hline
  \end{tabular}
\end{table*}

\begin{table*}[htbp]
  \centering
  \caption{Effect of re-ranker model over hybrid (BM25\,+\,vector, RRF) first-stage retrieval
  (APS-Bench; 10k-index diagnostic). Precision@10 is not reported because each gold answer spans
  2--3 passages. \label{tab:reranker-hybrid}}
  \small
  \begin{tabular}{l cccc}
    \hline
    Reranker Model        & Model Size  & MRR@10 (\%) & NDCG@10 (\%) & Recall@10 (\%) \\
    \hline
    (no rerank)           & \textemdash & 64.2 & 57.6 & 47.2 \\
    jina-reranker-v3      & 0.6B        & 86.1 & 73.7 & 64.8 \\
    bge-reranker-v2-m3    & 0.6B       & 83.0 & 71.3 & 65.3 \\
    qwen3-reranker-4B     & 4B       & 85.7 & 71.9 & 67.3 \\
    \hline
  \end{tabular}
\end{table*}

\clearpage

\section{Supplementary figures}\label{s:figs}

\begin{figure*}[htbp]
\begin{lstlisting}[style=jsonschema]
{
  "$schema": "corpus_v2",
  "$id": "aps-rag/corpus-record.json",
  "title": "APS-RAG unified corpus record",
  "$comment": "One indexed chunk. Every collection is normalized to CorpusRecord; work_requests adds chunk-provenance fields.",
  "$defs": {
    "CorpusRecord": {
      "type": "object",
      "additionalProperties": false,
      "required": ["doc_id", "content", "title", "author", "collection", "system", "creation_datetime", "file_id"],
      "properties": {
        "doc_id":   {"type": "string", "pattern": ".+__chunk_\\d+$", "description": "chunk id '<source>__chunk_<n>'"},
        "content":  {"type": "string", "minLength": 1, "description": "chunk body: BM25-analyzed and embedded"},
        "title":    {"type": "string"},
        "author":   {"type": "string", "description": "may be empty"},
        "collection": {
          "type": "string",
          "enum": ["bely_logbook", "icms_collection", "work_requests",
                   "teams_messages", "sdds_collection", "oag_collection"]
        },
        "system":   {"type": "string", "description": "accelerator system; may be empty"},
        "creation_datetime": {
          "anyOf": [{"type": "string", "format": "date-time"},
                    {"type": "string", "maxLength": 0}]
        },
        "file_id":  {"type": "string"}
      }
    },
    "WorkRequestRecord": {
      "$comment": "work_requests: CorpusRecord plus chunk-provenance fields",
      "allOf": [{"$ref": "#/$defs/CorpusRecord"}],
      "unevaluatedProperties": false,
      "properties": {
        "parent_id":   {"type": "string"},
        "chunk_index": {"type": "integer", "minimum": 0},
        "chunk_type":  {"type": "string"}
      },
      "required": ["parent_id", "chunk_index", "chunk_type"]
    }
  },
  "oneOf": [
    {"allOf": [{"$ref": "#/$defs/CorpusRecord"}],
     "properties": {"collection": {"not": {"const": "work_requests"}}}},
    {"allOf": [{"$ref": "#/$defs/WorkRequestRecord"}],
     "properties": {"collection": {"const": "work_requests"}}}
  ]
}
\end{lstlisting}
\caption{A unified corpus-record schema, based on JSON Schema, is implemented across all eight collections. Each indexed chunk is validated against \texttt{CorpusRecord}, while the \texttt{work\_requests} collection is further validated against \texttt{WorkRequestRecord}, constrained by the terminal \texttt{oneOf}. The eight required fields facilitate standardized filtering, temporal scoping, and source citation across otherwise heterogeneous sources. Among these, \texttt{content} is the only field subjected to BM25 analysis and embedding.}
\end{figure*}
\twocolumngrid

\begin{figure*}[t!]
\begin{lstlisting}[style=jsonschema]
{
  "$schema": "v2",
  "$id": "aps-rag/bely-logbook-source.json",
  "title": "BELY logbook native source schema (nested: file -> entry -> sub-entry)",
  "$comment": "24 files, 4536 entries, 29567 sub-entries. Each entry is flattened to CorpusRecord at index time.",
  "$ref": "#/$defs/BelyLogbookFile",
  "$defs": {
    "BelyLogbookFile": {
      "type": "object",
      "additionalProperties": false,
      "required": ["subfolder_name", "total_entries", "processing_timestamp", "entries"],
      "properties": {
        "subfolder_name":      {"type": "string", "description": "beamline/section, e.g. '25id', 'studies-sr'"},
        "total_entries":       {"type": "integer", "minimum": 0},
        "processing_timestamp":{"type": "string", "format": "date-time"},
        "entries": {"type": "array", "items": {"$ref": "#/$defs/BelyLogbookEntry"}}
      }
    },
    "BelyLogbookEntry": {
      "type": "object",
      "$comment": "Operational entries leave studies_plan/summary empty; SR studies entries populate them (subtype by value of 'logbook').",
      "additionalProperties": false,
      "required": [
        "BELY_ID", "version", "logbook", "machine/system", "bely_filename",
        "bely_creation_datetime", "created_by", "topic", "personnel_involved",
        "keywords", "studies_plan", "studies_summary", "bely_sub_entries",
        "processing_timestamp"
      ],
      "properties": {
        "BELY_ID":       {"type": "string", "description": "numeric-string id, e.g. '10617'"},
        "version":       {"type": "string"},
        "logbook":       {"type": "string", "description": "source logbook / discriminator, e.g. 'Operations', 'SR', '25-ID'"},
        "machine/system":{"type": "string"},
        "bely_filename": {"type": "string"},
        "bely_creation_datetime": {"type": "string", "format": "date-time"},
        "created_by":    {"type": "string"},
        "topic":         {"type": "string"},
        "personnel_involved": {"type": "array", "items": {"type": "string"}},
        "keywords":      {"type": "array", "items": {"type": "string"}},
        "studies_plan":    {"type": "string", "description": "populated only for SR studies; '' otherwise"},
        "studies_summary": {"type": "string", "description": "populated only for SR studies; '' otherwise"},
        "bely_sub_entries": {"type": "array", "items": {"$ref": "#/$defs/BelySubEntry"}},
        "processing_timestamp": {"type": "string", "format": "date-time"}
      }
    },
    "BelySubEntry": {
      "type": "object",
      "additionalProperties": false,
      "required": ["entry_text", "attachment_urls"],
      "properties": {
        "entry_text":      {"type": "string"},
        "attachment_urls": {"type": "array", "items": {"type": "string", "format": "uri"}}
      }
    }
  }
}
\end{lstlisting}
\caption{The BELY logbook native source schema is unique in its composite, three-level nested structure. Each group corresponds to a \texttt{BelyLogbookFile} template ($24$ files in the frozen index), which contains \texttt{BelyLogbookEntry} records ($4{,}536$). Each entry subsequently holds chronological \texttt{BelySubEntry} posts. During indexing, entries and their sub-entries are flattened into the unified corpus record.}
\end{figure*}

\begin{figure}[htbp]
\includegraphics[width=\columnwidth]{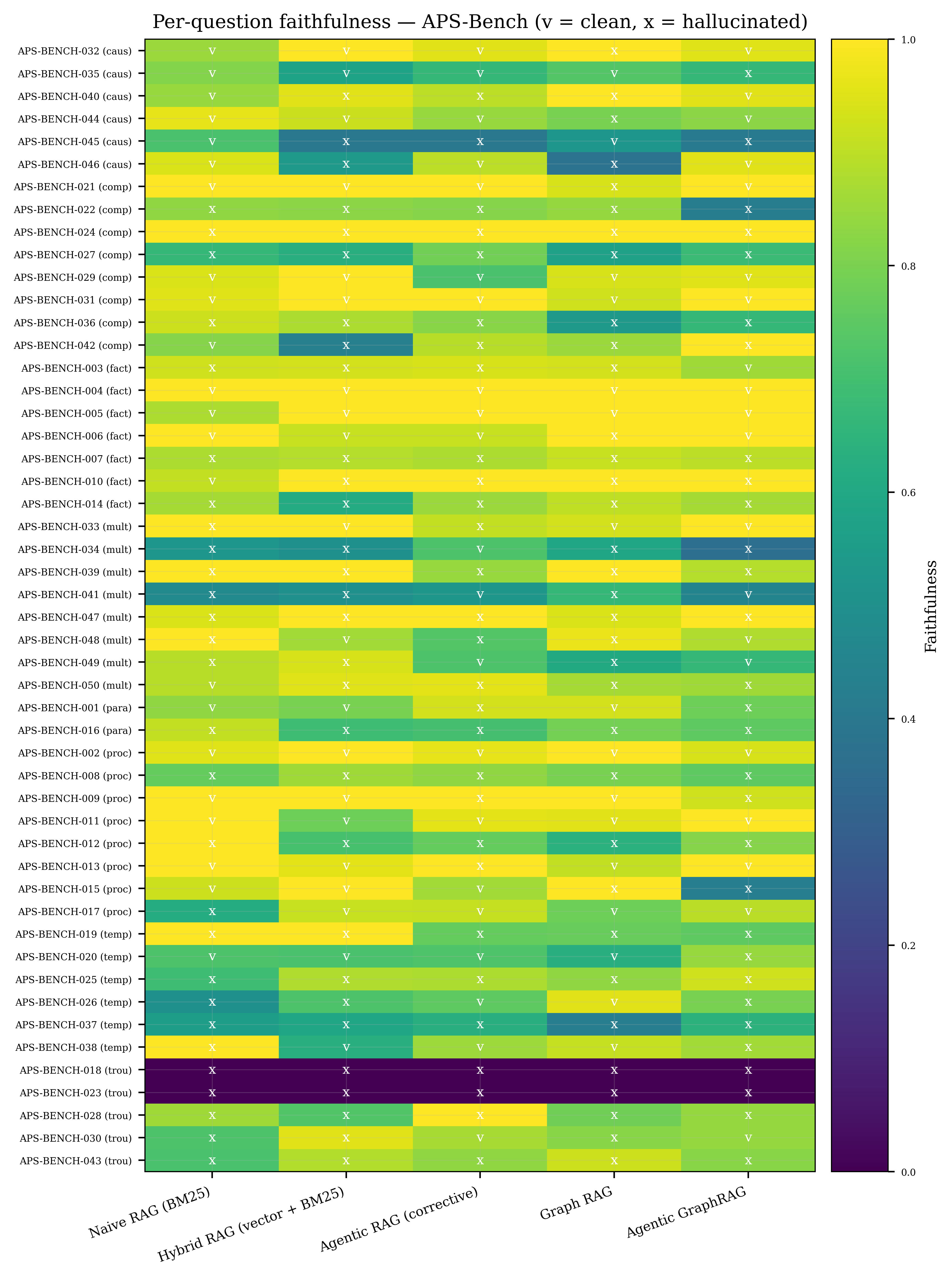}
\caption{\label{fig:s1}Per-question faithfulness on APS-Bench is reported, with each cell representing the per-question faithfulness ($1-$hallucination rate) of a system. Questions are grouped by query type along the vertical axis. The overlay indicates whether each cell is hallucination-free (v) or hallucinated (x), based on the HHEM-2.1 backstop with a threshold of $0.5$. Failures cluster by question rather than by system; a small subset of challenging items, particularly two troubleshooting questions, are hallucinated by all five variants. Consequently, the faithfulness tail is primarily determined by question difficulty rather than system architecture.}
\end{figure}

\begin{figure}[htbp]
\includegraphics[width=\columnwidth]{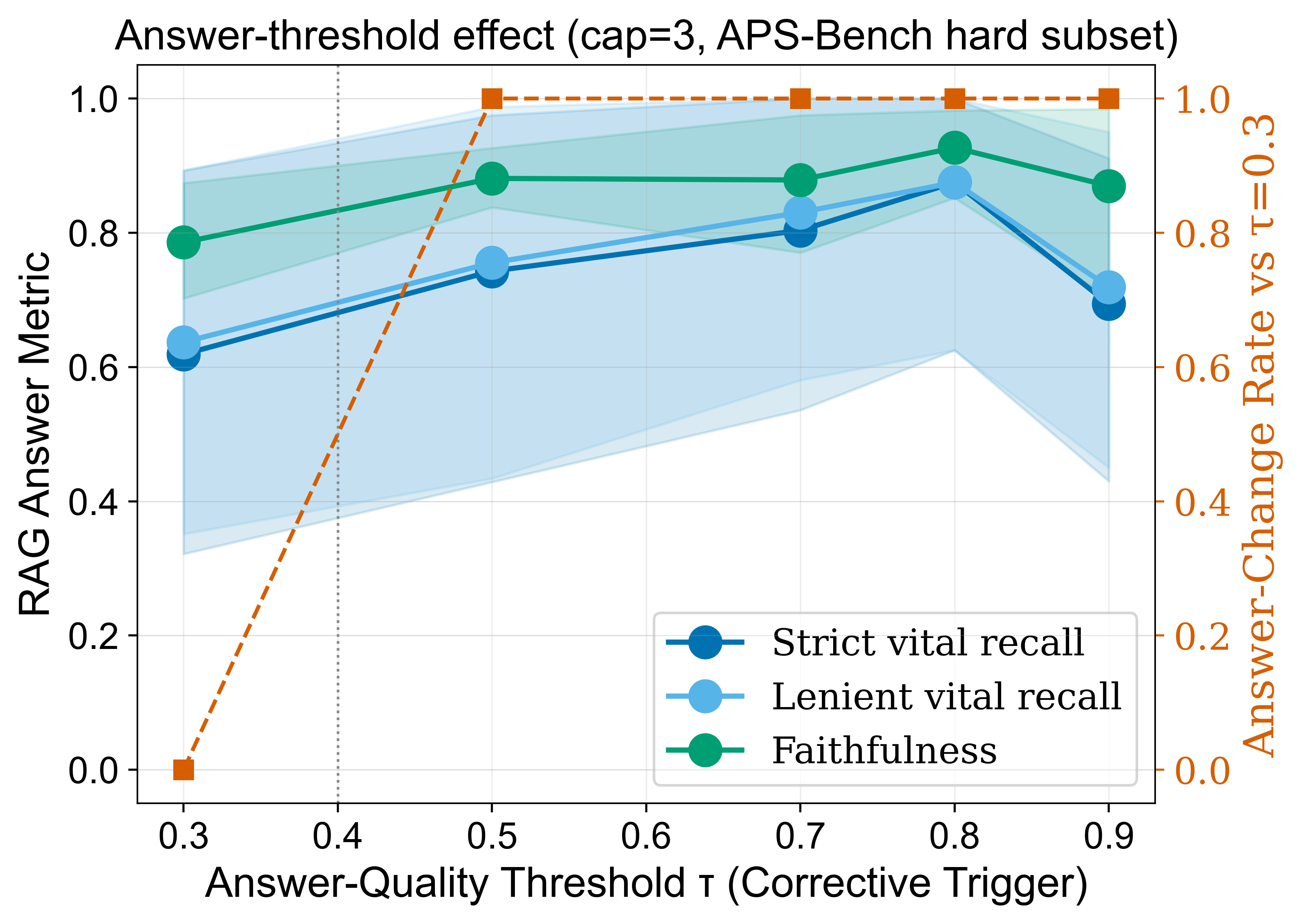}
\caption{The sensitivity of answer quality to the corrective-trigger threshold $\tau$ is evaluated on the APS-Bench ablation subset (main-text Sec.~V\,C), with the round cap fixed at three. The left axis presents strict vital recall, lenient vital recall, and claim-level faithfulness (with 95\% bootstrap confidence interval bands), while the right axis shows the fraction of answers differing from the $\tau=0.3$ baseline. The vertical dotted line indicates the deployed default $\tau=0.4$. All three answer metrics increase modestly from $\tau=0.3$, plateau near $\tau=0.7$ to $0.8$, and decrease at $\tau=0.9$. The answer-change rate reaches saturation at $1.0$ by $\tau=0.5$, as higher thresholds trigger re-retrieval for nearly every question. Given the wide and overlapping confidence intervals at this subset size, the sweep suggests low sensitivity around the deployed default, rather than providing justification for retuning.}
\end{figure}

\begin{figure}[htbp]
\includegraphics[width=\columnwidth]{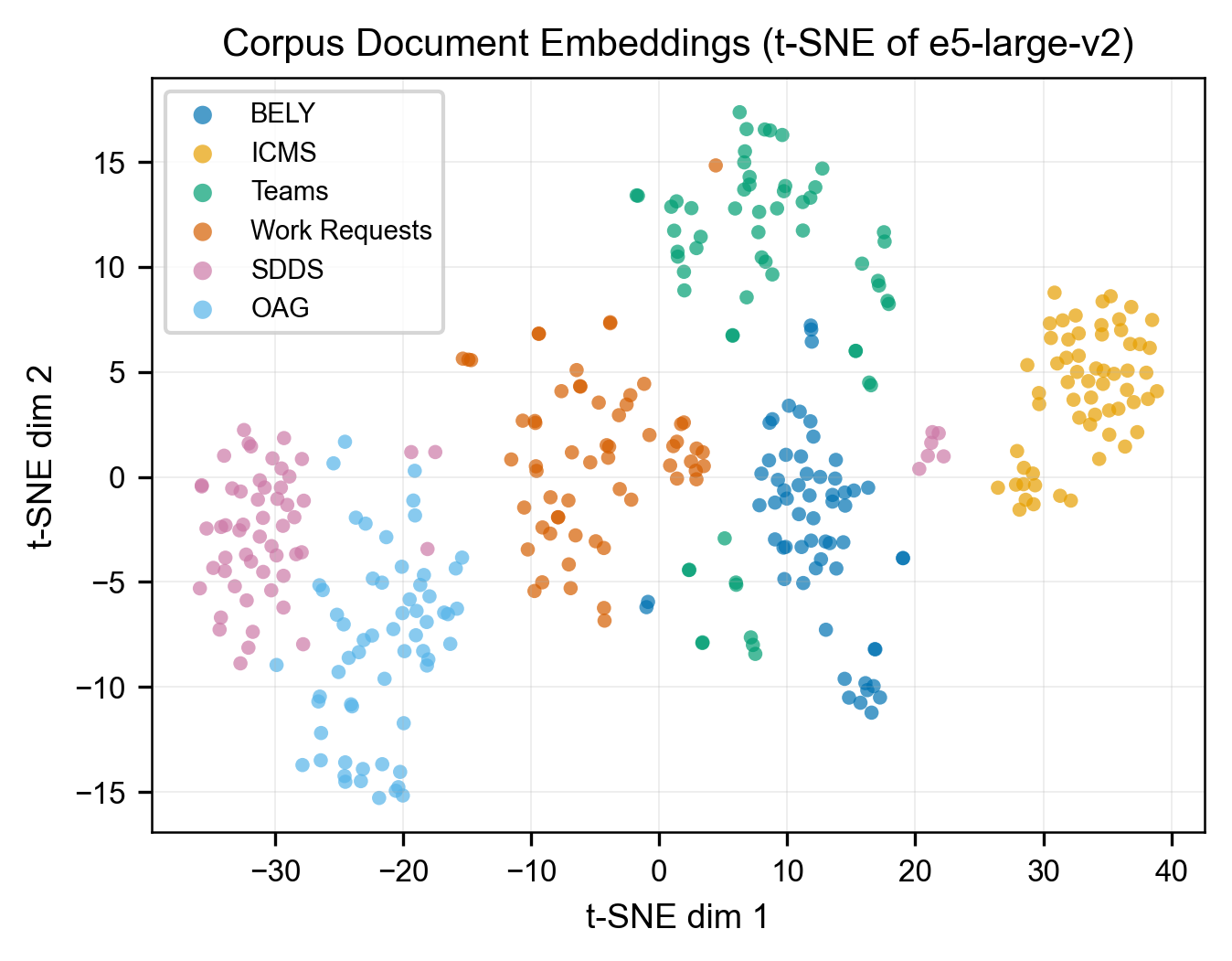}
\caption{\label{fig:s2}A two-dimensional t-SNE~\cite{vandermaaten2008tsne} projection of e5-large-v2 embeddings is presented for corpus documents from six APS collections (BELY, ICMS, Work Requests, Teams, SDDS, OAG) in APS-Bench. The collections occupy distinct but partially overlapping regions, with cross-collection overlap where the same devices and faults recur. This observed heterogeneity provides empirical motivation for the hybrid, graph-augmented design. The projection is performed at the corpus level and remains independent of the benchmark.}
\end{figure}

\begin{figure}[htbp]
\includegraphics[width=\columnwidth]{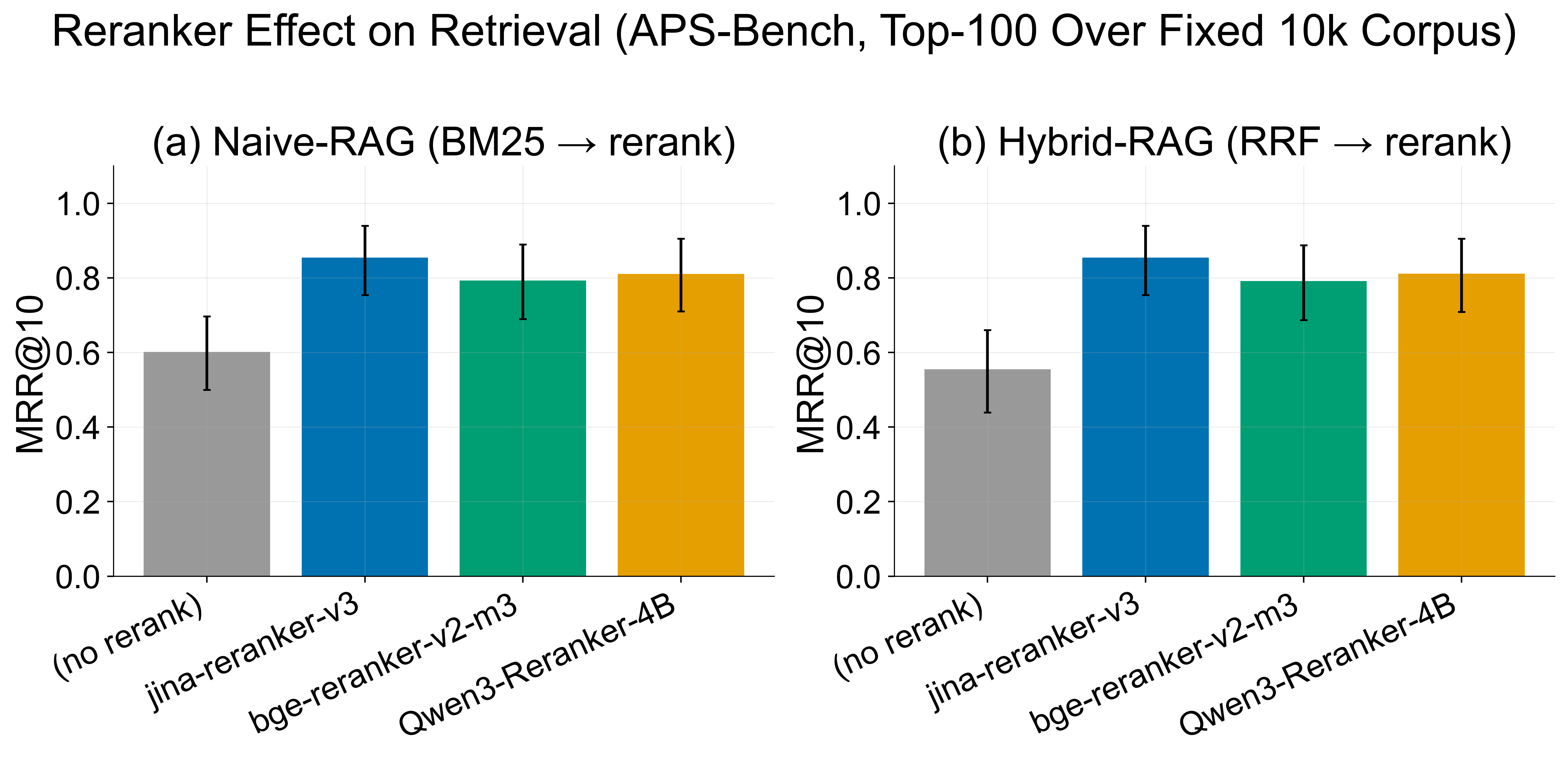}
\caption{\label{fig:reranker}The effect of the cross-encoder reranker on first-stage retrieval quality is evaluated on APS-Bench, with the top-100 candidates reranked over a fixed 10,000-document diagnostic corpus. MRR@10 is reported for (a) a BM25 (Naive-RAG) first stage and (b) an RRF hybrid (Hybrid-RAG) first stage, each evaluated without reranking and with three cross-encoders (jina-reranker-v3, bge-reranker-v2-m3, Qwen3-Reranker-4B). Error bars represent 95\% bootstrap confidence intervals. The jina-reranker-v3 achieves the highest point estimate in both settings, but the differences among the three rerankers are within single digits, and their confidence intervals overlap. This comparison is among functioning rerankers and should be interpreted alongside the component ablation (main-text Table~IV), where removing the cross-encoder with direct LLM assessment, rather than a competent alternative, results in a substantial performance drop.}
\end{figure}

\begin{figure}[htbp]
\includegraphics[width=\columnwidth]{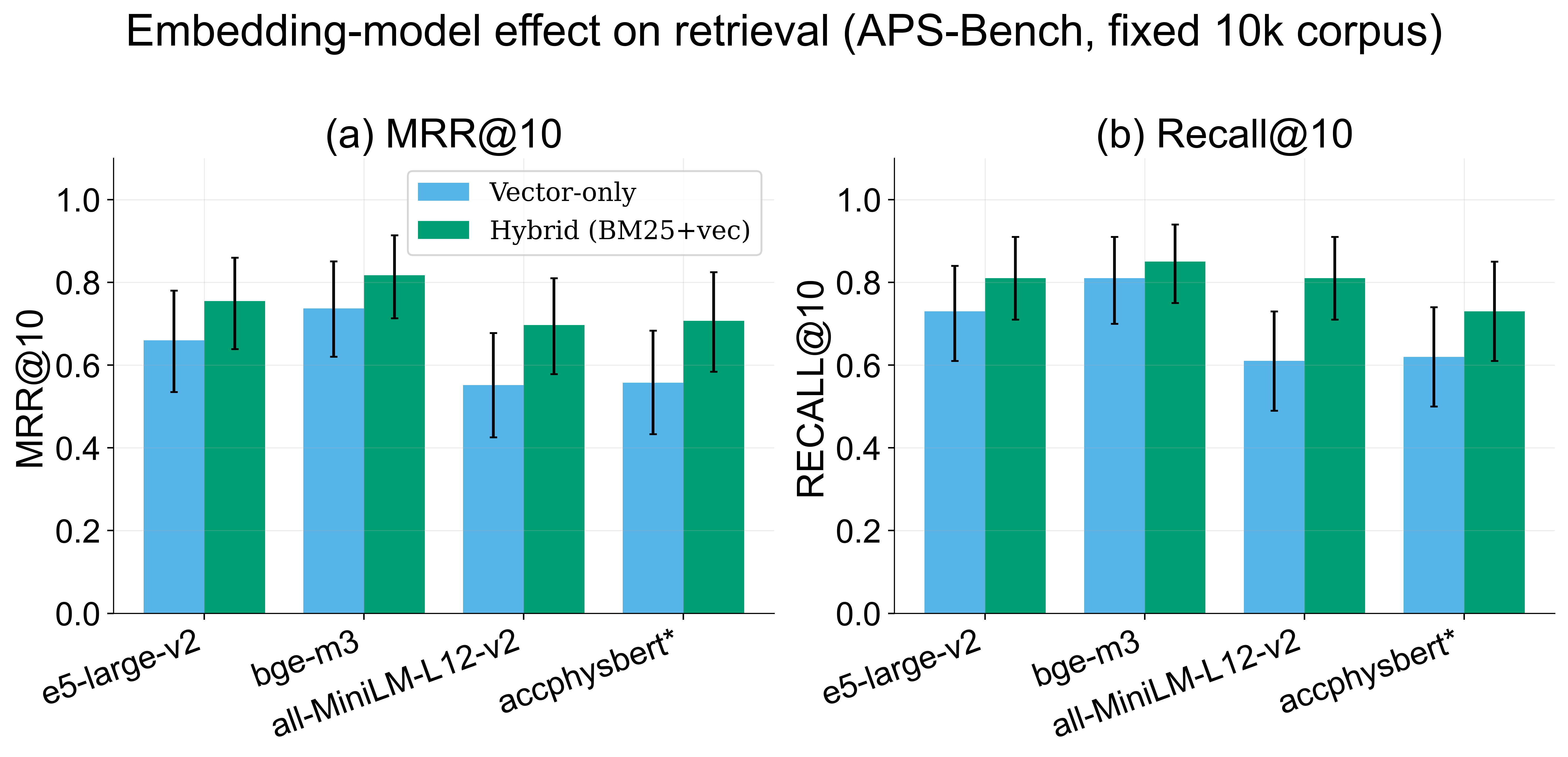}
\caption{\label{fig:retrievalquality}Retrieval quality by embedding model is assessed on APS-Bench using a fixed 10,000-document diagnostic corpus, with 95\% bootstrap confidence interval error bars, for both vector-only and hybrid (vector plus BM25) retrieval. Metrics reported include (a) MRR@10 and (b) Recall@10. Four models are compared: the deployed default e5-large-v2, bge-m3, all-MiniLM-L12-v2, and the domain-pretrained accphysbert. The dashed line indicates first-stage naive BM25. The results show an ordered spread (Table~\ref{tab:embedder-effect}): vector-only MRR@10 is highest for bge-m3 ($\approx$79\%), followed by e5-large-v2 ($\approx$70\%), with all-MiniLM-L12-v2 and accphysbert lower ($\approx$59--61\%). Hybrid fusion increases these values to approximately $75$--$85\%$, and first-stage BM25 alone is strong ($\approx$82\%). Notably, the 95\% confidence intervals overlap, so the ordering is suggestive rather than definitive. The domain-pretrained accphysbert does not outperform the general-purpose models; bge-m3 only marginally surpasses e5-large-v2, and the cross-encoder reranker remains the most significant retrieval improvement (main-text Table~IV).}
\end{figure}

\begin{figure*}[t]
\includegraphics[width=\textwidth]{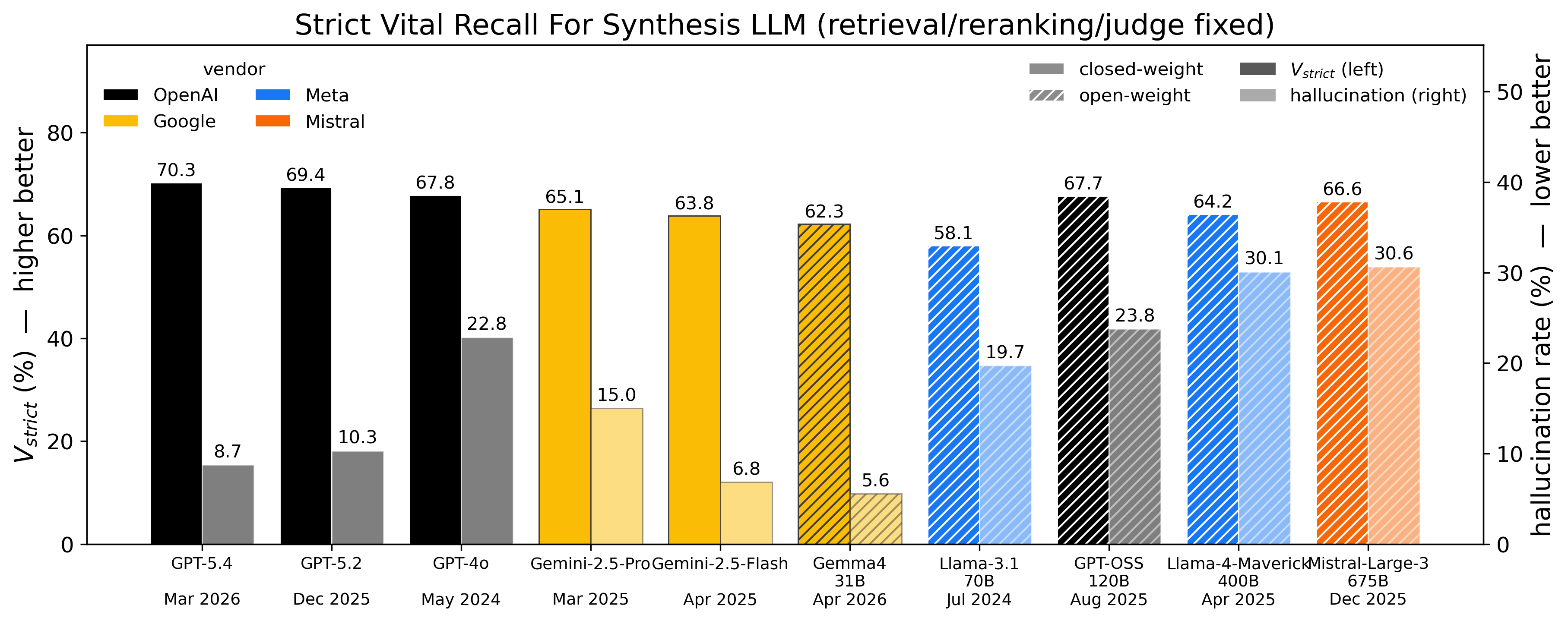}
\caption{\label{fig:synthesis-llm-compare} This figure examines the sensitivity of strict vital $V_{\text{strict}}$ and hallucination rate to the choice of synthesis large language model (LLM). The complete Agentic GraphRAG pipeline was re-executed on APS-Bench, systematically substituting the synthesis model across ten LLMs from four vendors, while keeping retrieval, fusion, reranking, and the cross-family judge constant. For each model, the dark bar represents $V_{\text{strict}}$ (left axis; higher values indicate better performance), and the light bar indicates the claim-level hallucination rate (right axis; lower values are preferable). Vendor is encoded by color; hatching differentiates open-weight from closed-weight models; release month is displayed below each model; and parameter counts are provided for open-weight models. Anthropic models are intentionally excluded because Claude serves as the fixed judge, ensuring that the generator and judge families are always distinct. The deployed generator (GPT-5.4, \texttt{gpt54} via ARGO) matches the full-system values reported in main-text Table II (70.3\%/8.7\%). Two primary patterns emerge from this analysis. First, strict vital recall demonstrates relative insensitivity to the generator; the ten-model range is 58.1--70.3\%, and the best open-weight model (GPT-OSS 120B, 67.7\%) is within 2.6 points of the closed-weight frontier. This finding aligns with the expectation that recall is primarily determined by the fixed retrieval and reranking front end. Second, hallucination rates vary by more than a factor of five (5.6--30.6\%) and do not correlate with either recall or parameter count. Notably, the 31B open-weight Gemma4 model exhibits the highest faithfulness (5.6\%, outperforming the deployed GPT-5.4 at 8.7\%), while the 400B--675B open-weight models are the least faithful (30.1--30.6\%). Additionally, faithfulness improves across successive generations within the OpenAI model line (22.8\% to 10.3\% to 8.7\%). The decoupling of recall and faithfulness observed in the component ablation (main-text Sec. V C) is thus also evident at the generator level. All values are point estimates based on a single frozen answer set; recall differences of a few points fall within the bootstrap confidence interval widths reported in main-text Table II, so the precise ranking should be interpreted as suggestive rather than definitive.}
\end{figure*}

\begin{figure*}[t]
\includegraphics[width=0.95\textwidth]{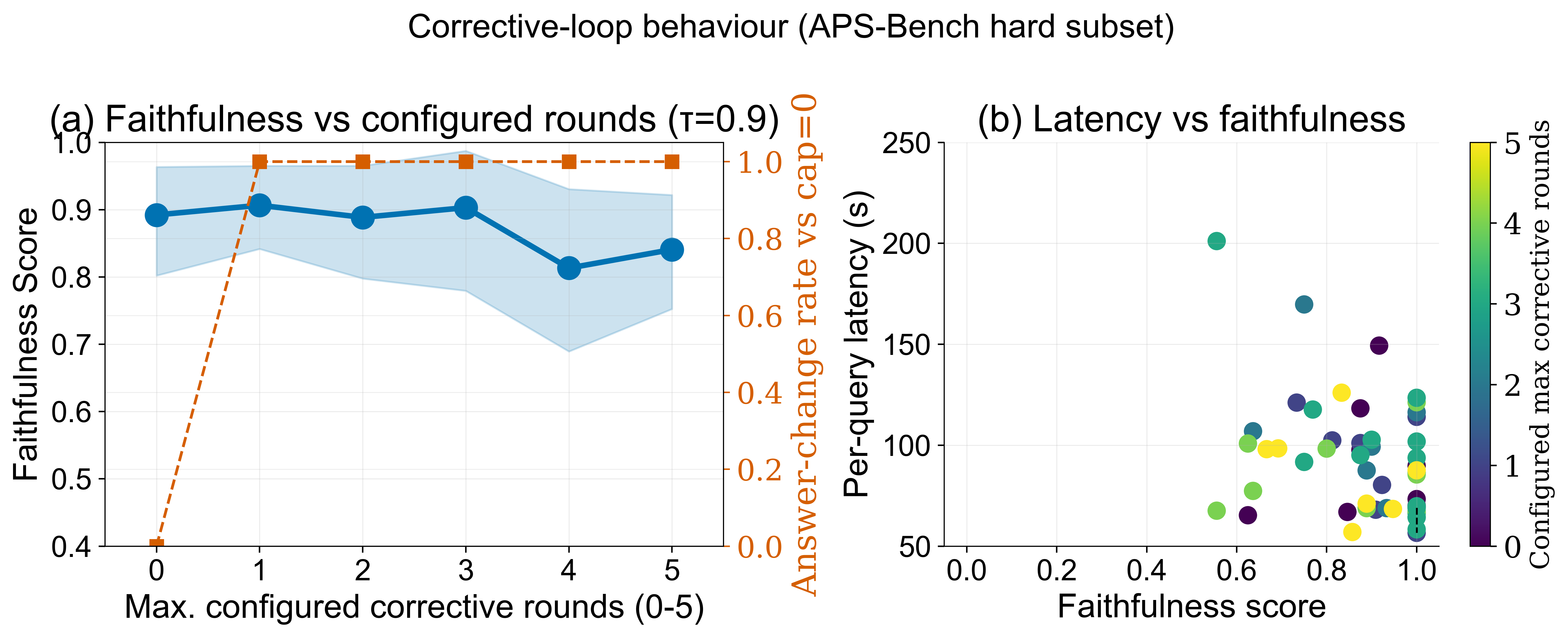}
\caption{\label{fig:s_corrective-rounds} Corrective-loop behavior is analyzed on the APS-Bench ablation subset (main-text Sec.~V\,C). (a) Claim-level faithfulness score (left axis; mean with 95\% bootstrap confidence interval band) and the fraction of answers differing from the single-pass baseline (answer-change rate vs. $\text{cap}=0$; right axis, dashed) are reported as a function of the maximum number of configured corrective rounds. The sweep is conducted at a diagnostic trigger threshold of $\tau=0.9$, intentionally set above the deployed $\tau=0.4$ so that the corrective loop activates for nearly every question; absolute levels are therefore not directly comparable to the deployed-configuration tables. Allowing a single round results in nearly every answer being revised (answer-change rate saturates at ${\approx}1.0$ from round 1 onward), while faithfulness remains statistically stable from zero to three rounds (${\approx}0.89\rightarrow0.91$ at round 1, a gain of ${\approx}0.02$ within the confidence interval band) and declines slightly at four to five rounds (${\approx}0.81$--$0.84$). The effect of the loop is concentrated in the first corrective round, with additional rounds altering answers without improving grounding. (b) Per-query wall-clock latency versus faithfulness for the same runs, colored by the configured round cap, shows latency concentrated at $50$--$150$~s with a tail to ${\approx}200$~s and no systematic dependence on either the round cap or achieved faithfulness. These results support the deployed two-round cap: rounds beyond the first increase latency and revision churn without measurable faithfulness improvement.}
\end{figure*}

\clearpage
\section{Corpus collections and document schema}\label{s:corpus}
The retrieval corpus consists of $130{,}663$ indexed records obtained from eight operational sources. For sparse retrieval, records are indexed in Elasticsearch using BM25, while dense retrieval is performed in Qdrant using the e5-large-v2 model~\cite{wang2022e5} with $1024$-dimensional embeddings. Main text Table~I summarizes the collections and their respective sizes. Each indexed chunk, irrespective of its source, is normalized into a unified corpus record comprising eight fields: \texttt{doc\_id}, \texttt{content}, \texttt{title}, \texttt{author}, \texttt{collection}, \texttt{system}, \texttt{creation\_datetime}, and \texttt{file\_id}. The \texttt{content} field contains the body text. Among all collections, only the \texttt{work\_requests} collection splits records into multiple chunks per parent record. For this collection, three additional chunk-provenance fields are included: \texttt{parent\_id}, \texttt{chunk\_index}, and \texttt{chunk\_type}.

Five of the eight sources exhibit a flat structure. In the case of ICMS, SDDS, and OAG, each file contains a single JSON document. For Teams and Work Requests, each file contains a single list of records. These flat sources map directly onto the unified corpus record format. In contrast, the BELY collection employs a three-level nested schema, as illustrated in Fig.~\ref{fig:bely-schema}. Each beamline is represented by a file, resulting in a total of $24$ files. Each file contains logbook entries, with $4{,}536$ entries in total. Each entry comprises $14$ fields, including array-valued fields such as \texttt{personnel\_involved} and \texttt{keywords}. Within each entry, there are chronological sub-entries, totaling $29{,}567$, where each sub-entry consists of \texttt{entry\_text} and \texttt{attachment\_urls}. The schema includes two entry subtypes that share the same structure and are distinguished by the value of the \texttt{logbook} field. Routine operational entries have empty \texttt{studies\_plan} and \texttt{studies\_summary} fields, while storage-ring studies entries populate these fields. There are $1{,}020$ storage-ring studies entries among the $4{,}536$ total entries. During indexing, each BELY entry, together with its sub-entries, is serialized into a single unified corpus record.

\section{Benchmark construction details and datasheet}\label{s:datasheet}
APS-Bench is constructed as a corpus-derived benchmark rather than through manual annotation. Question–answer pairs are generated from sampled corpus passages following the InPars methodology~\cite{bonifacio_inpars_2022}, where a large language model, provided with a document and a small set of in-context exemplars, generates a query that is answerable by the document. While InPars employs such synthetic pairs for retriever training, we adapt this approach for evaluation purposes, aligning more closely with automated self-data curation protocols for domain adaptation in retrieval-augmented generation systems~\cite{mao_rag-studio_2024}. This adaptation introduces type conditioning, structure-based difficulty assignment, retrievability filtering, and a gold answer construction stage that is explicitly separated from the generation process. The pipeline comprises five stages, with each stage recording its outputs in the released provenance record (Sec.~\ref{s:benchdata}).

\paragraph{Stage 1: Seed discovery and passage sampling}. For each cell in the query type by collection grid, defined by eight operational query types (factual, temporal, multi-hop, procedural, causal, comparative, troubleshooting, parametric) and eight indexed collections, at least five BM25 seed queries are issued against the live corpus (2026-05-22). The top-ranked results are pooled, and seed queries are constructed using the operational vocabulary specific to each cell, including device and subsystem names, fault terminology, and date- or shift-specific phrasing. The pooled results are deduplicated at the document level. Candidate source passages are then sampled from each cell's pool, ensuring that every generated question is grounded in a specific, indexable passage rather than relying on the generator's parametric knowledge.

\paragraph{Stage 2: Type-conditioned question–answer generation (InPars-style)}. Each sampled passage is provided to a GPT-family generator (gpt54 via ARGO, temperature 0.7) using a type-conditioned prompt template. This template specifies the target query type and its structural requirements. For example, a multi-hop question must require at least two distinct evidence spans, a temporal question must include an explicit date or shift scope, and a troubleshooting question must request a symptom to cause to resolution chain. The template also supplies in-context exemplars of the target type and instructs the model to return a single JSON object containing the question, a draft reference answer grounded only in the supplied passage, and the supporting passage spans. Two key modifications to the original InPars protocol are introduced: first, full question–answer pairs are generated rather than queries alone, as the subsequent nugget annotation (Stage 5) requires a grounded answer for decomposition; second, generation is conditioned on the operational query-type taxonomy, ensuring that the benchmark's type distribution is determined by design. Each per-type template is fixed and hashed, and the hash is stored in every record it produces for inclusion in the released package.

\begin{lstlisting}[style=jsonschema]
SYSTEM: generate evaluation questions for an accelerator-operations
        knowledge base
INPUT:
  passage       <- sampled corpus passage (doc_id, title, date, text)
  query_type    <- one of the eight types + its structural requirements
  exemplars     <- in-context (passage, question, answer) examples
                   of the target type
OUTPUT (one JSON object):
  question      <- answerable from the passage; must satisfy the
                   type's structural requirements
  draft_answer  <- grounded only in the supplied passage, with an
                   inline [doc_id] citation
  support_spans <- character spans of the passage backing the answer
\end{lstlisting}

\paragraph{Stage 3: Structure-first difficulty assignment.} Difficulty (easy, medium, hard) is determined based on the structural properties of the question, including the number of required evidence spans, the need for temporal scoping or cross-collection linking, and the specificity of the requested quantity. Assignment of difficulty is never based on observed system performance. This structure-based approach ensures that difficulty labels remain independent of the five systems under evaluation, thereby preventing any potential feedback from evaluation outcomes into the benchmark design.

\paragraph{Stage 4: Over-generation and filtering.} Candidate question–answer pairs are over-generated for each cell and subsequently filtered. The primary filtering criterion is retrievability: a candidate is retained only if its source passage can be recovered from the full index by an independent retriever. This step eliminates questions for which the supporting evidence is not accessible at retrieval time. Remaining candidates are further deduplicated to remove near-duplicate questions within each cell and are screened for type conformance and answerability. The final pool is balanced across the type-by-collection grid to yield the released set of n=50. It is important to note that retrieveability does not guarantee the source is present in the final LLM synthesis. 

\paragraph{Stage 5: Nugget annotation}. The generator's draft answer from Stage 2 is not used as the released gold answer; it serves solely to specify the intent of the question. The gold answer is reconstructed using an anti-circular, multi-retriever protocol. Evidence is retrieved independently by BM25, dense, and graph retrievers, and the answer is synthesized based on the pooled independent support with grounded inline [doc\_id] citations. Each answer is decomposed into 3 to 10 atomic nuggets labeled as vital or okay, with at least two vital nuggets per question, following the AutoNuggetizer convention~\cite{pradeep2024autonuggetizer}. Vital nuggets are required in any acceptable answer, while okay nuggets provide correct supporting detail. Each nugget is associated with a verified passage\_span and a chain\_role in {cause, effect, resolution, none}. Reconstructing the gold answer from independent retrieval ensures that no evaluated system's output, and no artifact of the generation step other than the question text, is included in the released gold or its nuggets.

\paragraph{Provenance.} Every stage writes to the record: seed queries and retrieved doc-ids (Stage 1), prompt-template hash and generation seed (Stage 2), the difficulty rationale (Stage 3), filter outcomes (Stage 4), and the gold-discovery trace (\texttt{regold\_provenance}, Stage 5), together with a package-level SHA-256 (Sec.~\ref{s:benchdata}).

\paragraph{Known Biases} Two biases are inherent to InPars-style construction and bound the claims made on this benchmark, as stated in the main-text Limitations. First, corpus-derived questions echo the vocabulary of their source passages, which makes them intrinsically easier to retrieve than verbatim operator questions; absolute scores should therefore be read as an upper bound on staff-question performance. Second, the questions, gold answers, and nugget labels are all LLM-produced, so the benchmark inherits the generator's stylistic and topical preferences within each cell even though the type mix is controlled. The expert-validation protocol is the gate on both: at least two APS experts adjudicate a stratified calibration sample for question validity and answerability, vital-nugget agreement against the auto-extracted nuggets, and inter-annotator agreement (Cohen's $\kappa$, Krippendorff's $\alpha$, percent agreement), using the risk-first audit sheets of Sec.~\ref{s:ocact} 

\section{Benchmark dataset: schema, provenance, and full listing}\label{s:benchdata}
We release the benchmark in full. Each record is one JSON object carrying the \texttt{question}, its \texttt{query\_type}, \texttt{difficulty}, and \texttt{source\_collection}; the \texttt{gold\_answer} with inline \texttt{[doc\_id]} citations; the supporting \texttt{evidence\_passages}; and the atomic \texttt{nuggets}, each labeled \texttt{vital} or \texttt{okay}. Questions were generated by a GPT-family model (\texttt{gpt54} via ARGO) from the live APS corpus using the above-defined InPars methodology.

\clearpage

\subsection{APS-Bench: full question listing ($n=50$, corpus-derived)}\label{s:listing-a}
{\footnotesize\raggedright
\smallskip\noindent\textbf{APS-BENCH-001}\quad {\footnotesize \textit{parametric} $\cdot$ ICMS $\cdot$ easy}\par
\noindent\emph{Q.}~For the PAR THz beamline measurement setup described in DIAG-TN-2024-018, what bunch-charge setpoint should I request if I want to run at the highest tested charge from the November 13 and December 9, 2024 studies?\par
\noindent\emph{Gold.}~The highest tested bunch charge in the November 13 and December 9, 2024 PAR THz beamline studies was 16 nC, so the bunch-charge setpoint to request is 16 nC [APS\_2294687\_\_chunk\_0].\par
\noindent\emph{Nuggets.}~\textbf{[V]}~The highest tested bunch charge in the November 13 and December 9, 2024 PAR THz beamline studies was 16 nC. \quad \textbf{[V]}~The bunch-charge setpoint to request is 16 nC. \quad [o]~The answer applies specifically to the PAR THz beamline measurement setup described in DIAG-TN-2024-018.\par
\noindent\emph{Evidence.}~\texttt{APS\_2294687\_\_chunk\_0} (primary)\par

\smallskip\noindent\textbf{APS-BENCH-002}\quad {\footnotesize \textit{procedural} $\cdot$ ICMS $\cdot$ easy}\par
\noindent\emph{Q.}~How would you set up the PAR to serve as the injector for the Linac Extension Area instead of the usual photocathode or RF electron gun source, including the basic sequence of using the PAR as a damping or accumulator ring before extracting and transporting the beam to LEA?\par
\noindent\emph{Gold.}~To set up the PAR as the injector for the Linac Extension Area instead of the usual photocathode or RF electron gun source, one would use the Particle Accumulator Ring (PAR) as a damping or accumulator ring, then transport the extracted beam to LEA [APS\_2189598\_\_chunk\_0]. The passage also states that the typical LEA source is the photocathode electron gun and that it is also possible to use electron beams from radiofrequency electron guns [APS\_2189598\_\_chunk\_0]. The passages do not provide further setup details beyond this basic sequence [APS\_2189598\_\_chunk\_0].\par
\noindent\emph{Nuggets.}~\textbf{[V]}~The Particle Accumulator Ring (PAR) can be used as the injector source for the Linac Extension Area (LEA). \quad \textbf{[V]}~In this setup, the PAR is used as a damping ring or accumulator ring before beam delivery to LEA. \quad \textbf{[V]}~After use in the PAR, the beam is extracted from the PAR and transported to LEA. \quad [o]~The usual source for LEA is the photocathode electron gun. \quad [o]~Electron beams from radiofrequency electron guns can also be used for LEA. \quad [o]~The provided passage does not give additional setup details beyond the basic sequence of using PAR, extracting the beam, and transporting it to LEA.\par
\noindent\emph{Evidence.}~\texttt{APS\_2189598\_\_chunk\_0} (primary)\par

\smallskip\noindent\textbf{APS-BENCH-003}\quad {\footnotesize \textit{factual} $\cdot$ OAG $\cdot$ easy}\par
\noindent\emph{Q.}~In the APS accelerator control system, what diagnostic GUI would an operator use to verify that OAG data loggers are actively writing files, that the RunControl PVs are running, and to identify any workstations or processes needing attention when data archiving problems are suspected?\par
\noindent\emph{Gold.}~The operator would use the checkDataLoggers diagnostic GUI, which checks whether data loggers are actively writing files, verifies RunControl PVs are running, and helps identify which workstations or processes need attention when data archiving problems are suspected [checkDataLoggers\_\_chunk\_0].\par
\noindent\emph{Nuggets.}~\textbf{[V]}~The diagnostic GUI to use is checkDataLoggers. \quad \textbf{[V]}~checkDataLoggers verifies whether OAG data loggers are actively writing files. \quad \textbf{[V]}~checkDataLoggers verifies that the RunControl PVs are running. \quad [o]~checkDataLoggers helps identify which workstations need attention when data archiving problems are suspected. \quad [o]~checkDataLoggers helps identify which processes need attention when data archiving problems are suspected.\par
\noindent\emph{Evidence.}~\texttt{checkDataLoggers\_\_chunk\_0} (primary)\par

\smallskip\noindent\textbf{APS-BENCH-004}\quad {\footnotesize \textit{factual} $\cdot$ ICMS $\cdot$ easy}\par
\noindent\emph{Q.}~What is the principal function of the PAR at the APS?\par
\noindent\emph{Gold.}~The principal function of the PAR at the APS is to act as an accumulator ring, damping multiple (< 30) injections from the linac into a single bunch [APS\_2170123\_\_chunk\_0].\par
\noindent\emph{Nuggets.}~\textbf{[V]}~At the APS, the PAR's principal function is to act as an accumulator ring. \quad \textbf{[V]}~The PAR damps injections from the linac. \quad [o]~The PAR handles multiple injections, fewer than 30, from the linac. \quad \textbf{[V]}~The PAR combines the multiple linac injections into a single bunch.\par
\noindent\emph{Evidence.}~\texttt{APS\_2170123\_\_chunk\_0} (primary)\par

\smallskip\noindent\textbf{APS-BENCH-005}\quad {\footnotesize \textit{factual} $\cdot$ ICMS $\cdot$ easy}\par
\noindent\emph{Q.}~According to DIAG-TN-2022-007, what ring is being considered as a damping or accumulator ring to inject beam into the Linac Extension Area?\par
\noindent\emph{Gold.}~According to DIAG-TN-2022-007, the ring being considered as a damping or accumulator ring to inject beam into the Linac Extension Area is the Particle Accumulator Ring (PAR) [APS\_2189598\_\_chunk\_0].\par
\noindent\emph{Nuggets.}~\textbf{[V]}~DIAG-TN-2022-007 considers the Particle Accumulator Ring (PAR) as the ring for injecting beam into the Linac Extension Area. \quad \textbf{[V]}~The ring is being considered in the role of a damping ring. \quad \textbf{[V]}~The ring is being considered in the role of an accumulator ring. \quad [o]~The acronym PAR stands for Particle Accumulator Ring.\par
\noindent\emph{Evidence.}~\texttt{APS\_2189598\_\_chunk\_0} (primary)\par

\smallskip\noindent\textbf{APS-BENCH-006}\quad {\footnotesize \textit{factual} $\cdot$ BELY $\cdot$ easy}\par
\noindent\emph{Q.}~For 2D scans on 9-BM using EPICS scan control, which after-scan PVs need to be set to 0 with "NoWait" to prevent the scan from hanging at the end of random lines?\par
\noindent\emph{Gold.}~For 2D scans on 9-BM using EPICS scan control, the after-scan fields must set 9bmc:scan1.AWAIT to 0 and 9bmc:scan2.AWAIT to 0 with "NoWait"; otherwise the scan can hang at the end of random lines [10635\_\_chunk\_0].\par
\noindent\emph{Nuggets.}~\textbf{[V]}~For 2D scans on 9-BM using EPICS scan control, the after-scan field 9bmc:scan1.AWAIT must be set to 0. \quad \textbf{[V]}~For 2D scans on 9-BM using EPICS scan control, the after-scan field 9bmc:scan2.AWAIT must be set to 0. \quad \textbf{[V]}~The after-scan settings for these PVs should use "NoWait". \quad [o]~If these after-scan PVs are not set this way, the scan can hang at the end of random lines.\par
\noindent\emph{Evidence.}~\texttt{10635\_\_chunk\_0} (primary)\par

\smallskip\noindent\textbf{APS-BENCH-007}\quad {\footnotesize \textit{factual} $\cdot$ SDDS $\cdot$ easy}\par
\noindent\emph{Q.}~When preparing an orbit response matrix for use in feedback, which SDDS Toolkit command would you use to transpose the tabular data so the output has one column for each input row?\par
\noindent\emph{Gold.}~The command is "sddstranspose()" because it "transposes the tabular data in the input file, so that the output file contains one column for each row in the input" and the example usage is to "transpose an orbit response matrix as part of preparing to use it for feedback" [SDDS\_OAG\_User's Guide for SDDS Toolkit Version 5.0\_chunk\_19\_\_chunk\_1].\par
\noindent\emph{Nuggets.}~\textbf{[V]}~The SDDS Toolkit command to use is sddstranspose(). \quad \textbf{[V]}~sddstranspose() transposes the tabular data in the input file. \quad \textbf{[V]}~After transposition, the output file contains one column for each row in the input. \quad [o]~A documented use of sddstranspose() is transposing an orbit response matrix when preparing it for feedback.\par
\noindent\emph{Evidence.}~\texttt{SDDS\_OAG\_User's Guide for SDDS Toolkit Version 5.0\_chunk\_19\_\_chunk\_1} (primary)\par

\smallskip\noindent\textbf{APS-BENCH-008}\quad {\footnotesize \textit{procedural} $\cdot$ SDDS $\cdot$ easy}\par
\noindent\emph{Q.}~How would you use the SDDS toolkit to prepare an orbit response matrix and then generate a correction vector for step-by-step feedback, including which commands you would run to transpose the response matrix and multiply it by the error vector?\par
\noindent\emph{Gold.}~Using the SDDS toolkit, you would prepare the orbit response matrix for feedback by running sddstranspose on the orbit response matrix file, since sddstranspose transposes tabular data and is used to transpose an orbit response matrix as part of preparing it for feedback [SDDS\_OAG\_User's Guide for SDDS Toolkit Version 5.0\_chunk\_19\_\_chunk\_1]. Then you would generate the correction vector for step-by-step feedback by running sddsmatrixmult to multiply the error vector by the correction matrix, since sddsmatrixmult multiplies the tabular data in two input files and is used to multiply a vector of errors with a correction matrix to obtain a vector of corrections to apply in a step-by-step feedback system [SDDS\_OAG\_User's Guide for SDDS Toolkit Version 5.0\_chunk\_19\_\_chunk\_1]. The passages do not provide the exact command-line syntax, only the toolkit commands to use [SDDS\_OAG\_User's Guide for SDDS Toolkit Version 5.0\_chunk\_19\_\_chunk\_1].\par
\noindent\emph{Nuggets.}~\textbf{[V]}~Use the SDDS command sddstranspose to transpose the orbit response matrix file when preparing it for feedback. \quad [o]~Transposing the orbit response matrix is part of preparing the matrix for feedback. \quad \textbf{[V]}~Use the SDDS command sddsmatrixmult to multiply the error vector by the correction matrix. \quad \textbf{[V]}~Multiplying the error vector by the correction matrix produces the correction vector to apply in step-by-step feedback. \quad [o]~The cited SDDS passages identify which toolkit commands to use but do not provide exact command-line syntax.\par
\noindent\emph{Evidence.}~\texttt{SDDS\_OAG\_User's Guide for SDDS Toolkit Version 5.0\_chunk\_19\_\_chunk\_1} (primary)\par

\smallskip\noindent\textbf{APS-BENCH-009}\quad {\footnotesize \textit{procedural} $\cdot$ SDDS $\cdot$ easy}\par
\noindent\emph{Q.}~How do I use sddsplot to display the Twiss functions for APS lattices so that each function is on its own plotting page while each lattice data page appears with a different line type?\par
\noindent\emph{Gold.}~Use: sddsplot -columnNames=s,'(beta?,etax)' APS.twi -graphic=line,vary -split=page -groupby=nameIndex -separate=nameIndex [SDDS\_OAG\_User's Guide for SDDS Toolkit Version 5.0\_chunk\_126\_\_chunk\_0]\par
\noindent\emph{Nuggets.}~\textbf{[V]}~Use the command `sddsplot -columnNames=s,'(beta?,etax)' APS.twi -graphic=line,vary -split=page -groupby=nameIndex -separate=nameIndex`. \quad \textbf{[V]}~Specify `-columnNames=s,'(beta?,etax)'` to plot the Twiss-function columns against `s`. \quad \textbf{[V]}~Use `-split=page` so each function is placed on its own plotting page. \quad \textbf{[V]}~Use `-graphic=line,vary` so different lattice data pages are distinguished by different line types. \quad [o]~Use `-groupby=nameIndex`. \quad [o]~Use `-separate=nameIndex`. \quad [o]~The input lattice file in the example is `APS.twi`.\par
\noindent\emph{Evidence.}~\texttt{SDDS\_OAG\_User's Guide for SDDS Toolkit Version 5.0\_chunk\_126\_\_chunk\_0} (primary)\par

\smallskip\noindent\textbf{APS-BENCH-010}\quad {\footnotesize \textit{factual} $\cdot$ SDDS $\cdot$ easy}\par
\noindent\emph{Q.}~Using sddsplot, what command would you use to plot the horizontal beta function versus s for the APS design lattice file APS0.twi?\par
\noindent\emph{Gold.}~The command is "sddsplot -columnNames=s,betax APS0.twi" [SDDS\_OAG\_User's Guide for SDDS Toolkit Version 5.0\_chunk\_126\_\_chunk\_0].\par
\noindent\emph{Nuggets.}~\textbf{[V]}~The required command uses the program `sddsplot`. \quad \textbf{[V]}~The command specifies the columns `s` and `betax` with `-columnNames=s,betax`. \quad \textbf{[V]}~The input file in the command is `APS0.twi`. \quad \textbf{[V]}~The full command is `sddsplot -columnNames=s,betax APS0.twi`. \quad [o]~In this command, `s` is the horizontal axis variable. \quad [o]~In this command, `betax` is the horizontal beta function being plotted.\par
\noindent\emph{Evidence.}~\texttt{SDDS\_OAG\_User's Guide for SDDS Toolkit Version 5.0\_chunk\_126\_\_chunk\_0} (primary)\par

\smallskip\noindent\textbf{APS-BENCH-011}\quad {\footnotesize \textit{procedural} $\cdot$ ICMS $\cdot$ easy}\par
\noindent\emph{Q.}~How is the PAR terahertz diagnostic intended to be used operationally to monitor longitudinal bunch instabilities during accumulation of multiple linac injections into a single bunch?\par
\noindent\emph{Gold.}~The PAR terahertz diagnostic is intended to be used to observe longitudinal electron bunch instabilities [APS\_2170123\_\_chunk\_0] during operation of the Particle Accumulator Ring, whose principal function is damping multiple (< 30) injections from the linac into a single bunch [APS\_2170123\_\_chunk\_0]. The passage does not fully specify the operational procedure for monitoring those instabilities during accumulation.\par
\noindent\emph{Nuggets.}~\textbf{[V]}~The PAR terahertz diagnostic is intended to observe longitudinal electron bunch instabilities. \quad \textbf{[V]}~The diagnostic is intended for use during operation of the Particle Accumulator Ring (PAR). \quad \textbf{[V]}~The PAR accumulates multiple injections from the linac into a single bunch. \quad [o]~The number of linac injections accumulated is fewer than 30. \quad [o]~A principal function of the PAR is to damp the accumulated injections. \quad \textbf{[V]}~The passage does not specify a detailed operational procedure for how the terahertz diagnostic monitors instabilities during accumulation.\par
\noindent\emph{Evidence.}~\texttt{APS\_2170123\_\_chunk\_0} (primary)\par

\smallskip\noindent\textbf{APS-BENCH-012}\quad {\footnotesize \textit{procedural} $\cdot$ BELY $\cdot$ easy}\par
\noindent\emph{Q.}~If you're trying to refine chromaticity during SR studies and the chromaticity knobs are not behaving as expected, what measurement setup should you use to get more reliable chromaticity results, and what tune-adjustment precaution should you take given the quadrupole current limits and the observed conditioning behavior?\par
\noindent\emph{Gold.}~Use a chromaticity measurement setup with 15 points, since that gave much more reliable chromaticity results when the chromaticity knobs were not behaving as expected [3640\_\_chunk\_0]. For tune adjustment, take care because some quadrupoles are already near the swap-out safety/current limit, which caused issues in tune adjustment in one case, and also because conditioning the quadrupoles after adjusting the tune knobs once per plane did not restore the tunes; the passages do not specify a complete corrective procedure beyond this precaution [3640\_\_chunk\_0].\par
\noindent\emph{Nuggets.}~\textbf{[V]}~Use a chromaticity measurement setup with 15 points to obtain more reliable chromaticity results. \quad \textbf{[V]}~The 15-point chromaticity measurement setup was specifically recommended when the chromaticity knobs were not behaving as expected. \quad \textbf{[V]}~When adjusting tune, be cautious because some quadrupoles are already near the swap-out safety/current limit. \quad [o]~Quadrupoles being near the safety/current limit caused tune-adjustment issues in at least one case. \quad \textbf{[V]}~Conditioning the quadrupoles after adjusting the tune knobs once per plane did not restore the tunes. \quad [o]~The source passage does not provide a complete corrective procedure beyond warning to take this precaution.\par
\noindent\emph{Evidence.}~\texttt{3640\_\_chunk\_0} (primary)\par

\smallskip\noindent\textbf{APS-BENCH-013}\quad {\footnotesize \textit{procedural} $\cdot$ SDDS $\cdot$ easy}\par
\noindent\emph{Q.}~How do I compute an inverse response matrix for trajectory correction from a previously measured response matrix, and what machine element subset can I select when doing it?\par
\noindent\emph{Gold.}~You compute an inverse response matrix by using the ``Compute Inverse Response Matrix'' function to compute a trajectory correction matrix from a previously measured response matrix [SDDS\_OAG\_Control System High-Level Applications\_chunk\_18\_\_chunk\_0]. When doing this, you can select a subset of correctors and BPMs from that previously measured response matrix [SDDS\_OAG\_Control System High-Level Applications\_chunk\_18\_\_chunk\_0].\par
\noindent\emph{Nuggets.}~\textbf{[V]}~An inverse response matrix is computed by using the "Compute Inverse Response Matrix" function. \quad \textbf{[V]}~The inverse response matrix is computed from a previously measured response matrix. \quad [o]~The computed inverse response matrix is a trajectory correction matrix. \quad \textbf{[V]}~When computing the inverse response matrix, you can select a subset of correctors from the previously measured response matrix. \quad \textbf{[V]}~When computing the inverse response matrix, you can select a subset of BPMs from the previously measured response matrix.\par
\noindent\emph{Evidence.}~\texttt{SDDS\_OAG\_Control System High-Level Applications\_chunk\_18\_\_chunk\_0} (primary)\par

\smallskip\noindent\textbf{APS-BENCH-014}\quad {\footnotesize \textit{factual} $\cdot$ SDDS $\cdot$ easy}\par
\noindent\emph{Q.}~In the OAG control system high-level applications, what does the ``Compute Inverse Response Matrix'' function let you calculate from a previously measured response matrix?\par
\noindent\emph{Gold.}~It lets you calculate a trajectory correction matrix from a subset of correctors and BPMs in a previously measured response matrix [SDDS\_OAG\_Control System High-Level Applications\_chunk\_18\_\_chunk\_0].\par
\noindent\emph{Nuggets.}~\textbf{[V]}~The ``Compute Inverse Response Matrix'' function calculates a trajectory correction matrix. \quad \textbf{[V]}~The calculation uses a previously measured response matrix as input. \quad \textbf{[V]}~The trajectory correction matrix can be calculated from a subset of correctors. \quad \textbf{[V]}~The trajectory correction matrix can be calculated from a subset of BPMs.\par
\noindent\emph{Evidence.}~\texttt{SDDS\_OAG\_Control System High-Level Applications\_chunk\_18\_\_chunk\_0} (primary)\par

\smallskip\noindent\textbf{APS-BENCH-015}\quad {\footnotesize \textit{procedural} $\cdot$ BELY $\cdot$ easy}\par
\noindent\emph{Q.}~If we can't inject into the new sextupole settings on first beam, what bootstrap procedure should we follow to recover stored beam and transition back to the new sextupoles while maintaining orbit control?\par
\noindent\emph{Gold.}~If we can't inject into the new sextupole settings on first beam, the bootstrap procedure is: restore the old sextupoles by merely setting them without standardizing [6944\_\_chunk\_1]; fill 10 mA beam and run orbit correction [6944\_\_chunk\_1]; ramp to the new sextupoles while the beam is stored and orbit correction is running [6944\_\_chunk\_1]; save correctors [6944\_\_chunk\_1]; measure and correct tunes [6944\_\_chunk\_1]; and save quadrupoles [6944\_\_chunk\_1]. The passage also indicates orbit control must actually be active during the sextupole ramp, since a partial beam loss occurred when x-plane orbit correction had been left in a suspend state [6944\_\_chunk\_1].\par
\noindent\emph{Nuggets.}~\textbf{[V]}~If injection into the new sextupole settings fails on first beam, first restore the old sextupole settings by setting them without standardizing. \quad \textbf{[V]}~After restoring the old sextupoles, fill 10 mA of beam. \quad \textbf{[V]}~Run orbit correction on the stored 10 mA beam. \quad \textbf{[V]}~Ramp from the old sextupoles to the new sextupoles while the beam remains stored. \quad \textbf{[V]}~Orbit correction must be actively running during the sextupole ramp. \quad [o]~After reaching the new sextupole settings, save the corrector settings. \quad [o]~Measure and correct the tunes after the sextupole transition. \quad [o]~Save the quadrupole settings after tune correction. \quad [o]~A partial beam loss occurred when x-plane orbit correction was left suspended during the sextupole ramp.\par
\noindent\emph{Evidence.}~\texttt{6944\_\_chunk\_1} (primary)\par

\smallskip\noindent\textbf{APS-BENCH-016}\quad {\footnotesize \textit{parametric} $\cdot$ ICMS $\cdot$ medium}\par
\noindent\emph{Q.}~For the PAR integrating current transformer absolute calibration, what bunch-current range in mA corresponds to the calibrated bunch-charge range at the 9.77 MHz repetition rate?\par
\noindent\emph{Gold.}~The calibrated bunch-charge range of 1-21.9 nC corresponds to a bunch-current range of 9.8-214 mA at the 9.77 MHz repetition rate [APS\_2032345\_\_chunk\_0].\par
\noindent\emph{Nuggets.}~\textbf{[V]}~The PAR integrating current transformer absolute calibration uses a calibrated bunch-charge range of 1 nC to 21.9 nC. \quad \textbf{[V]}~The repetition rate for this conversion is 9.77 MHz. \quad \textbf{[V]}~A bunch charge of 1 nC corresponds to a bunch current of 9.8 mA at 9.77 MHz. \quad \textbf{[V]}~A bunch charge of 21.9 nC corresponds to a bunch current of 214 mA at 9.77 MHz. \quad \textbf{[V]}~The calibrated bunch-current range is 9.8 mA to 214 mA.\par
\noindent\emph{Evidence.}~\texttt{APS\_2032345\_\_chunk\_0} (primary)\par

\smallskip\noindent\textbf{APS-BENCH-017}\quad {\footnotesize \textit{procedural} $\cdot$ OAG $\cdot$ medium}\par
\noindent\emph{Q.}~How do I start a cathode bake-in on RG1 or RG2, including which parameters I need to set before hitting Ramp Cathode and how to stop and reset the procedure if I need to abort?\par
\noindent\emph{Gold.}~To start a cathode bake-in, first select the RF Gun (RG1 or RG2), then set the maximum heater power in Watts, set the time to hold at maximum power, configure the current step size in Amps, and configure the pause time between steps [BakeInCathode\_\_chunk\_1]. After these parameters are set, use Ramp Cathode to start the bake-in procedure [BakeInCathode\_\_chunk\_1]. If you need to abort, use Abort to stop the bake-in procedure and reset [BakeInCathode\_\_chunk\_1].\par
\noindent\emph{Nuggets.}~\textbf{[V]}~Select the RF Gun to bake in, choosing either RG1 or RG2, before starting the procedure. \quad \textbf{[V]}~Set the maximum heater power in Watts before starting the cathode bake-in. \quad \textbf{[V]}~Set the time to hold at maximum power before starting the cathode bake-in. \quad \textbf{[V]}~Configure the current step size in Amps before starting the cathode bake-in. \quad \textbf{[V]}~Configure the pause time between steps before starting the cathode bake-in. \quad \textbf{[V]}~Start the cathode bake-in by using Ramp Cathode after the required parameters are set. \quad \textbf{[V]}~Use Abort to stop the cathode bake-in procedure if you need to abort. \quad [o]~Using Abort also resets the cathode bake-in procedure.\par
\noindent\emph{Evidence.}~\texttt{BakeInCathode\_\_chunk\_1} (primary)\par

\smallskip\noindent\textbf{APS-BENCH-018}\quad {\footnotesize \textit{troubleshooting} $\cdot$ WORK\_REQUESTS $\cdot$ medium}\par
\noindent\emph{Q.}~Injection is blocked because B:IS is showing negative current readbacks, and the only recovery so far was a power cycle. Given the reported MTime power supply failure, what component should we suspect and what corrective action should we take to restore reliable injection?\par
\noindent\emph{Gold.}~We should suspect the MTime power supply because a reported MTime power supply failure requires that the power supply be swapped [50362\_\_chunk\_0]. The corrective action indicated to restore reliable injection is to swap the power supply [50362\_\_chunk\_0]. The passages do not explicitly state that the MTime power supply failure is the confirmed cause of B:IS negative current readbacks; they only say B:IS showed negative current readbacks preventing injection and was recovered via power cycle [50694\_\_chunk\_0].\par
\noindent\emph{Nuggets.}~\textbf{[V]}~The component to suspect is the MTime power supply. \quad \textbf{[V]}~The corrective action is to swap the MTime power supply. \quad \textbf{[V]}~Swapping the power supply is indicated as the way to restore reliable injection. \quad [o]~The evidence does not explicitly confirm that the MTime power supply failure caused the B:IS negative current readbacks. \quad [o]~B:IS negative current readbacks were preventing injection. \quad [o] (resolution)~A power cycle had previously recovered the issue.\par
\noindent\emph{Evidence.}~\texttt{50362\_\_chunk\_0} (primary), \texttt{50694\_\_chunk\_0} (supporting)\par

\smallskip\noindent\textbf{APS-BENCH-019}\quad {\footnotesize \textit{temporal} $\cdot$ TEAMS $\cdot$ medium}\par
\noindent\emph{Q.}~On the 10/8/24 shift after the beam dump and before attempting a reset, did Operations have any way to verify whether the alignment issue still existed without first filling and dumping beam again?\par
\noindent\emph{Gold.}~Based on the 10/8/24 passage, Operations did not have a stated way to verify whether the alignment issue still existed after the beam dump and before attempting a reset without filling and dumping beam again; Elaine Chandler explicitly said, "I don't know how to check alignment without filling and dumping beam," and asked Weixing Cheng whether there was another way [teams\_Borland, Michael David\_e79c556151b8\_\_chunk\_0]. The passages do not provide any answer from Weixing Cheng or anyone else confirming an alternative method, so the passages do not fully answer whether such a way existed [teams\_Borland, Michael David\_e79c556151b8\_\_chunk\_0].\par
\noindent\emph{Nuggets.}~\textbf{[V]}~The 10/8/24 passage does not state any method Operations could use to verify whether the alignment issue still existed after the beam dump and before attempting a reset without filling and dumping beam again. \quad \textbf{[V]}~Elaine Chandler said, "I don't know how to check alignment without filling and dumping beam." \quad [o]~Elaine Chandler asked Weixing Cheng whether there was another way to check alignment. \quad \textbf{[V]}~The provided passages do not include any answer from Weixing Cheng or anyone else confirming an alternative verification method. \quad [o]~Because no alternative method is given in the passages, the evidence does not fully establish that such a way existed.\par
\noindent\emph{Evidence.}~\texttt{teams\_Borland, Michael David\_e79c556151b8\_\_chunk\_0} (primary), \texttt{teams\_Borland, Michael David\_d9889662890a\_\_chunk\_0} (supporting)\par

\smallskip\noindent\textbf{APS-BENCH-020}\quad {\footnotesize \textit{temporal} $\cdot$ TEAMS $\cdot$ medium}\par
\noindent\emph{Q.}~During the 10/8/24 beam-dump/reset troubleshooting shift, before trying a reset on the suspected alignment issue, who did Elaine suggest checking with for an alternate way to verify alignment without filling and dumping beam, and on the later 1/26/25 BESOCM support shift, who was Andrew's fallback contact if Rob Keane and a Diagnostics secondary contact could not be reached?\par
\noindent\emph{Gold.}~Before trying a reset on the suspected alignment issue during the 10/8/24 shift, Elaine suggested checking with Weixing Cheng for another way to verify alignment without filling and dumping beam [teams\_Borland, Michael David\_e79c556151b8\_\_chunk\_0]. On the 1/26/25 BESOCM support shift, Andrew's fallback contact if Rob Keane and a Diagnostics secondary contact could not be reached was Guobao [teams\_Borland, Michael David\_d9889662890a\_\_chunk\_0].\par
\noindent\emph{Nuggets.}~\textbf{[V]}~During the 10/8/24 beam-dump/reset troubleshooting shift, Elaine suggested checking with Weixing Cheng before trying a reset on the suspected alignment issue. \quad [o]~Elaine suggested Weixing Cheng as someone who might offer another way to verify alignment without filling and dumping beam. \quad \textbf{[V]}~During the 1/26/25 BESOCM support shift, Andrew's fallback contact was Guobao if Rob Keane could not be reached. \quad \textbf{[V]}~During the 1/26/25 BESOCM support shift, Andrew's fallback contact was Guobao if a Diagnostics secondary contact also could not be reached.\par
\noindent\emph{Evidence.}~\texttt{teams\_Borland, Michael David\_e79c556151b8\_\_chunk\_0} (primary), \texttt{teams\_Borland, Michael David\_d9889662890a\_\_chunk\_0} (primary)\par

\smallskip\noindent\textbf{APS-BENCH-021}\quad {\footnotesize \textit{comparative} $\cdot$ BELY $\cdot$ medium}\par
\noindent\emph{Q.}~At 9-BM, how did the relative ion-chamber noise behavior change between the Ni K-edge tests and the later 24 keV dummy scan after all SR570s were manually set to low-noise/no-filter---specifically, which signal chain or chamber was identified as noisier in the Ni-edge study, and what gain/gas configuration was then used for the He-filled I0 versus the N2-filled chambers in the 24 keV comparison?\par
\noindent\emph{Gold.}~In the Ni K-edge tests, the noisier behavior was identified in the scaler 3 signal chain rather than the ion chamber itself, and separately the I0 signal was noted to have higher noise than IRef, with I0 being He-filled and IRef the ADC ion chamber [7413\_\_chunk\_0]. In the later 24 keV dummy scan after all SR570s were manually set to low-noise/no-filter, the He-filled I0-1 used 1 pA/V, while the N2-filled chambers were adjusted to 50 pA/V for all N2-filled ion chambers [7582\_\_chunk\_0].\par
\noindent\emph{Nuggets.}~\textbf{[V]}~In the Ni K-edge tests, the noisier behavior was attributed to the scaler 3 signal chain rather than to the ion chamber itself. \quad [o]~In the Ni-edge study, the I0 signal was observed to be noisier than IRef. \quad [o]~In that Ni-edge comparison, I0 was the He-filled chamber. \quad [o]~In the later 24 keV dummy scan, all SR570s had been manually set to low-noise mode with no filter. \quad \textbf{[V]}~In the 24 keV comparison, the He-filled I0-1 was set to a gain of 1 pA/V. \quad \textbf{[V]}~In the 24 keV comparison, the N2-filled ion chambers were set to a gain of 50 pA/V.\par
\noindent\emph{Evidence.}~\texttt{7413\_\_chunk\_0} (primary), \texttt{7582\_\_chunk\_0} (primary)\par

\smallskip\noindent\textbf{APS-BENCH-022}\quad {\footnotesize \textit{comparative} $\cdot$ WORK\_REQUESTS $\cdot$ medium}\par
\noindent\emph{Q.}~Between the Booster dipole slave power supply SCR-failure trip and the ioclet1bpm reboot to add event timestamps for injection-cycle data collection, which issue is a hardware fault versus a controls/data-acquisition change, and how would their expected operational impact differ during Booster injection operations?\par
\noindent\emph{Gold.}~The Booster dipole slave power supply issue is a hardware fault: the power supply needed repair because it "tripped on SCR failure" [45927\_\_chunk\_0]. The ioclet1bpm issue is a controls/data-acquisition change: it required a reboot "to add event timestamps for injection cycle data collection" [44960\_\_chunk\_0]. Their exact operational impact during Booster injection operations is not fully stated in the passages, but the first implies a failed power-supply condition requiring repair [45927\_\_chunk\_0], whereas the second implies a data-collection/timestamping change for injection-cycle measurements [44960\_\_chunk\_0].\par
\noindent\emph{Nuggets.}~\textbf{[V]}~The Booster dipole slave power supply issue was a hardware fault. \quad \textbf{[V]}~The hardware fault was that the power supply tripped on SCR failure. \quad [o]~The Booster dipole slave power supply required repair. \quad \textbf{[V]}~The ioclet1bpm issue was a controls/data-acquisition change. \quad \textbf{[V]}~The ioclet1bpm reboot was done to add event timestamps for injection-cycle data collection. \quad [o]~The expected operational impact of the power-supply issue was a failed hardware condition affecting operation until repaired. \quad [o]~The expected operational impact of the ioclet1bpm change was on injection-cycle measurement data collection rather than a hardware failure.\par
\noindent\emph{Evidence.}~\texttt{45927\_\_chunk\_0} (primary), \texttt{44960\_\_chunk\_0} (primary)\par

\smallskip\noindent\textbf{APS-BENCH-023}\quad {\footnotesize \textit{troubleshooting} $\cdot$ WORK\_REQUESTS $\cdot$ medium}\par
\noindent\emph{Q.}~B:IS is showing negative current readbacks and injection is blocked---should the first recovery step be to power cycle the supply, and if that doesn't hold do we need to swap the MTime power supply?\par
\noindent\emph{Gold.}~Yes. The passages indicate that when B:IS showed negative current readbacks preventing injection, it was recovered via power cycle [50694\_\_chunk\_0]. Separately, they state that for an MTime power supply failure, the power supply needs to be swapped [50362\_\_chunk\_0]. However, the passages do not explicitly say that if the power-cycle recovery does not hold for B:IS, then the next step is to swap the MTime power supply [50694\_\_chunk\_0][50362\_\_chunk\_0].\par
\noindent\emph{Nuggets.}~\textbf{[V]} (resolution)~When B:IS shows negative current readbacks and injection is blocked, a power cycle recovered it. \quad \textbf{[V]}~The first recovery step for the B:IS negative-current-readback issue is to power cycle the supply. \quad \textbf{[V]}~For an MTime power supply failure, the documented action is to swap the power supply. \quad \textbf{[V]}~The passages do not explicitly state that if the B:IS power-cycle recovery does not hold, the next step is to swap the MTime power supply. \quad [o]~The evidence for power-cycling B:IS comes from source 50694\_\_chunk\_0. \quad [o]~The evidence for swapping the MTime power supply comes from source 50362\_\_chunk\_0.\par
\noindent\emph{Evidence.}~\texttt{50694\_\_chunk\_0} (primary), \texttt{50362\_\_chunk\_0} (supporting)\par

\smallskip\noindent\textbf{APS-BENCH-024}\quad {\footnotesize \textit{comparative} $\cdot$ ICMS $\cdot$ medium}\par
\noindent\emph{Q.}~Compared with normal operation without VESD, what happened to the PAR THz beamline CSR signal at 1 nC when the LTP:FL1 flag was inserted?\par
\noindent\emph{Gold.}~At 1 nC, inserting the LTP:FL1 flag attenuated the beam so much that the PAR THz beamline CSR signal became indistinguishable from the background, and no CSR signal was detected, compared with normal operation without VESD [APS\_2294687\_\_chunk\_6].\par
\noindent\emph{Nuggets.}~\textbf{[V]}~The condition discussed is at a bunch charge of 1 nC. \quad \textbf{[V]}~With the LTP:FL1 flag inserted, the beam was strongly attenuated. \quad \textbf{[V]}~Under that condition, the PAR THz beamline CSR signal became indistinguishable from background. \quad \textbf{[V]}~No PAR THz beamline CSR signal was detected when the LTP:FL1 flag was inserted. \quad [o]~This outcome is in contrast to normal operation without VESD.\par
\noindent\emph{Evidence.}~\texttt{APS\_2294687\_\_chunk\_6} (primary), \texttt{APS\_2294687\_\_chunk\_0} (supporting)\par

\smallskip\noindent\textbf{APS-BENCH-025}\quad {\footnotesize \textit{temporal} $\cdot$ WORK\_REQUESTS $\cdot$ medium}\par
\noindent\emph{Q.}~During the 2024-10-18 shift, was there any active PAR machine-repair work like the 2020 harmonic-system troubleshooting, or was the only relevant work request the Linac RG2 kicker interlock sum fault investigation?\par
\noindent\emph{Gold.}~During the 2024-10-18 shift, the passages show a completed work request to investigate the Linac RG2 kicker interlock sum fault [50660\_\_chunk\_0]. The only PAR machine-repair example provided is from 2020-09-28 to troubleshoot and repair the PAR harmonic system [43541\_\_chunk\_0]. The passages do not show any active PAR machine-repair work on 2024-10-18, so based only on these passages, the only relevant work request shown for 2024-10-18 was the Linac RG2 kicker interlock sum fault investigation [50660\_\_chunk\_0].\par
\noindent\emph{Nuggets.}~\textbf{[V]}~On 2024-10-18, the passages show a completed work request to investigate the Linac RG2 kicker interlock sum fault. \quad \textbf{[V]}~The passages do not show any active PAR machine-repair work on 2024-10-18. \quad [o]~The only PAR machine-repair example provided in the passages is from 2020-09-28. \quad [o]~The 2020-09-28 PAR machine-repair example was to troubleshoot and repair the PAR harmonic system. \quad \textbf{[V]}~Based only on these passages, the only relevant work request shown for 2024-10-18 was the Linac RG2 kicker interlock sum fault investigation.\par
\noindent\emph{Evidence.}~\texttt{50660\_\_chunk\_0} (primary), \texttt{43541\_\_chunk\_0} (supporting)\par

\smallskip\noindent\textbf{APS-BENCH-026}\quad {\footnotesize \textit{temporal} $\cdot$ BELY $\cdot$ medium}\par
\noindent\emph{Q.}~During the 2026-02-05 to 2026-02-06 pumpdown of the 9-ID Grand Tube, what was the GT vacuum reading just before GV11 was opened, and if we wanted to use the 9-BM Ge8 xspress3 setup for a synchronized scan in that same run window, which [main] section trigger/sync PVs would have failed the caget existence check?\par
\noindent\emph{Gold.}~Just before GV11 was opened during the 2026-02-05 to 2026-02-06 9-ID Grand Tube pumpdown, the GT vacuum reading was 3.6e04 at 2026-02-06\_14:34:06 [10119\_\_chunk\_0]. For the 9-BM Ge8 xspress3 setup, the [main] section trigger/sync PVs that would have failed the caget existence check were the d\#ScanStart2.OUTN scan-start PVs and the scalerd\#sync.INBN scan-sync PVs [10086\_\_chunk\_0].\par
\noindent\emph{Nuggets.}~\textbf{[V]}~Just before GV11 was opened during the 2026-02-05 to 2026-02-06 9-ID Grand Tube pumpdown, the GT vacuum reading was 3.6e04. \quad [o]~The timestamp for the GT vacuum reading just before GV11 was opened was 2026-02-06\_14:34:06. \quad \textbf{[V]}~For the 9-BM Ge8 xspress3 setup, the [main] section d\#ScanStart2.OUTN scan-start PVs would have failed the caget existence check. \quad \textbf{[V]}~For the 9-BM Ge8 xspress3 setup, the [main] section scalerd\#sync.INBN scan-sync PVs would have failed the caget existence check.\par
\noindent\emph{Evidence.}~\texttt{10119\_\_chunk\_0} (primary), \texttt{10086\_\_chunk\_0} (primary)\par

\smallskip\noindent\textbf{APS-BENCH-027}\quad {\footnotesize \textit{comparative} $\cdot$ WORK\_REQUESTS $\cdot$ medium}\par
\noindent\emph{Q.}~Which issue is tied to adding event timestamps for injection cycle data collection---the ioclet1bpm reboot request or the MTime power supply failure---and what corrective action is requested for each?\par
\noindent\emph{Gold.}~Adding event timestamps for injection cycle data collection is tied to the ioclet1bpm reboot request; the requested corrective action is to reboot ioclet1bpm [44960\_\_chunk\_0]. The MTime issue is a power supply failure; the requested corrective action is to swap the power supply [50362\_\_chunk\_0].\par
\noindent\emph{Nuggets.}~\textbf{[V]}~The issue tied to adding event timestamps for injection cycle data collection is the ioclet1bpm reboot request. \quad \textbf{[V]}~The requested corrective action for the injection cycle event timestamp issue is to reboot ioclet1bpm. \quad \textbf{[V]}~The MTime issue is a power supply failure. \quad \textbf{[V]}~The requested corrective action for the MTime power supply failure is to swap the power supply.\par
\noindent\emph{Evidence.}~\texttt{44960\_\_chunk\_0} (primary), \texttt{50362\_\_chunk\_0} (primary)\par

\smallskip\noindent\textbf{APS-BENCH-028}\quad {\footnotesize \textit{troubleshooting} $\cdot$ ICMS $\cdot$ medium}\par
\noindent\emph{Q.}~SPX0 keeps dropping RF and staying latched RF-OFF after a fast interlock event, with the equipment protection display showing an ARC-related trip and no automatic recovery. Based on the cryomodule and master fast interlock scheme, what fault path is most likely causing the shutdown, and what operator action is required to restore klystron drive?\par
\noindent\emph{Gold.}~The most likely fault path is an ARC-related cryomodule fast interlock fault---specifically an arc detector on the forward power coupler or LOM damper waveguide RF windows---being detected by the fast cryomodule interlock, which sends a latching trip command via fiber optic link to the master fast RF interlock in the amplifier [APSU\_1430567\_\_chunk\_1037]. That master fast interlock then interrupts RF drive to the klystron and stays latched in the RF-off condition [APSU\_1430567\_\_chunk\_1037]. To restore klystron drive, the operator must issue a manual reset command, because there is no automatic recovery described in the passage [APSU\_1430567\_\_chunk\_1037].\par
\noindent\emph{Nuggets.}~\textbf{[V]} (cause)~The shutdown is most likely caused by an ARC-related cryomodule fast interlock fault. \quad [o] (cause)~A likely source of the ARC fault is an arc detector on the forward power coupler RF window. \quad [o] (cause)~A likely source of the ARC fault is an arc detector on the LOM damper waveguide RF window. \quad \textbf{[V]} (cause)~The cryomodule fast interlock sends a latching trip command to the master fast RF interlock via a fiber optic link. \quad \textbf{[V]}~The master fast RF interlock interrupts RF drive to the klystron. \quad \textbf{[V]} (cause)~The master fast RF interlock remains latched in the RF-OFF condition after this trip. \quad \textbf{[V]} (resolution)~Restoring klystron drive requires the operator to issue a manual reset command. \quad [o] (cause)~There is no automatic recovery for this fault path described in the passage.\par
\noindent\emph{Evidence.}~\texttt{APSU\_1430567\_\_chunk\_1037} (primary), \texttt{APSU\_1430567\_\_chunk\_1034} (supporting)\par

\smallskip\noindent\textbf{APS-BENCH-029}\quad {\footnotesize \textit{comparative} $\cdot$ BELY $\cdot$ medium}\par
\noindent\emph{Q.}~Compared with the 20-BM AuPd nanoparticles setup where the HR mirror was backed off to 4 mrad, what HR mirror angle was used for the 9-BM 10 keV ion-chamber test on the Pt stripe, and how do the two mirror settings differ operationally?\par
\noindent\emph{Gold.}~For the 9-BM 10 keV ion-chamber test on the Pt stripe, the HR mirror was inserted to 6.0 mrad [8773\_\_chunk\_0]. Compared with the 20-BM AuPd nanoparticles setup, where the HR mirror was at 5 mrad and then backed off to 4 mrad [9300\_\_chunk\_0], the 9-BM test used a higher angle by 2.0 mrad relative to the final 4 mrad setting [8773\_\_chunk\_0][9300\_\_chunk\_0]. Operationally, the 20-BM setting involved backing the mirror off from 5 to 4 mrad [9300\_\_chunk\_0], while the 9-BM test involved inserting the HR mirror onto the Pt stripe at 6.0 mrad during the 10 keV test after removing it, setting energy to 10 keV, and detuning the mono by 25\% [8773\_\_chunk\_0].\par
\noindent\emph{Nuggets.}~\textbf{[V]}~For the 9-BM 10 keV ion-chamber test on the Pt stripe, the HR mirror angle was 6.0 mrad. \quad \textbf{[V]}~In the 20-BM AuPd nanoparticles setup, the HR mirror's final backed-off setting was 4 mrad. \quad \textbf{[V]}~The 9-BM 6.0 mrad setting is 2.0 mrad higher than the 20-BM final 4 mrad setting. \quad [o]~At 20-BM, the HR mirror was first at 5 mrad before being backed off to 4 mrad. \quad \textbf{[V]}~Operationally, the 20-BM configuration involved backing the HR mirror off from 5 mrad to 4 mrad. \quad \textbf{[V]}~Operationally, the 9-BM test involved inserting the HR mirror onto the Pt stripe at 6.0 mrad. \quad [o]~During the 9-BM procedure, the mirror insertion at 6.0 mrad occurred during a 10 keV test after the mirror had been removed, with the monochromator detuned by 25\%.\par
\noindent\emph{Evidence.}~\texttt{8773\_\_chunk\_0} (primary), \texttt{9300\_\_chunk\_0} (supporting)\par

\smallskip\noindent\textbf{APS-BENCH-030}\quad {\footnotesize \textit{troubleshooting} $\cdot$ ICMS $\cdot$ medium}\par
\noindent\emph{Q.}~The SPX0 cavity RF is latched off and won't come back after a coupler/window arc trip. What should I check in the fast cryomodule interlock path, and what operator action is required to restore RF?\par
\noindent\emph{Gold.}~You should check the fast cryomodule interlock path for a latching fault from the arc detectors on the forward power coupler and RF windows, and more generally for any fast cryomodule interlock fault being relayed by the fiber-optic link to the master fast RF interlock chassis in the amplifier [APSU\_1430567\_\_chunk\_1037]. The trip causes the master fast interlock to interrupt RF drive to the klystron and latch RF off until a manual reset command is issued by the system operator [APSU\_1430567\_\_chunk\_1037]. The passages do not provide any more specific SPX0-only recovery step beyond that manual operator reset [APSU\_1430567\_\_chunk\_1037].\par
\noindent\emph{Nuggets.}~\textbf{[V]} (cause)~Check the fast cryomodule interlock path for a latching fault from the arc detectors on the forward power coupler. \quad \textbf{[V]} (cause)~Check the fast cryomodule interlock path for a latching fault from the arc detectors on the RF windows. \quad \textbf{[V]} (cause)~Check for any fast cryomodule interlock fault being relayed by the fiber-optic link to the master fast RF interlock chassis in the amplifier. \quad [o] (cause)~The trip causes the master fast interlock to interrupt RF drive to the klystron. \quad \textbf{[V]} (resolution)~The master fast interlock latches RF off until a manual reset command is issued. \quad \textbf{[V]} (resolution)~The required recovery action is a manual reset by the system operator. \quad [o] (resolution)~No more specific SPX0-only recovery step is provided beyond the manual operator reset.\par
\noindent\emph{Evidence.}~\texttt{APSU\_1430567\_\_chunk\_1037} (primary), \texttt{APSU\_1430567\_\_chunk\_1034} (supporting)\par

\smallskip\noindent\textbf{APS-BENCH-031}\quad {\footnotesize \textit{comparative} $\cdot$ BELY $\cdot$ medium}\par
\noindent\emph{Q.}~Compared with the 9-BM Cu foil tests where switching the SR570s to low-noise/no filters and increasing the N2 IC HV to 450 V reduced glitches and improved the spectra, how did the later UW PSIC Mo foil runs compare operationally in terms of SR570 filter/sensitivity settings and the Dtheta-pitch setpoint used during the alternating knife-edge/XANES sequence?\par
\noindent\emph{Gold.}~In the later UW PSIC Mo foil runs, the SR570 settings were initially low-noise mode with no filters, no offsets, and 50 pA/V sensitivity [8468\_\_chunk\_9], then changed to 12 dB low-pass, 1 kHz filter frequency, low-noise mode, no offset, 50 pA/V sensitivity [8468\_\_chunk\_9], and later changed again to no filters, low-noise mode, no offset, 10 pA/V sensitivity [8468\_\_chunk\_9]. During the alternating knife-edge/XANES sequence, the Dtheta-pitch setpoint was 7965 steps on 2025/10/17 and 2025/10/18, and 7964 steps on 2025/10/19 [8468\_\_chunk\_9]. By comparison, the earlier 9-BM Cu foil tests reported setting all SR570s to low-noise, no filters and increasing the N2-filled IC HV to 450 V, which reduced glitches and improved the spectra [7323\_\_chunk\_0].\par
\noindent\emph{Nuggets.}~\textbf{[V]}~In the later UW PSIC Mo foil runs, the SR570s were initially set to low-noise mode with no filters, no offset, and 50 pA/V sensitivity. \quad \textbf{[V]}~Later in the UW PSIC Mo foil runs, the SR570s were changed to low-noise mode with a 12 dB low-pass filter at 1 kHz, no offset, and 50 pA/V sensitivity. \quad \textbf{[V]}~Later again in the UW PSIC Mo foil runs, the SR570s were changed to low-noise mode with no filters, no offset, and 10 pA/V sensitivity. \quad \textbf{[V]}~During the alternating knife-edge/XANES sequence, the Dtheta-pitch setpoint was 7965 steps on 2025/10/17. \quad [o]~During the alternating knife-edge/XANES sequence, the Dtheta-pitch setpoint was 7965 steps on 2025/10/18. \quad \textbf{[V]}~During the alternating knife-edge/XANES sequence, the Dtheta-pitch setpoint was 7964 steps on 2025/10/19.\par
\noindent\emph{Evidence.}~\texttt{8468\_\_chunk\_9} (primary), \texttt{7323\_\_chunk\_0} (supporting)\par

\smallskip\noindent\textbf{APS-BENCH-032}\quad {\footnotesize \textit{causal} $\cdot$ ICMS $\cdot$ medium}\par
\noindent\emph{Q.}~Why would using the Particle Accumulator Ring as the injector for the Linac Extension Area change LEA beam delivery compared with the usual photocathode or RF gun source?\par
\noindent\emph{Gold.}~Using the Particle Accumulator Ring (PAR) as the injector would change LEA beam delivery because the usual LEA source is the photocathode electron gun, with RF electron guns also possible, whereas the proposed PAR-based scheme would instead use PAR as a damping or accumulator ring before transporting the extracted beam to LEA [APS\_2189598\_\_chunk\_0]. The passage does not provide further details on exactly how the delivered beam properties or delivery mode would change compared with the usual photocathode or RF gun source [APS\_2189598\_\_chunk\_0].\par
\noindent\emph{Nuggets.}~\textbf{[V]}~The usual injector source for LEA is the photocathode electron gun. \quad [o]~RF electron guns are also possible injector sources for LEA. \quad \textbf{[V]}~In the proposed scheme, the Particle Accumulator Ring would be used as a damping or accumulator ring before beam extraction to LEA. \quad \textbf{[V]}~Using PAR would therefore replace the usual direct photocathode or RF gun injection path with a PAR-based injection path. \quad [o]~The cited passage does not specify exactly how the delivered beam properties would change relative to the usual sources. \quad [o]~The cited passage does not specify exactly how the beam delivery mode would change relative to the usual sources.\par
\noindent\emph{Evidence.}~\texttt{APS\_2189598\_\_chunk\_0} (primary), \texttt{APS\_2189598\_\_chunk\_7} (supporting)\par

\smallskip\noindent\textbf{APS-BENCH-033}\quad {\footnotesize \textit{multi\_hop} $\cdot$ ICMS $\cdot$ medium}\par
\noindent\emph{Q.}~For PAR photon diagnostics, if I need the monitor with the larger light-gathering acceptance, which instrument should I use and what are the accepted synchrotron-radiation half-angles for it and for the streak camera?\par
\noindent\emph{Gold.}~Use SLM1 for the larger light-gathering acceptance, because it accepts a larger synchrotron-radiation half-angle than the streak camera [APS\_2181087\_\_chunk\_0]. The accepted synchrotron-radiation half-angle is 5.3 mrad for SLM1 [APS\_2181087\_\_chunk\_0] and 1.9 mrad for the streak camera [APS\_2181087\_\_chunk\_0].\par
\noindent\emph{Nuggets.}~\textbf{[V]}~The instrument to use for larger light-gathering acceptance is SLM1. \quad \textbf{[V]}~SLM1 accepts a synchrotron-radiation half-angle of 5.3 mrad. \quad \textbf{[V]}~The streak camera accepts a synchrotron-radiation half-angle of 1.9 mrad. \quad [o]~SLM1 has a larger accepted synchrotron-radiation half-angle than the streak camera.\par
\noindent\emph{Evidence.}~\texttt{APS\_2181087\_\_chunk\_0} (primary), \texttt{APS\_2181087\_\_chunk\_6} (supporting)\par

\smallskip\noindent\textbf{APS-BENCH-034}\quad {\footnotesize \textit{multi\_hop} $\cdot$ TEAMS $\cdot$ medium}\par
\noindent\emph{Q.}~If the FID pulsers are locked out and the IOC is showing timeout errors because the devices look unresponsive, and you also can't reach Rob Keane and don't have a secondary Diagnostics contact for the BESOCM system, who should you call next for help?\par
\noindent\emph{Gold.}~If the FID pulsers are locked out, the IOC will see an unresponsive device and encounter timeout errors [teams\_Fors, Thomas L.\_f01fd5c1b495\_\_chunk\_0]. If you can't reach Rob Keane and do not have a secondary Diagnostics contact for the BESOCM system, you should call Guobao next for help [teams\_Johnson, Andrew N.\_16bc9ca6ec27\_\_chunk\_0].\par
\noindent\emph{Nuggets.}~\textbf{[V]}~Locked-out FID pulsers appear unresponsive to the IOC. \quad [o]~When the FID pulsers appear unresponsive, the IOC shows timeout errors. \quad \textbf{[V]}~If Rob Keane cannot be reached, another person should be contacted for BESOCM help. \quad \textbf{[V]}~If there is no secondary Diagnostics contact for the BESOCM system, the next person to call is Guobao.\par
\noindent\emph{Evidence.}~\texttt{teams\_Fors, Thomas L.\_f01fd5c1b495\_\_chunk\_0} (supporting), \texttt{teams\_Johnson, Andrew N.\_16bc9ca6ec27\_\_chunk\_0} (primary)\par

\smallskip\noindent\textbf{APS-BENCH-035}\quad {\footnotesize \textit{causal} $\cdot$ TEAMS $\cdot$ medium}\par
\noindent\emph{Q.}~Why did the IOC start showing timeout errors on the FID pulsers after the 33ID PSS trip, if the trip was caused by the BPLD from the MPS side?\par
\noindent\emph{Gold.}~The 33ID PSS trip was caused by the BPLD from the MPS side [teams\_Shen, Guobao\_b50b74042679\_\_chunk\_0]. The IOC started showing timeout errors on the FID pulsers because the FID pulsers do not have separate control power, so when they are locked out, the IOC sees them as unresponsive devices and encounters timeout errors [teams\_Fors, Thomas L.\_f01fd5c1b495\_\_chunk\_0].\par
\noindent\emph{Nuggets.}~\textbf{[V]} (cause)~The 33ID PSS trip was caused by the BPLD from the MPS side. \quad \textbf{[V]}~The FID pulsers do not have separate control power. \quad \textbf{[V]}~When the FID pulsers are locked out, the IOC sees them as unresponsive devices. \quad \textbf{[V]} (cause)~The IOC timeout errors occurred because it was trying to communicate with locked-out, unresponsive FID pulsers.\par
\noindent\emph{Evidence.}~\texttt{teams\_Shen, Guobao\_b50b74042679\_\_chunk\_0} (primary), \texttt{teams\_Fors, Thomas L.\_f01fd5c1b495\_\_chunk\_0} (primary)\par

\smallskip\noindent\textbf{APS-BENCH-036}\quad {\footnotesize \textit{comparative} $\cdot$ ICMS $\cdot$ medium}\par
\noindent\emph{Q.}~Compared with a 2M undulator, how many additional motor power and motor feedback cables should ops expect for a 4M undulator when checking the HPMU cable/port organization?\par
\noindent\emph{Gold.}~A 2M undulator has 2 motor power cables and 2 motor feedback cables, while a 4M undulator has 4 motor power cables and 4 motor feedback cables [APSU\_2185843\_\_chunk\_3]. Therefore, compared with a 2M undulator, ops should expect 2 additional motor power cables and 2 additional motor feedback cables for a 4M undulator [APSU\_2185843\_\_chunk\_3].\par
\noindent\emph{Nuggets.}~\textbf{[V]}~A 2M undulator has 2 motor power cables. \quad \textbf{[V]}~A 2M undulator has 2 motor feedback cables. \quad \textbf{[V]}~A 4M undulator has 4 motor power cables. \quad \textbf{[V]}~A 4M undulator has 4 motor feedback cables. \quad \textbf{[V]}~Compared with a 2M undulator, a 4M undulator has 2 additional motor power cables. \quad \textbf{[V]}~Compared with a 2M undulator, a 4M undulator has 2 additional motor feedback cables.\par
\noindent\emph{Evidence.}~\texttt{APSU\_2185843\_\_chunk\_3} (primary), \texttt{APSU\_2185843\_\_chunk\_14} (supporting)\par

\smallskip\noindent\textbf{APS-BENCH-037}\quad {\footnotesize \textit{temporal} $\cdot$ ICMS $\cdot$ medium}\par
\noindent\emph{Q.}~During the June 2021 PAR diagnostics review for APS-U single-bunch accumulation, which photon monitor had the smaller accepted synchrotron-radiation half-angle, and why would that matter operationally when trying to observe longitudinal bunch instabilities at the high-charge condition called out for PAR?\par
\noindent\emph{Gold.}~The streak camera had the smaller accepted synchrotron-radiation half-angle: 1.9 mrad versus 5.3 mrad for SLM1 [APS\_2181087\_\_chunk\_0]. Operationally, this would matter because PAR needs diagnostics to observe longitudinal electron bunch instabilities at the high-charge single-bunch condition of up to 20 nC, where measurement and control of beam instabilities may be important [APS\_2170123\_\_chunk\_0]. The passages do not explicitly state the operational consequence of the smaller half-angle beyond identifying it as a smaller accepted synchrotron-radiation half-angle [APS\_2181087\_\_chunk\_0].\par
\noindent\emph{Nuggets.}~\textbf{[V]}~The streak camera had the smaller accepted synchrotron-radiation half-angle. \quad \textbf{[V]}~The streak camera's accepted synchrotron-radiation half-angle was 1.9 mrad. \quad \textbf{[V]}~SLM1's accepted synchrotron-radiation half-angle was 5.3 mrad. \quad \textbf{[V]}~PAR diagnostics were needed to observe longitudinal electron bunch instabilities in single-bunch operation. \quad \textbf{[V]}~The high-charge single-bunch condition called out for PAR was up to 20 nC. \quad [o]~At that high-charge condition, measurement and control of beam instabilities may be important operationally. \quad [o]~The cited passages do not explicitly state any more specific operational consequence of the smaller half-angle than that it is smaller.\par
\noindent\emph{Evidence.}~\texttt{APS\_2181087\_\_chunk\_0} (primary), \texttt{APS\_2170123\_\_chunk\_0} (supporting)\par

\smallskip\noindent\textbf{APS-BENCH-038}\quad {\footnotesize \textit{temporal} $\cdot$ BELY $\cdot$ medium}\par
\noindent\emph{Q.}~During the 2026-02-03 9-BM Ge8 with xspress3 setup, which [main] xsp3.ini scan-trigger PVs should we flag as missing with caget, and how does that differ from the 2025-01-10 ISN\_Vortex\_ME7 note where the EPICS IOC was running but still showed high reset counts and pileup on all channels except ch 7?\par
\noindent\emph{Gold.}~For the 2026-02-03 9-BM Ge8 with xspress3 setup, the [main] xsp3.ini scan-trigger PVs to flag as missing with caget are the "d\#ScanStart2.OUTN" PVs and the "scalerd\#sync.INBN" PVs, i.e. the configured d1/d2/d3 scan-start and scan-sync PVs do not exist [10086\_\_chunk\_0]. This differs from the 2025-01-10 ISN\_Vortex\_ME7 note because there the EPICS IOC was running ("Using EPICS IOC (ioc-7Channel)") but, despite that, all channels except ch 7 showed high reset counts and pileup in the last channel [6297\_\_chunk\_0].\par
\noindent\emph{Nuggets.}~\textbf{[V]}~In the 2026-02-03 9-BM Ge8 with xspress3 setup, the [main] xsp3.ini scan-trigger PVs to flag as missing with caget are the d\#ScanStart2.OUTN PVs. \quad \textbf{[V]}~In the 2026-02-03 9-BM Ge8 with xspress3 setup, the [main] xsp3.ini scan-trigger PVs to flag as missing with caget are the scalerd\#sync.INBN PVs. \quad [o]~The missing PVs correspond to the configured d1, d2, and d3 scan-start and scan-sync PVs not existing. \quad \textbf{[V]}~In the 2025-01-10 ISN\_Vortex\_ME7 note, the EPICS IOC was running. \quad [o]~The 2025-01-10 ISN\_Vortex\_ME7 note specifically says "Using EPICS IOC (ioc-7Channel)". \quad \textbf{[V]}~In the 2025-01-10 ISN\_Vortex\_ME7 note, all channels except channel 7 showed high reset counts. \quad \textbf{[V]}~In the 2025-01-10 ISN\_Vortex\_ME7 note, pileup was present in the last channel.\par
\noindent\emph{Evidence.}~\texttt{10086\_\_chunk\_0} (primary), \texttt{6297\_\_chunk\_0} (supporting)\par

\smallskip\noindent\textbf{APS-BENCH-039}\quad {\footnotesize \textit{multi\_hop} $\cdot$ ICMS $\cdot$ medium}\par
\noindent\emph{Q.}~For the new PAR terahertz diagnostic, if we add the in-vacuum mirror assembly and transmission viewport on the visible light port extension, would that vacuum hardware be expected to fall outside ASME pressure-vessel code scope, and what design limits in the spec determine that?\par
\noindent\emph{Gold.}~Yes. The specification says vacuum vessels designed under this specification are expected to be outside the scope of the ASME Boilers and Pressure Vessel Code because of the design pressure differential being less than 15 psi and the size being less than 6 inches diameter regardless of length [APS\_2176697\_\_chunk\_19]. The new PAR terahertz diagnostic includes an in-vacuum mirror assembly and transmission viewport as deliverables for the visible light diagnostics port extension [APS\_2176697\_\_chunk\_0].\par
\noindent\emph{Nuggets.}~\textbf{[V]}~The in-vacuum mirror assembly and transmission viewport for the new PAR terahertz diagnostic would be expected to fall outside ASME pressure-vessel code scope. \quad \textbf{[V]}~The specification states that vacuum vessels under this specification are expected to be outside the scope of the ASME Boilers and Pressure Vessel Code. \quad \textbf{[V]}~One design limit in the specification is a design pressure differential of less than 15 psi. \quad \textbf{[V]}~Another design limit in the specification is a vessel size of less than 6 inches in diameter. \quad [o]~The less-than-6-inches diameter limit applies regardless of vessel length. \quad [o]~The new PAR terahertz diagnostic includes an in-vacuum mirror assembly as a deliverable for the visible light diagnostics port extension. \quad [o]~The new PAR terahertz diagnostic includes a transmission viewport as a deliverable for the visible light diagnostics port extension.\par
\noindent\emph{Evidence.}~\texttt{APS\_2176697\_\_chunk\_19} (primary), \texttt{APS\_2176697\_\_chunk\_0} (supporting)\par

\smallskip\noindent\textbf{APS-BENCH-040}\quad {\footnotesize \textit{causal} $\cdot$ ICMS $\cdot$ medium}\par
\noindent\emph{Q.}~Why would accumulating up to 20 nC in a single PAR bunch during APS-U make it important to add a terahertz diagnostic for observing longitudinal bunch instabilities?\par
\noindent\emph{Gold.}~During APS-U, the PAR is required to accumulate charge in a single bunch up to 20 nC [APS\_2170123\_\_chunk\_0]. At such high charges, measurement and control of electron beam instabilities may be important [APS\_2170123\_\_chunk\_0]. The terahertz diagnostic is intended to be used to observe longitudinal electron bunch instabilities [APS\_2170123\_\_chunk\_0]. Therefore, accumulating up to 20 nC in one PAR bunch makes adding a terahertz diagnostic important so those longitudinal bunch instabilities can be observed, measured, and controlled [APS\_2170123\_\_chunk\_0].\par
\noindent\emph{Nuggets.}~\textbf{[V]}~During APS-U, the PAR must accumulate up to 20 nC of charge in a single bunch. \quad \textbf{[V]}~At such high single-bunch charge, electron beam instabilities become important to measure and control. \quad \textbf{[V]}~The terahertz diagnostic is intended to observe longitudinal electron bunch instabilities. \quad \textbf{[V]} (cause)~Adding the terahertz diagnostic is important because it enables observation of the longitudinal instabilities that can arise at 20 nC single-bunch operation. \quad [o]~Observing those instabilities supports their measurement and control.\par
\noindent\emph{Evidence.}~\texttt{APS\_2170123\_\_chunk\_0} (primary), \texttt{APS\_2170123\_\_chunk\_22} (supporting)\par

\smallskip\noindent\textbf{APS-BENCH-041}\quad {\footnotesize \textit{multi\_hop} $\cdot$ WORK\_REQUESTS $\cdot$ hard}\par
\noindent\emph{Q.}~If the Booster Dipole Slave power supply tripped on an SCR failure and, separately, the MTime power supply has failed and needs to be swapped, which two power-supply issues should operations coordinate to address before returning the affected systems to service?\par
\noindent\emph{Gold.}~Operations should coordinate (1) repairing the Booster Dipole Slave power supply because it tripped on an SCR failure [45927\_\_chunk\_0] and (2) swapping the failed MTime power supply [50362\_\_chunk\_0].\par
\noindent\emph{Nuggets.}~\textbf{[V]}~The Booster Dipole Slave power supply tripped on an SCR failure. \quad \textbf{[V]}~The Booster Dipole Slave power supply needs to be repaired before the affected system is returned to service. \quad \textbf{[V]}~The MTime power supply has failed. \quad \textbf{[V]}~The failed MTime power supply needs to be swapped before the affected system is returned to service. \quad [o]~Operations should coordinate addressing both the Booster Dipole Slave power supply issue and the MTime power supply issue.\par
\noindent\emph{Evidence.}~\texttt{45927\_\_chunk\_0} (primary), \texttt{50362\_\_chunk\_0} (primary)\par

\smallskip\noindent\textbf{APS-BENCH-042}\quad {\footnotesize \textit{comparative} $\cdot$ ICMS $\cdot$ hard}\par
\noindent\emph{Q.}~For PAR commissioning readiness, how did the status of the injector-side ESH/HP shield design review implementation compare with the post-installation vacuum checkout sign-offs, specifically in terms of whether each was accepted and what operational controls or interlocks were demonstrated on the vacuum system?\par
\noindent\emph{Gold.}~The injector-side ESH/HP shield design review implementation was accepted as implemented per the Injector Safety Assessment document [APS\_1271753\_\_chunk\_170]. The post-installation vacuum checkout sign-offs were also accepted [APS\_1271753\_\_chunk\_137]. On the vacuum system, the demonstrated/accepted operational controls and interlocks included manual override demonstrated, system control of pumps and valves accepted, over-pressure interlock demonstrated, ground-fault interlock demonstrated, and local control and override demonstrated [APS\_1271753\_\_chunk\_137].\par
\noindent\emph{Nuggets.}~\textbf{[V]}~The injector-side ESH/HP shield design review implementation was accepted as implemented. \quad \textbf{[V]}~The post-installation vacuum checkout sign-offs were accepted. \quad \textbf{[V]}~A manual override on the vacuum system was demonstrated. \quad \textbf{[V]}~System control of vacuum pumps was accepted. \quad \textbf{[V]}~System control of vacuum valves was accepted. \quad \textbf{[V]}~An over-pressure interlock on the vacuum system was demonstrated. \quad \textbf{[V]}~A ground-fault interlock on the vacuum system was demonstrated. \quad [o]~Local control on the vacuum system was demonstrated. \quad [o]~Local override on the vacuum system was demonstrated.\par
\noindent\emph{Evidence.}~\texttt{APS\_1271753\_\_chunk\_170} (primary), \texttt{APS\_1271753\_\_chunk\_137} (primary)\par

\smallskip\noindent\textbf{APS-BENCH-043}\quad {\footnotesize \textit{troubleshooting} $\cdot$ ICMS $\cdot$ hard}\par
\noindent\emph{Q.}~On SPX0, after an arc detector or cavity vacuum/over-field/quench trip from the cryomodule interlock, the klystron RF drive drops out and stays latched RF-off even though the protection status looks normal again. Based on the SPX0 interlock architecture, what should the operator check and do to restore RF, and where are the trip thresholds set?\par
\noindent\emph{Gold.}~After an arc detector trip or a cavity vacuum/over-field/quench trip from the cryomodule interlock, the fault sends a latching trip command to the master fast RF interlock in the amplifier, which interrupts RF drive to the klystron and latches RF off [APSU\_1430567\_\_chunk\_1037]. Even if protection status appears normal again, the operator should check whether the master fast interlock is still latched RF-off and issue a manual reset command to restore RF [APSU\_1430567\_\_chunk\_1037]. The trip thresholds for the fast interlock signals are set manually by potentiometers in the fast interlock chassis [APSU\_1430567\_\_chunk\_1037].\par
\noindent\emph{Nuggets.}~\textbf{[V]} (cause)~An arc detector trip sends a latching trip command to the master fast RF interlock in the amplifier. \quad \textbf{[V]} (cause)~A cavity vacuum trip from the cryomodule interlock sends a latching trip command to the master fast RF interlock in the amplifier. \quad \textbf{[V]} (cause)~A cavity over-field trip from the cryomodule interlock sends a latching trip command to the master fast RF interlock in the amplifier. \quad \textbf{[V]} (cause)~A cavity quench trip from the cryomodule interlock sends a latching trip command to the master fast RF interlock in the amplifier. \quad \textbf{[V]}~The master fast RF interlock interrupts RF drive to the klystron and latches RF off. \quad \textbf{[V]}~The operator should check whether the master fast RF interlock is still latched RF-off. \quad \textbf{[V]} (resolution)~The operator must issue a manual reset command to the master fast RF interlock to restore RF. \quad [o] (cause)~The trip thresholds for the fast interlock signals are set manually by potentiometers in the fast interlock chassis.\par
\noindent\emph{Evidence.}~\texttt{APSU\_1430567\_\_chunk\_1037} (primary), \texttt{APSU\_1430567\_\_chunk\_1034} (supporting)\par

\smallskip\noindent\textbf{APS-BENCH-044}\quad {\footnotesize \textit{causal} $\cdot$ BELY $\cdot$ hard}\par
\noindent\emph{Q.}~Why did the jumps in the I\_T/scaler 3 signal disappear after the SR570 settings were manually corrected and the ini files were later updated to turn all SR570 filters off and align the PV/channel mapping with beamline.ini, even though LabVIEW had been showing different ion chamber settings?\par
\noindent\emph{Gold.}~The jumps disappeared because the problem was in the scaler 3/SR570 signal chain settings rather than the ion chamber itself: after swapping cables, scaler 3 remained noisy, which indicated an issue with the signal chain, not the IC [7413\_\_chunk\_0]. LabVIEW was not actually controlling the ion chamber mode and filter settings, so the settings shown there could be wrong [7413\_\_chunk\_0][7583\_\_chunk\_0]. When all SR570s were manually set to low-noise with no filters, the jumps in scaler 3 disappeared [7413\_\_chunk\_0]. The later ini-file changes then made the configured SR570 settings consistent by turning all SR570 filters off in sr570.ini and changing beamline.ini channels to match sr570.ini [7583\_\_chunk\_0]. Beamline server reads the PVs to set the SR570, while LabVIEW is not updated when PVs change, so LabVIEW could still show different ion chamber settings even though the actual PV/channel configuration had been corrected [7583\_\_chunk\_0].\par
\noindent\emph{Nuggets.}~\textbf{[V]}~After swapping cables, scaler 3 remained noisy, showing the problem was in the scaler 3/SR570 signal chain rather than in the ion chamber itself. \quad \textbf{[V]}~LabVIEW was not actually controlling the ion chamber mode and filter settings. \quad \textbf{[V]} (cause)~Because LabVIEW was not controlling those settings, the ion chamber settings displayed in LabVIEW could be wrong. \quad \textbf{[V]}~When all SR570 units were manually set to low-noise mode with no filters, the jumps in scaler 3 disappeared. \quad [o]~The later sr570.ini update turned all SR570 filters off. \quad [o]~The later beamline.ini update changed channel assignments to match sr570.ini. \quad [o]~The beamline server reads PVs to set the SR570 configuration. \quad [o]~LabVIEW is not updated when the PVs change. \quad \textbf{[V]} (resolution)~The jumps disappeared because the actual SR570/PV/channel configuration was corrected, even though LabVIEW still showed different ion chamber settings.\par
\noindent\emph{Evidence.}~\texttt{7413\_\_chunk\_0} (primary), \texttt{7583\_\_chunk\_0} (primary)\par

\smallskip\noindent\textbf{APS-BENCH-045}\quad {\footnotesize \textit{causal} $\cdot$ SDDS $\cdot$ hard}\par
\noindent\emph{Q.}~Why would the predicted BPM change come out incorrect if we first used sddsmatrixmult with the response matrix and correctorChange file and then reviewed the result in sddsplot, but the matrix multiplication had the left- and right-hand inputs reversed instead of using the response matrix on the left?\par
\noindent\emph{Gold.}~The predicted BPM change would be incorrect because sddsmatrixmult interprets the first input file as the left-hand matrix and the second input file as the right-hand matrix [SDDS\_OAG\_User's Guide for SDDS Toolkit Version 5.0\_chunk\_106\_\_chunk\_1]. In the stated BPM example, the correct computation is the response matrix multiplied by the columns of the correctorChange file to produce bpmExpectedChange [SDDS\_OAG\_User's Guide for SDDS Toolkit Version 5.0\_chunk\_106\_\_chunk\_1]. If the inputs were reversed without using the option that swaps left and right roles, then the multiplication order would not match the documented BPM example, so the resulting predicted BPM change would come out incorrect [SDDS\_OAG\_User's Guide for SDDS Toolkit Version 5.0\_chunk\_106\_\_chunk\_1]. The passages do not provide any additional explanation about sddsplot causing the error; they only show that sddsplot is used for plotting data [SDDS\_OAG\_User's Guide for SDDS Toolkit Version 5.0\_chunk\_126\_\_chunk\_0].\par
\noindent\emph{Nuggets.}~\textbf{[V]}~sddsmatrixmult interprets its first input file as the left-hand matrix. \quad \textbf{[V]}~sddsmatrixmult interprets its second input file as the right-hand matrix. \quad \textbf{[V]}~In the BPM example, the correct computation is response matrix multiplied by the columns of the correctorChange file. \quad \textbf{[V]}~Reversing the response matrix and correctorChange inputs changes the multiplication order from the documented BPM computation. \quad \textbf{[V]}~If the inputs are reversed without using the option that swaps left and right roles, the predicted BPM change result is incorrect. \quad [o]~The cited passages do not attribute the error to sddsplot. \quad [o]~The cited passages only show sddsplot being used to plot data.\par
\noindent\emph{Evidence.}~\texttt{SDDS\_OAG\_User's Guide for SDDS Toolkit Version 5.0\_chunk\_106\_\_chunk\_1} (primary), \texttt{SDDS\_OAG\_User's Guide for SDDS Toolkit Version 5.0\_chunk\_126\_\_chunk\_0} (supporting)\par

\smallskip\noindent\textbf{APS-BENCH-046}\quad {\footnotesize \textit{causal} $\cdot$ OAG $\cdot$ hard}\par
\noindent\emph{Q.}~Why would a new VSWR or ArcTrip event on one of the linac RF stations show up in LinacRFFaultReview for operator annotation and printing only after LinacRFFaultLogger detects a fault-count change and writes the timestamped snapshot to the historical SDDS file?\par
\noindent\emph{Gold.}~A new VSWR or ArcTrip event would show up in LinacRFFaultReview only after LinacRFFaultLogger detects a fault-count change because the logger is the component that automatically detects new faults by monitoring count PVs and then saves fault data with timestamps, including fault snapshots, to historical SDDS files [LinacRFFaultLogger\_\_chunk\_0]. LinacRFFaultReview is a reviewer application that displays RF fault events with timestamps, allows operators to annotate them, supports Review/Print reports, and saves comment updates to SDDS files, but the passages do not say that it detects faults itself [LinacRFFaultReview\_\_chunk\_0]. Therefore, the event appears there for operator annotation and printing after the logger has written the timestamped fault record to the historical SDDS file that the review workflow uses [LinacRFFaultLogger\_\_chunk\_0][LinacRFFaultReview\_\_chunk\_0].\par
\noindent\emph{Nuggets.}~\textbf{[V]} (cause)~LinacRFFaultLogger detects new RF faults by monitoring fault-count PVs for count changes. \quad \textbf{[V]} (cause)~When LinacRFFaultLogger detects a fault-count change, it writes a timestamped fault snapshot to a historical SDDS file. \quad \textbf{[V]} (cause)~LinacRFFaultReview displays RF fault events with timestamps from the stored fault records. \quad [o] (cause)~LinacRFFaultReview supports operator annotation of RF fault events. \quad [o] (cause)~LinacRFFaultReview supports review and print reporting for RF fault events. \quad \textbf{[V]}~The documentation does not indicate that LinacRFFaultReview detects VSWR or ArcTrip faults itself. \quad \textbf{[V]}~Therefore, a new VSWR or ArcTrip event appears in LinacRFFaultReview only after LinacRFFaultLogger has created the timestamped historical SDDS record used by the review workflow.\par
\noindent\emph{Evidence.}~\texttt{LinacRFFaultLogger\_\_chunk\_0} (primary), \texttt{LinacRFFaultReview\_\_chunk\_0} (supporting)\par

\smallskip\noindent\textbf{APS-BENCH-047}\quad {\footnotesize \textit{multi\_hop} $\cdot$ SDDS $\cdot$ hard}\par
\noindent\emph{Q.}~If you need to verify an orbit-response prediction by first multiplying the response matrix with a corrector change vector to generate expected BPM changes, and then visually compare the result page-by-page in an interactive X-windows plot, which SDDS tool would you use for the matrix product, what is the required left-hand/right-hand file ordering for that multiplication, and which plotting tool provides the zoom/pan and cursor-readout GUI for reviewing the output?\par
\noindent\emph{Gold.}~Use sddsmatrixmult to multiply the response matrix by the corrector change vector to generate expected BPM changes [SDDS\_OAG\_User's Guide for SDDS Toolkit Version 5.0\_chunk\_106\_\_chunk\_1]. For that multiplication, the left-hand matrix must be in the first file (file1) and the right-hand matrix must be in the second file (file2); in the example, response is first and correctorChange is second [SDDS\_OAG\_User's Guide for SDDS Toolkit Version 5.0\_chunk\_106\_\_chunk\_1]. The plotting tool is sddsplot, whose X-windows GUI provides zoom/pan and cursor readout for interactive review page-by-page [SDDS\_OAG\_User's Guide for SDDS Toolkit Version 5.0\_chunk\_126\_\_chunk\_0].\par
\noindent\emph{Nuggets.}~\textbf{[V]}~The SDDS tool to multiply the response matrix by the corrector change vector is sddsmatrixmult. \quad \textbf{[V]}~In sddsmatrixmult, the left-hand matrix must be supplied in the first input file (file1). \quad \textbf{[V]}~In sddsmatrixmult, the right-hand matrix must be supplied in the second input file (file2). \quad \textbf{[V]}~For this specific multiplication, the response matrix is the left-hand matrix and therefore goes in file1. \quad \textbf{[V]}~For this specific multiplication, the corrector change vector is the right-hand matrix and therefore goes in file2. \quad \textbf{[V]}~The plotting tool for interactive visual review is sddsplot. \quad [o]~sddsplot provides an X-windows GUI for reviewing the output page-by-page. \quad [o]~The sddsplot GUI includes zoom and pan capabilities. \quad [o]~The sddsplot GUI includes cursor readout.\par
\noindent\emph{Evidence.}~\texttt{SDDS\_OAG\_User's Guide for SDDS Toolkit Version 5.0\_chunk\_106\_\_chunk\_1} (primary), \texttt{SDDS\_OAG\_User's Guide for SDDS Toolkit Version 5.0\_chunk\_126\_\_chunk\_0} (primary)\par

\smallskip\noindent\textbf{APS-BENCH-048}\quad {\footnotesize \textit{multi\_hop} $\cdot$ ICMS $\cdot$ hard}\par
\noindent\emph{Q.}~Before we install and commission the new PAR terahertz synchrotron-radiation diagnostic, which APS Tech Note documents should operations review to cover both the engineering specification and the physics requirements, and how do those documents relate to the measurement capability being proposed for PAR?\par
\noindent\emph{Gold.}~Operations should review the PAR terahertz diagnostic engineering specification, DIAG-TN-2021-009 (Rev. 3), titled "Particle Accumulator Ring Terahertz Diagnostic Engineering Specification Document" [APS\_2176698\_\_chunk\_7], and the physics requirements document, DIAG-TN-2020-049 (Rev. 1), titled "Particle Accumulator Ring Terahertz Diagnostic Physics Requirements Document" [APS\_2176698\_\_chunk\_7]. These documents relate to the proposed PAR measurement capability because the design-review response states that "a new measurement capability is proposed for the Particle Accumulator Ring (PAR)" and that "Terahertz frequency synchrotron radiation is proposed" [APS\_2176698\_\_chunk\_0]. The passages do not further explain how each of the two tech notes maps onto specific measurement capabilities beyond identifying them as the engineering specification and physics requirements documents for the PAR terahertz diagnostic [APS\_2176698\_\_chunk\_7].\par
\noindent\emph{Nuggets.}~\textbf{[V]}~Operations should review APS Tech Note DIAG-TN-2021-009, Revision 3. \quad \textbf{[V]}~DIAG-TN-2021-009 (Rev. 3) is titled "Particle Accumulator Ring Terahertz Diagnostic Engineering Specification Document." \quad \textbf{[V]}~Operations should review APS Tech Note DIAG-TN-2020-049, Revision 1. \quad \textbf{[V]}~DIAG-TN-2020-049 (Rev. 1) is titled "Particle Accumulator Ring Terahertz Diagnostic Physics Requirements Document." \quad \textbf{[V]}~The proposed capability is a new measurement capability for the Particle Accumulator Ring (PAR). \quad \textbf{[V]}~The proposed measurement uses terahertz-frequency synchrotron radiation. \quad [o]~The cited passages identify the two tech notes as the engineering-specification and physics-requirements documents for the PAR terahertz diagnostic. \quad [o]~The cited passages do not further explain how each tech note maps to specific measurement capabilities beyond those document roles.\par
\noindent\emph{Evidence.}~\texttt{APS\_2176698\_\_chunk\_7} (primary), \texttt{APS\_2176698\_\_chunk\_0} (supporting)\par

\smallskip\noindent\textbf{APS-BENCH-049}\quad {\footnotesize \textit{multi\_hop} $\cdot$ BELY $\cdot$ hard}\par
\noindent\emph{Q.}~Before relying on 9-BM's new DTcorrI0 from the vortex ME4, what beam-based check should I run to confirm the DXP deadtime-correction PVs are actually working, and if the result is not flat as I change the incident flux, which Wenachee2 deadtime-correction parameters would I need to adjust?\par
\noindent\emph{Gold.}~Run a 1D dummy scan with about 0.3 s integration time per point and plot Element\_total / deadtime corrected i0 while changing the incident flux by putting in attenuators of 10\% to 50\%; if that plot stays constant, the deadtime-corrected I0 is working well [7096\_\_chunk\_0]. Before relying on 9-BM's new DTcorrI0, this beam-based check is needed because the setup was only tested without beam and the DXP PVs still need to be verified with beam [10660]. If the result is not flat as I0 changes, adjust the Wenachee2 DT-calculation parameters: change the 2.5e-6 value so the plot stays constant when i0 changes, and also change the divided-by-12 factor if you are not using 12 elements [7096\_\_chunk\_0].\par
\noindent\emph{Nuggets.}~\textbf{[V]}~Run a 1D dummy scan as the beam-based check before relying on 9-BM's new DTcorrI0. \quad [o]~Use about 0.3 s integration time per point in that scan. \quad \textbf{[V]}~Vary the incident flux during the scan by inserting attenuators spanning roughly 10\% to 50\%. \quad \textbf{[V]}~Plot Element\_total divided by deadtime-corrected I0 to evaluate whether the DXP deadtime-correction PVs are working. \quad \textbf{[V]}~If that plot remains constant as incident flux changes, the deadtime-corrected I0 is working well. \quad [o]~This beam-based verification is needed because the setup had only been tested without beam. \quad \textbf{[V]}~If the plot is not flat as I0 changes, adjust the Wenachee2 deadtime-calculation parameter currently set to 2.5e-6. \quad \textbf{[V]}~If the plot is not flat, also adjust the Wenachee2 factor that divides by 12 when the detector is not using 12 elements.\par
\noindent\emph{Evidence.}~\texttt{7096\_\_chunk\_0} (primary), \texttt{10660} (supporting)\par

\smallskip\noindent\textbf{APS-BENCH-050}\quad {\footnotesize \textit{multi\_hop} $\cdot$ BELY $\cdot$ hard}\par
\noindent\emph{Q.}~At 9-BM, once beam is back and you need to verify the new caqtdm DTcorrI0 setup for the vortex ME4, what beam-based check should you run and which parameter(s) would you adjust if the deadtime-corrected I0 does not stay constant as attenuation is changed?\par
\noindent\emph{Gold.}~Once beam is back, run a 1D dummy scan with about 0.3 sec integration time per point, plot Element\_total / deadtime corrected I0, and insert attenuators from about 10\% to 50\% to see whether the plot stays constant [7096\_\_chunk\_0]. If the deadtime-corrected I0 does not stay constant as attenuation is changed, adjust the DT calculation parameters on the Wenachee2 computer: change the 2.5e-6 value so the plot stays constant when I0 changes, and also change the divide-by-12 factor if you are not using 12 elements [7096\_\_chunk\_0]. The 9-BM note says the setup was tested without beam and that with beam you still need to test that the dxp PVs are updating [10660].\par
\noindent\emph{Nuggets.}~\textbf{[V]}~Once beam is back, run a 1D dummy scan to verify the new caqtdm DTcorrI0 setup for the vortex ME4. \quad [o]~Use about 0.3 seconds integration time per point for the dummy scan. \quad \textbf{[V]}~Plot Element\_total divided by deadtime-corrected I0 during the check. \quad \textbf{[V]}~Insert attenuators over roughly the 10\% to 50\% range during the scan. \quad \textbf{[V]}~The verification criterion is that Element\_total / deadtime-corrected I0 should stay constant as attenuation is changed. \quad \textbf{[V]}~If deadtime-corrected I0 does not stay constant with attenuation, adjust the DT calculation parameters on the Wenachee2 computer. \quad \textbf{[V]}~Change the 2.5e-6 parameter until the plot stays constant when I0 changes. \quad \textbf{[V]}~Change the divide-by-12 factor if the detector is not using 12 elements. \quad [o]~With beam, also verify that the dxp PVs are updating.\par
\noindent\emph{Evidence.}~\texttt{7096\_\_chunk\_0} (primary), \texttt{10660} (supporting)\par

}

\end{document}